\pdfoutput=1
\documentclass{quantumarticle}
\usepackage{graphicx, color}% Include figure files
\usepackage[amsthm,thmmarks]{ntheorem}
\usepackage{amsmath}
\usepackage{enumitem}
\usepackage{amssymb}
\usepackage{pstricks}
\usepackage{float}
\usepackage{multirow}
\usepackage[colorlinks=true,citecolor=blue,linkcolor=magenta]{hyperref}
\usepackage[T1]{fontenc}
\usepackage{bbm}
\usepackage{thmtools,thm-restate}
\usepackage{verbatim}
\usepackage{mathtools}
\usepackage[numbers,sort&compress]{natbib}
\usepackage[linesnumbered,ruled,vlined]{algorithm2e}
\SetKwInput{kwInit}{Init}


\makeatletter
\g@addto@macro\bfseries{\boldmath}
\makeatother

\long\def\ca#1\cb{} %Use for commenting out: \ca...\cb

\newcommand{\ketbra}[2]{| \hspace{1pt} #1 \rangle \langle #2 \hspace{1pt} |}

\newcommand{\bramatket}[3]{\langle #1 \hspace{1pt} | #2 | \hspace{1pt} #3 \rangle}
\newcommand{\bramatketq}[2]{\bramatket{#1}{#2}{#1}}
\newcommand{\norm}[2][]{#1| \! #1| #2 #1| \! #1|}

\newcommand{\dya}[1]{\ket{#1}\!\bra{#1}}
\newcommand{\ip}[2]{\langle #1|#2\rangle}      %quantum inner product
\newcommand{\HST}{\ensuremath{\mathsf{HST}}\xspace}
\newcommand{\LHST}{\ensuremath{\mathsf{LHST}}\xspace}

\newcommand{\BQP}{\ensuremath{\mathsf{BQP}}\xspace}
\newcommand{\PH}{\ensuremath{\mathsf{PH}}\xspace}
\newcommand{\DQC}{\ensuremath{\mathsf{DQC1}}\xspace}
\newcommand{\AM}{\ensuremath{\mathsf{AM}}\xspace}

\newcommand{\QIP}{\ensuremath{\mathsf{QIP}}\xspace}

\newcommand{\poly}{\operatorname{poly}}

\newcommand{\Tr}{{\rm Tr}}

\renewcommand{\geq}{\geqslant}
\renewcommand{\leq}{\leqslant}
\newcommand{\mte}[2]{\langle#1|#2|#1\rangle }

\renewcommand{\Re}{\text{Re}}
\renewcommand{\Im}{\text{Im}}

\DeclareMathOperator*{\argmin}{arg\,min}
\renewcommand{\vec}[1]{\boldsymbol{#1}}  % Bold vectors instead of arrow vectors
\newcommand{\ot}{\otimes}
\newcommand{\ad}{^\dagger}
\newcommand*{\id}{\openone}

{}%         Body font
{}%         Indent amount (empty = no indent, \parindent 
\newtheorem{example}{Example}%[subsection]

\newtheorem{theorem}{Theorem}

\newenvironment{specialproof}{\textit{Proof:}}{\hfill$\square$}

\input{Qcircuit}

\begin{document}

\title{Quantum-assisted quantum compiling}

\author{Sumeet Khatri}
%\thanks{The first three authors contributed equally to this work.}
\affiliation{Theoretical Division, Los Alamos National Laboratory, Los Alamos, NM USA.}
\affiliation{Hearne Institute for Theoretical Physics and Department of Physics and Astronomy, Louisiana State University, Baton Rouge, LA USA.}

\author{Ryan LaRose}
\affiliation{Theoretical Division, Los Alamos National Laboratory, Los Alamos, NM USA.}
\affiliation{Department of Computational Mathematics, Science, and Engineering and Department of Physics and Astronomy, Michigan State University, East Lansing, MI USA.}

\author{Alexander Poremba}
%\thanks{The first three authors contributed equally to this work.}
\affiliation{Theoretical Division, Los Alamos National Laboratory, Los Alamos, NM USA.}
\affiliation{Computing and Mathematical Sciences, California Institute of Technology, Pasadena, CA USA.}

\author{Lukasz Cincio} 
\affiliation{Theoretical Division, Los Alamos National Laboratory, Los Alamos, NM USA.}

\author{Andrew T. Sornborger} 
\affiliation{Information Sciences, Los Alamos National Laboratory, Los Alamos, NM USA.}

\author{Patrick J. Coles} 
\affiliation{Theoretical Division, Los Alamos National Laboratory, Los Alamos, NM USA.}

\date{\today}

\begin{abstract}
Compiling quantum algorithms for near-term quantum computers (accounting for connectivity and native gate alphabets) is a major challenge that has received significant attention both by industry and academia. Avoiding the exponential overhead of classical simulation of quantum dynamics will allow compilation of larger algorithms, and a strategy for this is to evaluate an algorithm's cost on a quantum computer. To this end, we propose a variational hybrid quantum-classical algorithm called quantum-assisted quantum compiling (QAQC). In QAQC, we use the overlap between a target unitary $U$ and a trainable unitary $V$ as the cost function to be evaluated on the quantum computer. More precisely, to ensure that QAQC scales well with problem size, our cost involves not only the global overlap $\Tr(V\ad U)$ but also the local overlaps with respect to individual qubits. We introduce novel short-depth quantum circuits to quantify the terms in our cost function, and we prove that our cost cannot be efficiently approximated with a classical algorithm under reasonable complexity assumptions. We present both gradient-free and gradient-based approaches to minimizing this cost. As a demonstration of QAQC, we compile various one-qubit gates on IBM's and Rigetti's quantum computers into their respective native gate alphabets. Furthermore, we successfully simulate QAQC up to a problem size of 9 qubits, and these simulations highlight both the scalability of our cost function as well as the noise resilience of QAQC. Future applications of QAQC include algorithm depth compression, black-box compiling, noise mitigation, and benchmarking.
\end{abstract}

\maketitle

%\tableofcontents

\section{Introduction}\label{sctintro}

Factoring \cite{shor1997factoring}, approximate optimization \cite{farhi2014QAOA}, and simulation of quantum systems \cite{feynman1982simulating} are some of the applications for which quantum computers have been predicted to provide speedups over classical computers. Consequently, the prospect of large-scale quantum computers has generated interest from various sectors, such as the financial and pharmaceutical industries. Currently available quantum computers are not large-scale but rather have been called noisy intermediate-scale quantum (NISQ) computers \cite{preskill2018quantum}. A proof-of-principle demonstration of quantum supremacy with a NISQ device may be coming soon \cite{preskill2012quantum, neill2017blueprint}. Nevertheless, demonstrating the practical utility of NISQ computers appears to be a more difficult task.

While improvements to NISQ hardware are continuously being made by experimentalists, quantum computing theorists can contribute to the utility of NISQ devices by developing software. This software would aim to adapt textbook quantum algorithms (e.g., for factoring or quantum simulation) to NISQ constraints. NISQ constraints include: (1) limited numbers of qubits, (2) limited connectivity between qubits, (3) restricted (hardware-specific) gate alphabets, and (4) limited circuit depth due to noise. Algorithms adapted to these constraints will likely look dramatically different from their textbook counterparts.

These constraints have increased the importance of the field of quantum compiling. In classical computing, a compiler is a program that converts instructions into assembly language so that they can be read and executed by a computer. Similarly, a quantum compiler would take a high-level algorithm and convert it into a lower-level form that could be executed on a NISQ device. Already, a large body of literature exists on classical approaches for quantum compiling, e.g., using temporal planning \cite{venturelli2018compiling,booth2018comparing}, machine learning \cite{cincio2018learning}, and other techniques \cite{maslov2008quantum, fowler2011,booth2012quantum,nam2017automated, chong2017programming, heyfron2017efficient, haner2018software, oddi2018greedy}.

A recent exciting idea is to use quantum computers themselves to train parametrized quantum circuits, as proposed in Refs. \cite{farhi2014QAOA,peruzzo2014VQE,johnson2017qvector,benedetti2018generative, mitarai2018quantum,verdon2018universal, Romero17, Romero18, Dive17}. The cost function to be minimized essentially defines the application. For example, in the variational quantum eigensolver (VQE) \cite{peruzzo2014VQE} and the quantum approximate optimization algorithm (QAOA) \cite{farhi2014QAOA}, the application is ground state preparation, and hence the cost is the expectation value of the associated Hamiltonian. Another example is training error-correcting codes \cite{johnson2017qvector}, where the cost is the average code fidelity. In light of these works, it is natural to ask: what is the relevant cost function for the application of quantum compiling?

In this work, we introduce quantum-assisted quantum compiling (QAQC, pronounced ``Quack''). The goal of QAQC is to compile a (possibly unknown) target unitary to a trainable quantum gate sequence. A key feature of QAQC is the fact that the cost is computed directly on the quantum computer. This leads to an exponential speedup (in the number of qubits involved in the gate sequence) over classical methods to compute the cost, since classical simulation of quantum dynamics is exponentially slower than quantum simulation. Consequently, one should be able to optimally compile larger-scale gate sequences using QAQC, whereas classical approaches to optimal quantum compiling will be limited to smaller gate sequences.\footnote{We note that classical compilers may be applied to large-scale quantum algorithms, but they are limited to local compiling. We thus emphasize the distinction between translating the algorithm to the native alphabet with simple, local compiling and optimal compiling. Local compiling may reach partial optimization but in order to discover the shortest circuit one may need to use a holistic approach, where the entire algorithm is considered, which requires a quantum computer for compiling.}
    
We carefully define a cost function for QAQC that satisfies the following criteria:
\begin{enumerate}
    \item It is faithful (vanishing if and only if the compilation is exact);
    \item It is efficient to compute on a quantum computer;
    \item It has an operational meaning;
    \item It scales well with the size of the problem.
\end{enumerate}
A potential candidate for a cost function satisfying these criteria is the Hilbert-Schmidt inner product between a target unitary $U$ and a trainable unitary $V$:
\begin{align}
\label{eqn1}
\langle V, U\rangle = \Tr(V\ad U).
\end{align}
It turns out, however, that this cost function does not satisfy the last criterion. We thus use Eq.~\eqref{eqn1} only for small-scale problems. For general, large-scale problems, we define a cost function satisfying all criteria. This cost involves a weighted average of the global overlap in \eqref{eqn1} with localized overlaps, which quantify the overlap between $U$ and $V$ with respect to individual qubits.

We prove that computing our cost function is $\DQC$-hard, where $\DQC$ is the class of problems that can be efficiently solved in the one-clean-qubit model of computation \cite{knill1998POOQ}. Since $\DQC$ is classically hard to simulate \cite{DBLP:journals/corr/FujiiKMNTT14}, this implies that no classical algorithm can efficiently compute our cost function. We remark that an alternative cost function might be a worst-case distance measure (such as diamond distance), but such measures are known to be \QIP-complete \cite{DBLP:journals/corr/cs-CC-0407056} and hence would violate criterion 2 in our list above. In this sense, our cost function appears to be ideal.

Furthermore, we present novel short-depth quantum circuits for efficiently computing the terms in our cost function. Our circuits achieve short depth by avoiding implementing controlled versions of $U$ and $V$, and by implementing $U$ and $V$ in parallel. We also present, in Appendix~\ref{sctGB}, circuits that compute the gradient of our cost function. One such circuit is a generalization of the well-known Power of One Qubit \cite{knill1998POOQ} that we call the Power of Two Qubits.

As a proof-of-principle, we implement QAQC on both IBM's and Rigetti's quantum computers, and we compile various one-qubit gates to the native gate alphabets used by these hardwares. To our knowledge, this is the first compilation of a target unitary with cost evaluation on actual NISQ hardware. In addition, we successfully implement QAQC on both a noiseless and noisy simulator for problems as large as 9-qubit unitaries. These larger scale implementations illustrate the scalability of our cost function, and in the case of the noisy simulator, show a somewhat surprising resilience to noise.

\begin{figure}[t]
\centering
\includegraphics[width = \columnwidth]{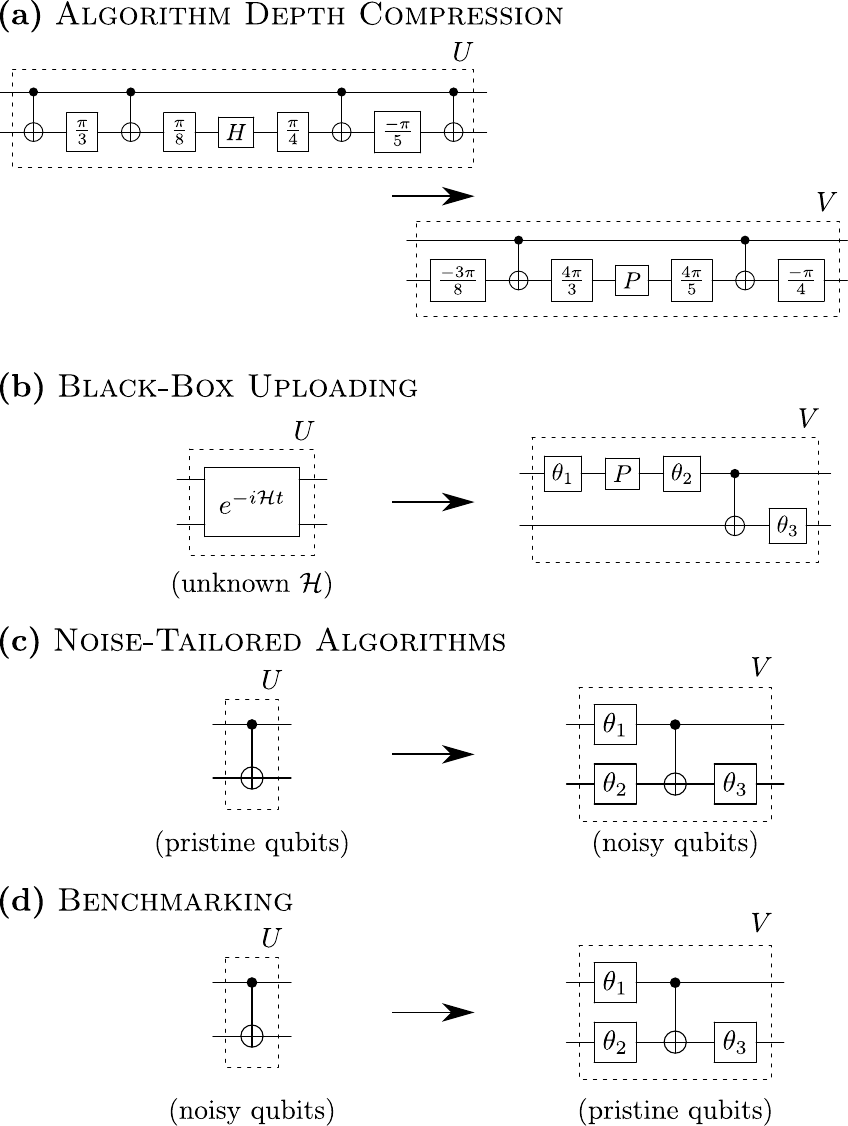}
\caption{Potential applications of QAQC. Here,  {\protect\includegraphics[height=0.35cm]{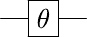}} denotes the $z$-rotation gate $R_z(\theta)$, while {\protect\includegraphics[height=0.35cm]{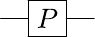}} represents the $\pi/2$-pulse given by the $x$-rotation gate $R_x(\pi/2)$. Both gates are natively implemented on commercial hardware \cite{rigetti,cross17ibm}. {\bf (a)} Compressing the depth of a given gate sequence $U$ to a shorter-depth gate sequence $V$ in terms of native hardware gates. {\bf (b)} Uploading a black-box unitary. The black box could be an analog unitary $U = e^{-i \mathcal{H} t}$, for an unknown Hamiltonian $\mathcal{H}$, that one wishes to convert into a gate sequence to be run on a gate-based quantum computer. {\bf (c)} Training algorithms in the presence of noise to learn noise-resilient algorithms (e.g., via gates that counteract the noise). Here, the unitary $U$ is performed on high-quality, pristine qubits and $V$ is performed on noisy ones. {\bf (d)} Benchmarking a quantum computer by compiling a unitary $U$ on noisy qubits and learning the gate sequence $V$ on high-quality qubits.}
\label{fig:applications}
\end{figure}

In what follows, we first discuss several applications of interest for QAQC. Section~\ref{sctQAQC} provides a general outline of the QAQC algorithm. Section~\ref{sctcircuits} presents our short-depth circuits for cost evaluation on a quantum computer. Section~\ref{sctComplexity} states that our cost function is classically hard to simulate. Sections~\ref{sctimplementations} and \ref{sctLargeImplementations}, respectively, present small-scale and larger-scale implementations of QAQC.

\section{Applications of QAQC}\label{sctapps}

Figure~\ref{fig:applications} illustrates four potential applications of QAQC. Suppose that there exists a quantum algorithm to perform some task, but its associated gate sequence is longer than desired. As shown in Fig. \ref{fig:applications}(a), it is possible to use QAQC to shorten the gate sequence by accounting for the NISQ constraints of the specific computer. This depth compression goes beyond the capabilities of classical compilers.

As a simple example, consider the quantum Fourier transform on $n$ qubits. Its textbook algorithm is written in terms of Hadamard gates and controlled-rotation gates \cite{NC00}, which may need to be compiled into the native gate alphabet. The number of gates in the textbook algorithm is $O(n^2)$, so one could use a classical compiler to locally compile each gate. But this could lead to a sub-optimal depth since the compilation starts from the textbook structure. In contrast, QAQC is unbiased with respect to the structure of the gate sequence, taking a holistic approach to compiling as opposed to a local one. Hence, in principle, it can learn the optimal gate sequence for given hardware. Note that classical compilers cannot take this holistic approach for large $n$ due to the exponential scaling of the matrix representations of the gates.

Alternatively, consider the problem of simulating the dynamics of a given quantum system with an unknown Hamiltonian $\mathcal{H}$ (via $e^{-i\mathcal{H} t}$) on a quantum computer. We call this problem black-box uploading because by simulating the black-box, i.e., the unitary $e^{-i\mathcal{H}t}$, we are ``uploading'' the unitary onto the quantum computer. This scenario is depicted in Fig.~\ref{fig:applications}(b). QAQC could be used to convert an analog black-box unitary into a gate sequence on a digital quantum computer.

Finally, we highlight two additional applications that are the opposites of each other. These two applications can be exploited when the quantum computer has some pristine qubits (qubits with low noise) and some noisy qubits. We emphasize that, in this context, ``noisy qubits'' refers to coherent noise such as systematic gate biases, where the gate rotation angles are biased in a particular direction. In contrast, we consider incoherent noise (e.g., $T_1$ and $T_2$ noise) later in this article, see Section~\ref{sct-noisy}. 

Consider Fig.~\ref{fig:applications}(c). Here, the goal is to implement a CNOT gate on two noisy qubits. Due to the noise, to actually implement a true CNOT, one has to physically implement a dressed CNOT, i.e., a CNOT surrounded by one-qubit unitaries. QAQC can be used to learn the parameters in these one-qubit unitaries. By choosing the target unitary $U$ to be a CNOT on a pristine (i.e., noiseless) pair of qubits, it is possible to learn the unitary $V$ that needs to be applied to the noisy qubits in order to effectively implement a CNOT. We call this application noise-tailored algorithms, since the learned algorithms are robust to the noise process on the noisy qubits.

Figure~\ref{fig:applications}(d) depicts the opposite process, which is benchmarking. Here, the unitary $U$ acts on a noisy set of qubits, and the goal is to determine what the equivalent unitary $V$ would be if it were implemented on a pristine set of qubits. This essentially corresponds to learning the noise model, i.e., benchmarking the noisy qubits.

\section{The QAQC Algorithm}\label{sctQAQC}

\subsection{Approximate compiling}\label{sctapprox}

The goal of QAQC is to take a (possibly unknown) unitary $U$ and return a gate sequence $V$, executable on a quantum computer, that has approximately the same action as $U$ on any given input state (up to possibly a global phase factor). The notion of approximate compiling \cite{K97,DN06,PVH13,KMM13,KBS14,ZAM18} requires an operational figure-of-merit that quantifies how close the compilation is to exact. A natural candidate is the probability for the evolution under $V$ to mimic the evolution under $U$. Hence, consider the overlap between $\ket{\psi(U)}\coloneqq U \ket{\psi}$ and $\ket{\psi(V)}\coloneqq V \ket{\psi}$, averaged over all input states $\ket{\psi}$. This is the fidelity averaged over the Haar distribution,
\begin{equation}
    \overline{F}(U,V)\coloneqq\int_{\psi}| \ip{\psi(V)}{\psi(U)}|^2~\text{d}\psi\,.
%    \overline{F}(U,V)\coloneqq\int_{\psi}|\bra{\psi}V^\dagger U\ket{\psi}|^2~\text{d}\psi,
\end{equation}
We call $V$ an \textit{exact compilation of $U$} if $\overline{F}(U,V) = 1$. If $\overline{F}(U,V)\geq 1-\varepsilon$, where $\varepsilon\in[0,1]$, then we call $V$ an \textit{$\varepsilon$-approximate compilation of $U$}, or simply an \textit{approximate compilation of $U$}. 

As we will see, the quantity $\overline{F}(U,V)$ has a connection to our cost function, defined below, and hence our cost function has operational relevance to approximate compiling. Minimizing our cost function is related to maximizing $\overline{F}(U,V)$, and thus is related to compiling to a better approximation. 

QAQC achieves approximate compiling by training a gate sequence $V$ of a fixed length $L$, which may even be shorter than the length required to exactly compile $U$. As one increases $L$, one can further minimize our cost function. The length $L$ can therefore be regarded as a parameter that can be tuned to obtain arbitrarily good approximate compilations of $U$. %This parameter $L$ is also directly related the algorithm depth compression application of QAQC, as described in Sec. \ref{sctapps} and illustrated in Fig. \ref{fig:applications}.

\begin{figure}
    \centering
    \includegraphics[width=\columnwidth]{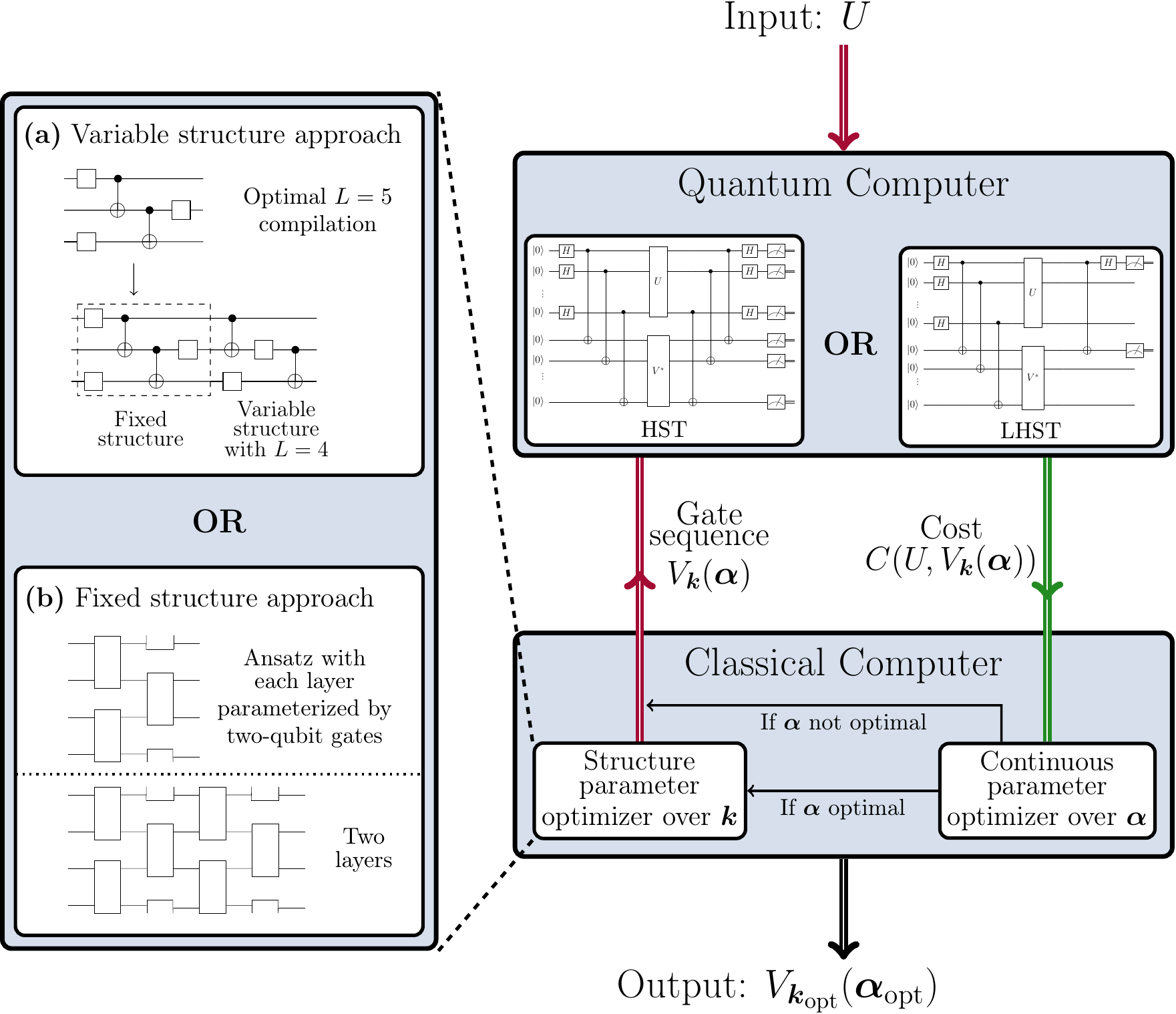}
    \caption{Outline of our variational hybrid quantum-classical algorithm, in which we optimize over gate structures and continuous gate parameters in order to perform QAQC for a given input unitary $U$. We take two approaches towards structure optimization: \textbf{(a)} For small problem sizes, we allow the gate structure to vary for a given gate sequence length $L$, which in general leads to an approximate compilation of $U$. To obtain a better approximate compilation, the best structure obtained can be concatenated with a new sequence of a possibly different length, whose structure can vary. For each iteration of the continuous parameter optimization, we calculate the cost using the Hilbert-Schmidt Test (HST); see Sec. \ref{sctHST}. \textbf{(b)} For large problem sizes, we fix the gate structure using an ansatz consisting of layers of two-qubit gates. By increasing the number of layers, we can obtain better approximate compilations of $U$. For each iteration of the continuous parameter optimization, we calculate the cost using the Local Hilbert-Schmidt Test (LHST); see Sec. \ref{sctlocalHST}. }
    %The optimization starts with a random choice of gate structure, denoted by $\vec{k}$, followed by a continuous optimization over the internal gate parameters $\vec{\alpha}$ in the trainable unitary $V_{\vec{k}}(\vec{\alpha})$. After obtaining the optimal gate parameters, as determined by minimizing the cost function $C(U,V_{\vec{k}}(\vec{\alpha}))$, the structure is updated and the internal gate parameters are optimized again, followed by another structure update. This process repeats until the cost reaches its minimum.[FIGURE AND CAPTION TO BE UPDATED]}
    \label{fig:optimization_flow}
\end{figure}

\subsection{Discrete and continuous parameters}\label{sctparameters}

The gate sequence $V$ should be expressed in terms of the native gates of the quantum computer being used. Consider an alphabet $\mathcal{A}=\{G_k(\alpha)\}_k$ of gates $G_k(\alpha)$ that are native to the quantum computer of interest. Here, $\alpha\in\mathbb{R}$ is a continuous parameter, and $k$ is a discrete parameter that identifies the type of gate and which qubits it acts on. For a given quantum computer, the problem of compiling $U$ to a gate sequence of length $L$ is to determine
\begin{equation}\label{eq:opt_params}
(\vec{\alpha}_{\text{opt}},\vec{k}_{\text{opt}})\coloneqq \argmin_{(\vec{\alpha},\vec{k})} C(U,V_{\vec{k}}(\vec{\alpha})),
\end{equation}
where
\begin{equation}
    V_{\vec{k}}(\vec{\alpha})=G_{k_L}(\alpha_L)G_{k_{L-1}}(\alpha_{L-1})\dotsb G_{k_1}(\alpha_1)
\end{equation}
is the trainable unitary. Here, $V_{\vec{k}}(\vec{\alpha})$ is a function of the sequence $\vec{k}=(k_1,\dotsc, k_L)$ of parameters describing which gates from the native gate set are used and of the continuous parameters $\vec{\alpha}=(\alpha_1,\dotsc,\alpha_L)$ associated with each gate. The function $C(U,V_{\vec{k}}(\vec{\alpha}))$ is the cost, which quantifies how close the trained unitary is to the target unitary. We define the cost below to have the properties: $0\leq C(U,V)\leq 1$ for all unitaries $U$ and $V$, and $C(U,V)=0$ if and only if $U=V$ (possibly up to a global phase factor). %The minimum value of the cost is therefore zero.

The optimization in \eqref{eq:opt_params} contains two parts: discrete optimization over the finite set of gate structures parameterized by $\vec{k}$, and continuous optimization over the parameters $\vec{\alpha}$ characterizing the gates within the structure. Our quantum-classical hybrid strategy to perform the optimization in \eqref{eq:opt_params} is illustrated in Fig. \ref{fig:optimization_flow}. In the next subsection, we present a general, ansatz-free approach to optimizing our cost function, which may be useful for systems with a small number of qubits. In the subsection following that, we present an ansatz-based approach that would allow the extension to larger system sizes. In each case, we perform the continuous parameter optimization using gradient-free methods as described in Appendix \ref{sctGF}. We also discuss a method for gradient-based continuous parameter optimization in Appendix \ref{sctGB}.

\subsection{Small problem sizes}\label{sctsmall}

Suppose $U$ and $V$ act on a $d$-dimensional space of $n$ qubits, so that $d=2^n$. To perform the continuous parameter optimization in \eqref{eq:opt_params}, we define the cost function
\begin{equation}\label{eq:GF-cost}
    \begin{aligned}
    C_{\text{HST}}(U,V)&\coloneqq 1-\frac{1}{d^2}\left|\left<V,U\right>\right|^2\\
    &=1-\frac{1}{d^2}|\Tr(V^\dagger U)|^2,
    \end{aligned}
\end{equation}
where HST stands for ``Hilbert-Schmidt Test'' and refers to the circuit used to evaluate the cost, which we introduce in Sec.~\ref{sctHST}. Note that the quantity $\frac{1}{d^2}\left|\left<V,U\right>\right|^2$ is simply the fidelity between the pure states obtained by applying $U$ and $V$ to one half of a maximally entangled state. Consequently, it has an operational meaning in terms of $\overline{F}(U,V)$. Indeed, it can be shown \cite{HHH99,nielsen02} that
\begin{equation}
\label{eq:HST_Fbar}
C_{\text{HST}}(U,V) = \frac{d+1}{d}\left(1-\overline{F}(U,V)\right)\,.
%    \frac{1}{d^2}|\Tr(V^\dagger U)|^2=\frac{(d+1)\overline{F}(U,V)-1}{d}.
\end{equation}
Also note that for any two unitaries $U$ and $V$, $C_{\text{HST}}(U,V)=0$ if and only if $U$ and $V$ differ by a global phase factor, i.e., $V=e^{i\varphi}U$ for some $\varphi\in\mathbb{R}$. By minimizing $C_{\text{HST}}$, we thus learn an equivalent unitary $V$ up to a global phase.

Now, to perform the optimization over gate structures in \eqref{eq:opt_params}, one strategy is to search over all possible gate structures for a gate sequence length $L$, which can be allowed to vary during the optimization. As the set of gate structures grows exponentially with the number of gates $L$, such a brute force search over all gate structures in order to obtain the best one is intractable in general. To efficiently search through this exponentially large space, we adopt an approach based on simulated annealing. (An alternative approach is genetic optimization, which has been implemented previously to classically optimize quantum gate sequences \cite{Gepp2009}.)

Our simulated annealing approach starts with a random gate structure, then performs continuous optimization over the parameters $\vec{\alpha}$ that characterize the gates in order to minimize the cost function. We then perform a structure update that involves randomly replacing a subset of gates in the sequence with new gates (which can be done in a way such that the sequence length can increase or decrease) and re-optimizing the cost function over the continuous parameters $\vec{\alpha}$. If this structure change produces a lower cost, then we accept the change. If the cost increases, then we accept the change with probability decreasing exponentially in the magnitude of the cost difference. We iterate this procedure until the cost converges or until a maximum number of iterations is reached.

With a fixed gate sequence length $L$, the approach outlined above will in general lead to an approximate compilation of $U$, which in many cases is sufficient. One strategy for obtaining better and better approximate compilations of $U$ is a layered approach illustrated in Fig. \ref{fig:optimization_flow}(a). In this approach, we consider a particular gate sequence length $L$ and perform the full structure optimization, as outlined above, to obtain an (approximate) length-$L$ compilation of $U$. The optimal gate sequence structure thus obtained can then be concatenated with a new sequence of a possibly different (but fixed) length, whose structure can vary. By performing the continuous parameter optimization over the entire longer gate sequence, and performing the structure optimization over the new additional segment of the gate sequence, we can obtain a better approximate compilation of $U$. Iterating this procedure can then lead to increasingly better approximate compilations of $U$.

\subsection{Large problem sizes}\label{sctlarge}

We emphasize two potential issues with scaling the above approach to large problem sizes.

First, one may want a guarantee that there exists an exact compilation of $U$ within a polynomial size search space for $V$. When performing full structure optimization, as above, the search space size grows exponentially in the length $L$ of the gate sequence. This implies that the search space size grows exponentially in $n$, if one chooses $L$ to grow polynomially in $n$. Indeed, one would typically require $L$ to grow polynomially in $n$ if one is interested in exact compilation, since the number of gates in $U$ itself grows polynomially in $n$ for many applications. (Note that this issue arises if one insists on exact, instead of approximate, compiling.)

Second, and arguably more importantly, the cost $C_{\text{HST}}(U,V)$ is exponentially fragile. The inner product between $U$ and $V$ will be exponentially suppressed for random choices of $V$, which means that $C_{\text{HST}}(U,V)$ will be very close to one for most unitaries $V$. Hence, for random unitaries $V$, the number of calls to the quantum computer needed to resolve differences in the cost $C_{\text{HST}}(U,V)$ to a given precision will grow exponentially.

The first issue can be addressed with an efficiently parameterized ansatz for $V$. With an ansatz, only the continuous parameters $\vec{\alpha}$ need to be optimized in $V$. The $\vec{k}$ parameters are fixed, which means that structure updates are not required. This fixed structure approach is depicted in Fig.~\ref{fig:optimization_flow}(b). One can choose an ansatz such that the number of parameters needed to represent the target unitary $U$ is only a polynomial function of $n$. Hence, one should allow the ansatz $A(U)$ to be application specific, i.e., to be a function of $U$. As an example, if $U = e^{-i\mathcal{H}t}$ for a local Hamiltonian $\mathcal{H}$, one could choose the ansatz to involve a polynomial number of local interactions. Due to the application-specific nature of the ansatz, the problem is a complex one, hence we leave the issue of finding efficient ansatzes for future work.

\begin{figure}
    \centering
    \includegraphics[scale=0.45]{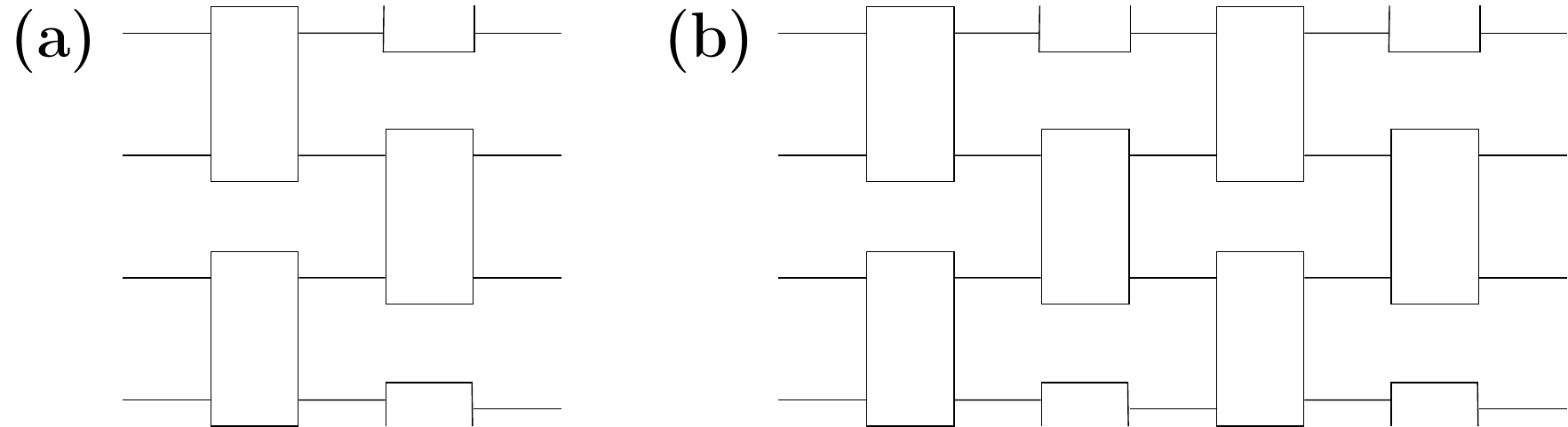}
    \caption{\textbf{(a)} One layer of the ansatz for the trainable unitary $V$ in the case of four qubits. The gate sequence in the layer consists of a two-qubit gate acting on the first and second qubits, the third and fourth qubits, the second and third qubits, and the first and fourth qubits. \textbf{(b)} The full ansatz defining the trainable unitary $V$ consists of a particular number $\ell$ of the layer in (a). Shown is two layers in the case of four qubits.}
    \label{fig:ansatz}
\end{figure}

Nevertheless, we show a concrete example of a potential ansatz for $V$ in Fig.~\ref{fig:ansatz}. The ansatz is defined by a number $\ell$ of layers, with each layer being a gate sequence of depth two consisting of two-qubit gates acting on neighboring qubits. Consider the following argument. In QAQC, the unitary $U$ to be compiled is executed on the quantum computer, so it must be efficiently implementable, i.e., the gate count is polynomial in $n$. Next, note that the gate sequence used to implement $U$ can be compiled into in the ansatz in Fig.~\ref{fig:ansatz} with only polynomial overhead. This implies that the ansatz in Fig.~\ref{fig:ansatz} could exactly describe $U$ in only a polynomial number of layers and would hence eliminate the need to search through an exponentially large space. We remark that the ansatz in Fig.~\ref{fig:ansatz} may be particularly useful for applications involving compiling quantum simulations of physically relevant systems, as the structure resembles that of the Suzuki-Trotter decomposition \cite{SUZUKI1990319} for nearest-neighbor Hamiltonians.

Let us now consider the second issue mentioned above: the exponentially suppressed inner product between $U$ and $V$ for large $n$. To address this, we propose an alternative cost function involving a weighted average between the function in \eqref{eq:GF-cost} and a ``local'' cost function:
\begin{equation}\label{eq-cost_weighted}
    C_q(U,V)\coloneqq qC_{\text{HST}}(U,V)+(1-q)C_{\text{LHST}}(U,V),
\end{equation}
where $0\leq q \leq 1$ and
\begin{equation}\label{eq-LHST}
    C_{\text{LHST}}(U,V)\coloneqq \frac{1}{n}\sum_{j=1}^nC_{\text{LHST}}^{(j)}(U,V) = 1 - \overline{F_e}.
\end{equation}
Here, LHST stands for ``Local Hilbert-Schmidt Test'', referring to the circuit discussed in Sec.~\ref{sctlocalHST} that is used to compute this function. Also, $\overline{F_e}\coloneqq \frac{1}{n}\sum_{j=1}^n F_e^{(j)}$, where the quantities $F_e^{(j)}$ are entanglement fidelities (hence the notation $F_e$) of local quantum channels $\mathcal{E}_j$ defined in Sec.~\ref{sctlocalHST}. Hence, $C_{\text{LHST}}(U,V)$ is a sum of local costs, where each local cost is written as a local entanglement fidelity: $C_{\text{LHST}}^{(j)}(U,V) = 1-F_e^{(j)} $. Expressing the overall cost as sum of local costs is analogous to what is done in the variational quantum eigensolver \cite{peruzzo2014VQE}, where the overall energy is expressed as a sum of local energies. The functions $C_{\text{LHST}}^{(j)}$ are local in the sense that only two qubits need to be measured in order to calculate each one of them. This is unlike the function $C_{\text{HST}}$, whose calculation requires the simultaneous measurement of $2n$ qubits.

The cost function $C_q$ in \eqref{eq-cost_weighted} is a weighted average between the ``global'' cost function $C_{\text{HST}}$ and the local cost function $C_{\text{LHST}}$, with $q$ representing the weight given to the global cost function. The weight $q$ can be chosen according to the size of the problem: for a relatively small number of qubits, we would let $q=1$. As the number of qubits increases, we would slowly decrease $q$ to mitigate the suppression of the inner product between $U$ and $V$.

To see why $C_{\text{LHST}}$ can be expected to deal with the issue of an exponentially suppressed inner product for large $n$, consider the following example. Suppose the unitary $U$ to be compiled is the tensor product $U=U_1\otimes U_2\otimes\dotsb\otimes U_n$ of unitaries $U_j$ acting on qubit $j$, and suppose we take the tensor product $V=V_1\otimes V_2\otimes\dotsb\otimes V_n$ as the trainable unitary. We get that $C_{\text{HST}}(U,V)=1-\prod_{j=1}^n r_j$, where $r_j =(1/4)|\Tr(V_j^\dagger U_j)|^2$. Since each $r_j$ will likely be less than one for a random choice of $V_j$, then their product will be small for large $n$. Consequently a very large portion of the cost landscape will have $C_{\text{HST}}(U,V)\approx 1$ and hence will have a vanishing gradient. However, the cost function $C_{\text{LHST}}$ is defined such that $C_{\text{LHST}}(U,V)=1-\frac{1}{n}\sum_{j=1}^n r_j$, so that we obtain an average of the $r_j$ quantities rather than a product. Taking the average instead of the product leads to a gradient that is not suppressed for large $n$. 

More generally, for any $U$ and $V$, the quantity $\overline{F_e}$, which is responsible for the variability in $C_{\text{LHST}}$, can be made non-vanishing by adding local unitaries to $V$. In particular, for a given $U$ and $V$, it is straightforward to show that for all $j\in\{1,2,\dotsc,n\}$ there exists a unitary $V_j$ acting on qubit $j$ such that $F_e^{(j)}\geq\frac{1}{4}$ for the gate sequence given by $V' = V_j V$. In other words, there exists a local unitary $V_j$ that can be added to the trainable gate sequence $V$ such that $C_{\text{LHST}}^{(j)}(U,V_jV)\leq\frac{3}{4}$. This implies that, with the appropriate local unitary applied to each qubit at the end of the trainable gate sequence, the local cost function $C_{\text{LHST}}$ can always be decreased to no greater than $\frac{3}{4}$. Note that local unitaries cannot be used in this way to decrease the global cost function $C_{\text{HST}}$, i.e., to make the second term in \eqref{eq:GF-cost} non-vanishing.

Finally, one can show (See Appendix~\ref{sctConverseBound}) that $C_{\text{LHST}}\geq (1/n)C_{\text{HST}}$. Combining this with Eq.~\eqref{eq:HST_Fbar} gives
\begin{align}
    C_q(U,V)&\geq \left(\frac{1-q+nq}{n}\right)\left(\frac{d+1}{d}\right)(1-\overline{F}(U,V))\,,
\end{align}
which implies that
\begin{equation}
    \overline{F}(U,V)\geq 1-\left(\frac{n}{1-q+nq}\right)\left(\frac{d}{d+1}\right)C_q(U,V).
\end{equation}
Hence, the cost function $C_q$ retains an operational meaning for the task of approximate compiling, since it provides a bound on the average fidelity between $U$ and $V$.

\subsection{Special case of a fixed input state}

An important special case of quantum compiling is when the target unitary $U$ happens to appear at the beginning of one's quantum algorithm, and hence the state that one inputs to $U$ is fixed. For many quantum computers, this input state is $\ket{\psi_0} = \ket{0}^{\otimes n}$. We emphasize that many use cases of QAQC do not fall under this special case, since one is often interested in compiling unitaries that do not appear at the beginning of one's algorithm. For example, one may be interested in the optimal compiliation of a controlled-unitary, but such a unitary would never appear at the beginning of an algorithm since its action would be trivial. Nevertheless we highlight this special case because QAQC can potentially be simplified in this case. In addition, this special case was very recently explored in Ref.~\cite{jones2018quantum} after the completion of our article.

In this special scenario, a natural cost function would be
\begin{equation}
\label{eqnFixedInput1}
    C_{\text{fixed input}} = 1 - |\mte{\psi_0}{UV\ad}|^2\,.
\end{equation}
This could be evaluated on a quantum computer in two possible ways. One way is to apply $U$ and then $V\ad$ to the $\ket{\psi_0}$ state and then measure the probability to be in the $\ket{\psi_0}$ state. Another way is to apply $U$ to one copy of $\ket{\psi_0}$ and $V$ to another copy of $\ket{\psi_0}$, and then measure the overlap \cite{cincio2018learning, garcia2013swap} between these two states.

However, this cost function would not scale well for the same reason discussed above that our $C_{\text{HST}}$ cost does not scale well, i.e., its gradient can vanish exponentially. Again, one can fix this issue with a local cost function. Assuming $\ket{\psi_0} = \ket{0}^{\otimes n}$, this local cost can take the form:
\begin{equation}
\label{eqnFixedInput2}
    C_{\text{fixed input}}^{\text{local}} = 1 - \frac{1}{n}\sum_{j=1}^n p_0^{(j)}\,,
\end{equation}
where
\begin{equation}
\label{eqnFixedInput3}
    p_0^{(j)} = \Tr[(\dya{0}_j \ot \id) V\ad U \dya{\psi_0}U\ad V]
\end{equation}
is the probability to obtain the zero measurement outcome on qubit $j$ for the state $V\ad U \ket{\psi_0}$. 

We remark that the two cost functions in \eqref{eqnFixedInput1} and \eqref{eqnFixedInput2} can each be evaluated with quantum circuits on only $n$ qubits. This is in contrast to $C_{\text{HST}}$ and $C_{\text{LHST}}$, whose evaluation involves quantum circuits with $2n$ qubits (see the next section for the circuits). This reduction in resource requirements is the main reason why we highlight this special case.

\section{Cost evaluation circuits}\label{sctcircuits}

In this section, we present short-depth circuits for evaluating the functions in \eqref{eq:GF-cost} and \eqref{eq-LHST} and hence for evaluating the overall cost in \eqref{eq-cost_weighted}. We note that these circuits are also interesting outside of the scope of QAQC, and they likely have applications in other areas.

In addition, in Appendix~\ref{sctGB}, we present circuits for computing the gradient of the cost function, including a generalization of the Power-of-one-qubit circuit \cite{knill1998POOQ} that computes both the real and imaginary parts of $\langle U , V \rangle$.

\subsection{Hilbert-Schmidt Test}\label{sctHST}

\begin{figure}
\centering
\includegraphics[scale=0.80]{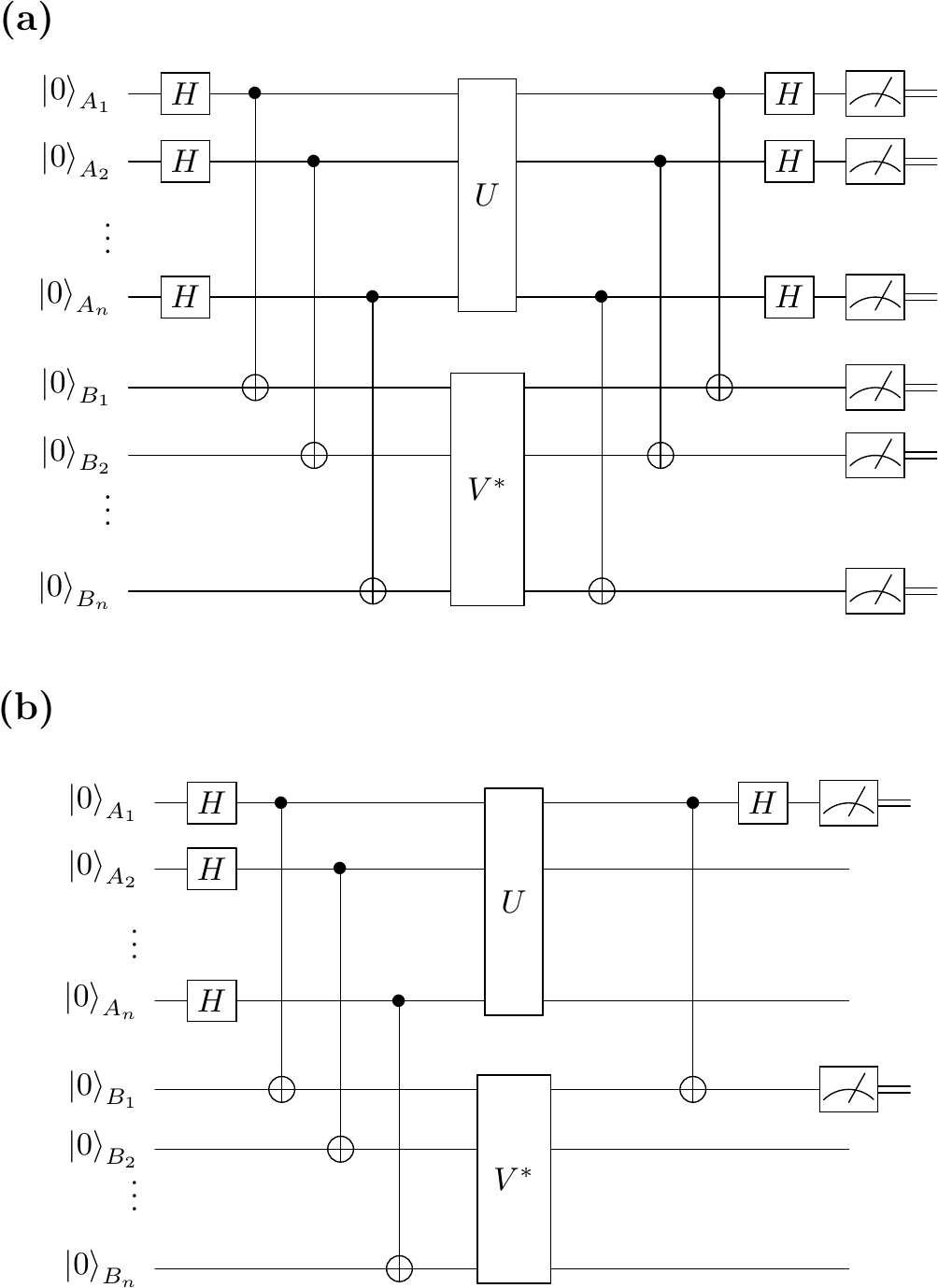}~
\caption{\textbf{(a)} The Hilbert-Schmidt Test. For this circuit, the probability to obtain the measurement outcome in which all $2n$ qubits are in the $\ket{0}$ state is equal to $(1/d^2)|\Tr(V^\dagger U)|^2$. Hence, this circuit computes the magnitude of the Hilbert-Schmidt inner product, $|\langle V,U\rangle|$, between $U$ and $V$. \textbf{(b)} The Local Hilbert-Schmidt Test, which is the same as the Hilbert-Schmidt Test except that only two of the $2n$ qubits are measured at the end. Shown is the measurement of the qubits $A_1$ and $B_1$, and the probability that both qubits are in the state $\ket{0}$ is given by \eqref{eq-LHST_prob} with $j=1$.}
\label{fig:hilbert-schmidt-inner-product-circuit}
\end{figure}

Consider the circuit in Fig.~\ref{fig:hilbert-schmidt-inner-product-circuit}(a). Below we show that this circuit computes $|\Tr(V\ad U)|^2$, where $U$ and $V$ are $n$-qubit unitaries. The circuit involves $2 n$ qubits, where we call the first (second) $n$-qubit system $A$ ($B$).

The first step in the circuit is to create a maximally entangled state between $A$ and $B$, namely, the state
\begin{align}
\label{eqn2}
\ket{\Phi^+}_{AB} = \frac{1}{\sqrt{d}}\sum_{\vec{j}} \ket{\vec{j}}_A \otimes \ket{\vec{j}}_B\,,
\end{align}
where $\vec{j} = (j_1, j_2,...,j_n)$ is a vector index in which each component $j_k$ is chosen from $\{0,1\}$. The first two gates in Fig.~\ref{fig:hilbert-schmidt-inner-product-circuit}(a)---the Hadamard gates and the CNOT gates (which are performed in parallel when acting on distinct qubits)---create the $\ket{\Phi^+}$ state.

The second step is to act with $U$ on system $A$ and with $V^*$ on system $B$. ($V^*$ is the complex conjugate of $V$, where the complex conjugate is taken in the standard basis.) Note that these two gates are performed in parallel. This gives the state
\begin{align}
\label{eqn3}
(U\otimes V^*)\ket{\Phi^+}_{AB} = \frac{1}{\sqrt{d}}\sum_{\vec{j}} U\ket{\vec{j}}_A \otimes V^*\ket{\vec{j}}_B\,.
\end{align}
We emphasize that the unitary $V^*$ is implemented on the quantum computer, not $V$ itself. (See Appendix~\ref{app:Vstar} for elaboration on this point.)

The third and final step is to measure in the Bell basis. This corresponds to undoing the unitaries (the CNOTs and Hadamards) used to prepare $\ket{\Phi^+}$ and then measuring in the standard basis. At the end, we are only interested in estimating a single probability: the probability for the Bell-basis measurement to give the $\ket{\Phi^+}$ outcome, which corresponds to the all-zeros outcome in the standard basis. The amplitude associated with this probability is
\begin{align}
\bramatketq{\Phi^+}{U\otimes V^*}&=\bramatketq{\Phi^+}{UV^\dagger\otimes \id}\\&=\frac{1}{d}\Tr(V^\dagger U)\,.\label{eq-HS_inner_prod}
\end{align}
To obtain the first equality we used the ricochet property:
\begin{equation}\label{eq:ricochet_prop}
    \id\otimes X\ket{\Phi^+}=X^T\otimes \id\ket{\Phi^+},
\end{equation}
which holds for any operator $X$ acting on a $d$-dimensional space. The probability of the $\ket{\Phi^+}$ outcome is then the absolute square of the amplitude, i.e., $(1/d^2)|\Tr(V^\dagger U)|^2$. Hence, this probability gives us the absolute value of the Hilbert-Schmidt inner product between $U$ and $V$. We therefore call the circuit in Fig.~\ref{fig:hilbert-schmidt-inner-product-circuit}(a) the Hilbert-Schmidt Test (HST).

Consider the depth of this circuit. Let $D(G)$ denote the depth of a gate sequence $G$ for a fully-connected quantum computer whose native gate alphabet includes the CNOT gate and the set of all one-qubit gates. Then, for the HST, we have
\begin{align}
\label{eqn6}
D(\text{HST}) = 4 + \max\{D(U),D(V^*)\}\,.
\end{align}
The first term of 4 is associated with the Hadamards and CNOTs in Fig.~\ref{fig:hilbert-schmidt-inner-product-circuit}(a), and this term is negligible when the depth of $U$ or $V^*$ is large. The second term results from the fact that $U$ and $V^*$ are performed in parallel. Hence, whichever unitary, $U$ or $V^*$, has the larger depth will determine the overall depth of the HST.

\subsection{Local Hilbert-Schmidt Test}\label{sctlocalHST}

Let us now consider a slightly modified form of the HST, shown in Fig.~\ref{fig:hilbert-schmidt-inner-product-circuit}(b). We call this the Local Hilbert-Schmidt Test (LHST) because, unlike the HST in Fig.~\ref{fig:hilbert-schmidt-inner-product-circuit}(a), only two of the total number $2n$ of qubits are measured: one qubit from system $A$, say $A_j$, and the corresponding qubit $B_j$ from system $B$, where $j\in\{1,2,\dotsc,n\}$.

The state of systems $A$ and $B$ before the measurements is given by Eq. \eqref{eqn3}. Using the ricochet property in \eqref{eq:ricochet_prop} as before, we obtain  
\begin{align}
    (U\otimes V^*)\ket{\Phi^+}_{AB}&=(UV^\dagger\otimes\id)\ket{\Phi^+}_{AB}\\
    &=(W\otimes\id)\ket{\Phi^+}_{AB},\label{eq-LHST_pf1}
\end{align}
where $W:=UV^\dagger$. Let $\bar{A_j}$ denote all systems $A_k$ except for $A_j$, and let $\bar{B_j}$ denote all systems $B_k$ except for $B_j$. Taking the partial trace over $\bar{A_j}$ and $\bar{B_j}$ on the state in \eqref{eq-LHST_pf1} gives us the following state on the qubits $A_j$ and $B_j$ that are being measured:

\begin{align}
    &\Tr_{\bar{A_j}\bar{B_j}}((W_A\otimes\id_B)\ket{\Phi^+}\bra{\Phi^+}_{AB}(W_A^\dagger\otimes\id_B))\nonumber\\
    &\quad =\Tr_{\bar{A_j}}\left((W_A\otimes\id_{B_j})\left(\ket{\Phi^+}\bra{\Phi^+}_{A_jB_j}\otimes\frac{\id_{\bar{A_j}}}{2^{n-1}}\right)\right.\nonumber\\
\label{eq-local_channel1}    &\qquad\qquad\qquad\qquad\qquad\qquad\qquad\left.\times(W_A^\dagger\otimes\id_{B_j})\right)\\
\label{eq-local_channel2}
&\quad =(\mathcal{E}_j\otimes\mathcal{I}_{B_j})(\ket{\Phi^+}\bra{\Phi^+}_{A_jB_j})\,.
\end{align}
In \eqref{eq-local_channel1}, $\ket{\Phi^+}_{A_jB_j}$ is a 2-qubit maximally entangled state of the form in \eqref{eqn2}. In \eqref{eq-local_channel2}, we have defined the channel $\mathcal{E}_j$ by
\begin{equation}\label{eq-HST_local_channel}
    \mathcal{E}_j(\rho_{A_j})\coloneqq\Tr_{\bar{A_j}}\left(W_A\left(\rho_{A_j}\otimes\frac{\id_{\bar{A_j}}}{2^{n-1}}\right)W_A^\dagger\right).
\end{equation}
The probability of obtaining the $(0,0)$ outcome in the measurement of $A_j$ and $B_j$ is the overlap of the state in \eqref{eq-local_channel2} with the $\ket{\Phi^+}_{A_jB_j}$ state, given by
\begin{equation}\label{eq-LHST_prob}
 F_e^{(j)} :=  \Tr\left(\ket{\Phi^+}\bra{\Phi^+}_{A_jB_j}(\mathcal{E}_{j}\otimes\mathcal{I}_{B_j})(\ket{\Phi^+}\bra{\Phi^+}_{A_jB_j})\right).
\end{equation}
Note that this is the entanglement fidelity of the channel $\mathcal{E}_j$. We use these entanglement fidelities (for each $j$) to define the local cost function $C_{\text{LHST}}(U,V)$ as
\begin{equation}\label{eq-LHST_2}
    C_{\text{LHST}}(U,V)= \frac{1}{n}\sum_{j=1}^nC_{\text{LHST}}^{(j)}(U,V),
\end{equation}
where
\begin{equation}\label{eq-local_cost_j}
    \begin{aligned}
    C_{\text{LHST}}^{(j)}(U,V)&\coloneqq 1- F_e^{(j)}\,.
    \end{aligned}
\end{equation}
Note that for all $j\in\{1,2,\dotsc, n\}$, the maximum value of $F_e^{(j)}$ is one, which occurs when $\mathcal{E}_j$ is the identity channel. This means that the minimum value of $C_{\text{LHST}}(U,V)$ is zero. In Appendix~\ref{app:LHST_faithful}, we show that $C_{\text{LHST}}$ is indeed a faithful cost function:

\begin{restatable}{proposition}{PropositionLHSTfaithfulness}\label{prop-LHST-faithfulness}
For all unitaries $U$ and $V$, it holds that $C_{\text{LHST}}(U,V)=0$ if and only if $U=V$ (up to a global phase).
\end{restatable}

The cost function $C_{\text{LHST}}$ is simply the average of the probabilities that the two qubits $A_j B_j$ are not in the $\ket{00}$ state, while the cost function $C_{\text{HST}}$ is the probability that all qubits are not in the $\ket{0}^{\ot 2n}$ state. Since the probability of an intersection of events is never greater than the average of the probabilities of the individual events, we find that
\begin{equation}\label{eqnDirectBound}  
    C_{\text{LHST}}(U,V)\leq C_{\text{HST}}(U,V)
\end{equation}
for all unitaries $U$ and $V$. Furthermore, we can also formulate a bound in the reverse direction 
\begin{equation}\label{eqnConverseBound}  
   n C_{\text{LHST}}(U,V) \geq C_{\text{HST}}(U,V).
\end{equation}
In Appendix~\ref{sctConverseBound}, we offer a proof for the above bounds.
\begin{restatable}{proposition}{TheoremConverseBound}  \label{TheoremConverseBound} 
Let $U$ and $V$ be $2^n \times 2^n$ unitaries. Then,
\begin{equation*}
   C_{\text{LHST}}(U,V) \leq  C_{\text{HST}}(U,V) \leq n C_{\text{LHST}}(U,V)\,.
\end{equation*}
\end{restatable}
The depth of the circuit in Fig. \ref{fig:hilbert-schmidt-inner-product-circuit}(b) used to compute the cost function $C_{\text{LHST}}$ is the same as the depth of the circuit in Fig. \ref{fig:hilbert-schmidt-inner-product-circuit}(a) used to compute $C_{\text{HST}}$, namely,
\begin{equation}
    D(\text{LHST})=4+\max\{D(U),D(V^*)\}.
\end{equation}

\begin{comment}
We now split the state $\ket{\Phi^+}_{AB}$ as $\ket{\Phi^+}_{AB}=\ket{\Phi^+}_{B_jA_j}\otimes\ket{\Phi^+}_{\bar{A_j}\bar{B_j}}$, where $\bar{A_j}$ denotes all qubits $A_1,A_2,\dotsc,A_n$ of system $A$ except for $A_j$, and similarly for $\bar{B_j}$. The state of the systems before the measurement is thus
\begin{equation}
    \begin{aligned}
    &(\id_{B_j}\otimes W_{A_j\bar{A_j}}\otimes\id_{\bar{B_j}})\\
    &\quad\times(\ket{\Phi^+}\bra{\Phi^+}_{B_jA_j}\otimes\ket{\Phi^+}\bra{\Phi^+}_{\bar{A_j}\bar{B_j}})\\
    &\quad\times(\id_{B_j}\otimes W_{A_j\bar{A_j}}^\dagger\otimes\id_{B_j}).
    \end{aligned}
\end{equation}
\end{comment}

\begin{comment}
\begin{align}
    &\Tr_{\bar{A_j}\bar{B_j}}\left((\id_{B_j}\otimes W_{A_j\bar{A_j}}\otimes\id_{\bar{B_j}})\right.\nonumber\\
    &\quad\qquad\qquad\times\left.(\ket{\Phi^+}\bra{\Phi^+}_{A_jB_j}\otimes\ket{\Phi^+}\bra{\Phi^+}_{\bar{A_j}\bar{B_j}})\right.\nonumber\\
    &\quad\qquad\qquad\times\left.(\id_{B_j}\otimes W_{A_j\bar{A_j}}^\dagger\otimes\id_{B_j})\right)\\
    &=\Tr_{\bar{A_j}}\left((\id_{B_j}\otimes W_{A_j\bar{A_j}})\right.\nonumber\\
    &\quad\qquad\qquad\times\left.\left(\ket{\Phi^+}\bra{\Phi^+}_{B_jA_j}\otimes\frac{\id_{\bar{A_j}}}{2^{n-1}}\right)\right.\nonumber\\
    &\quad\qquad\qquad\times\left.(\id_{B_j}\otimes W_{A_j\bar{A_j}}^\dagger)\right)\\
    &=(\mathcal{I}_{B_j}\otimes\mathcal{E}_j)(\ket{\Phi^+}\bra{\Phi^+}_{B_jA_j}),
\end{align}
\end{comment}

\section{Computational complexity of cost evaluation}\label{sctComplexity}

In this section, we state impossibility results for the efficient classical evaluation of both of our costs, $C_{\text{HST}}$ and $C_{\text{LHST}}$. To show this, we analyze our circuits in the framework of deterministic quantum computation with one clean qubit (\DQC) \cite{knill1998POOQ}. We then make use of known hardness results for the class \DQC, and establish that the efficient classical approximation of our cost functions is impossible under reasonable complexity assumptions.

\subsection{One-clean-qubit model of computation.}

The complexity class \DQC consists of all problems that can be efficiently solved with bounded error in the one-clean-qubit model of computation. 
Inspired by the early implementations of NMR quantum computing \cite{knill1998POOQ}, in the one-clean-qubit model of computation the input is specified by a single ``clean qubit'', together with a maximally mixed state on $n$ qubits:
\begin{equation}
\rho=\ketbra{0}{0} \otimes (\id/2)^{\otimes{n}}.
\end{equation}
A computation is then realized by applying a $\poly(n)$-sized quantum circuit $Q$ to the input. We then measure the clean qubit in the standard basis and consider the probability of obtaining the outcome ``0'', i.e.,
\begin{align}
\label{eqnDCQ1_probability}
 \Tr[ (\ketbra{0}{0} \otimes \id^{\otimes{n}}) Q \rho Q^\dagger].
\end{align}
The \DQC model of computation has been widely studied, and several natural problems have been found to be complete for \DQC. Most notably, Shor and Jordan \cite{DBLP:journals/qic/ShorJ08} showed that the problem of trace estimation for $2^n \times 2^n$ unitary matrices that specify $\poly(n)$-sized quantum circuits is \DQC-complete. Moreover, Fujii et al.~\cite{DBLP:journals/corr/FujiiKMNTT14} showed that classical simulation of \DQC is impossible, unless the polynomial hierarchy collapses to the second level. Specifically, it is shown that an efficient classical algorithm that is capable of weakly simulating the output probability distribution of any \DQC computation would imply a collapse of the polynomial hierarchy to the class of Arthur-Merlin protocols, which is not believed to be true. Rather, it is commonly believed that the class \DQC is strictly contained in \BQP, and thus provides a sub-universal model of quantum computation that is hard to simulate classically.
Finally, we point out that the complexity class \DQC is known to give rise to average-case distance measures, whereas worst-case distance measures (such as the diamond distance) are much harder to approximate, and known to be \QIP-complete \cite{DBLP:journals/corr/cs-CC-0407056}. Currently, it is not known whether there exists a distance measure that lies between the average-case and worst-case measures in \DQC and \QIP, respectively. However, we conjecture that only average-case distance measures are feasible for practical purposes.  We leave the task of finding a distance measure whose approximation is complete for the class \BQP as an interesting open problem.

Our contributions are the following. We adapt the proofs in \cite{DBLP:journals/qic/ShorJ08,DBLP:journals/corr/FujiiKMNTT14} and show that the problem of approximating our cost functions, $C_{\text{HST}}$ or $C_{\text{LHST}}$, up to inverse polynomial precision is \DQC-hard. Our results build on the fact that evaluating either of our cost functions is, in some sense, as hard as trace estimation. Using the results from \cite{DBLP:journals/corr/FujiiKMNTT14}, it then immediately follows that no classical algorithm can efficiently approximate our cost functions under certain complexity assumptions.

\subsection{Approximating $C_{\text{HST}}$ is \DQC-hard}

In Appendix \ref{sctComplexityProofs}, we prove the following:
\begin{restatable}{theorem}{TheoremDQCHST} \label{{TheoremDQCHST}}
Let $U$ and $V$ be $\poly(n)$-sized quantum circuits specified by $2^n \times 2^n$ unitary matrices, and let $\epsilon = O(1/\poly(n))$. Then, the problem of approximating $C_\HST(U,V)$ up to $\epsilon$-precision is \DQC-hard.
\end{restatable}

\subsection{Approximating $C_{\text{LHST}}$ is \DQC-hard}

In Appendix \ref{sctComplexityProofs}, we also prove the following:
\begin{restatable}{theorem}{TheoremDQCLHST} \label{{TheoremDQCLHST}}
Let $U$ and $V$ be $\poly(n)$-sized quantum circuits specified by $2^n \times 2^n$ unitary matrices, and let $\epsilon = O(1/\poly(n))$. Then, the problem of approximating $C_\LHST(U,V)$ up to $\epsilon$-precision is \DQC-hard.
\end{restatable}

As a consequence of these results, it then follows from \cite{DBLP:journals/corr/FujiiKMNTT14} that there is no classical algorithm to efficiently approximate our cost functions, $C_{\text{HST}}$ or $C_{\text{LHST}}$, with inverse polynomial precision, unless the polynomial hierarchy collapses to the second level.

\begin{comment}
\subsection{The impossibility of classically efficiently approximating the costs $C_{\text{HST}}$ and $C_{\text{LHST}}$ }

In order to establish that the classical evaluation of our cost functions is impossible under reasonable complexity assumptions, we use a result by Fuji et al. \cite{DBLP:journals/corr/FujiiKMNTT14}. It is shown that an efficient classical algorithm that is capable of weakly simulating the output probability of any \DQC computation would imply a collapse of the polynomial hierarchy (\PH) to the class of Arthur-Merlin protocols (\AM), which is not believed to be true.

\begin{theorem}[\cite{DBLP:journals/corr/FujiiKMNTT14}]\ \\
\DQC is not classically  simulatable, unless $\PH = \AM$. 
\end{theorem}

As a consequence, we immediately recover the main result of this section.

\begin{corollary}
Let $C$ be either $C_{\text{HST}}$ or $C_{\text{LHST}}$. Then,
there is no classical algorithm to efficiently approximate $C$ with inverse polynomial precision, unless the polynomial hierarchy collapses to the second level.
\end{corollary}
\end{comment}

\section{Small-scale implementations}\label{sctimplementations}

This section presents the results of implementing QAQC, as described in Sec. \ref{sctQAQC}, for well-known one- and two-qubit unitaries. Some of these implementations were done on actual quantum hardware, while others were on a simulator. In each case, we performed gradient-free continuous parameter optimization in order to minimize the cost function $C_{\text{HST}}$ in \eqref{eq:GF-cost}, evaluating this cost function using the circuit in Fig. \ref{fig:hilbert-schmidt-inner-product-circuit}(a). For full details on the optimization procedure, see Appendix \ref{sctGF}.

\subsection{Quantum hardware}\label{sctquantum}

We implement QAQC on both IBM's and Rigetti's quantum computers. In what follows, the depth of a gate sequence is defined relative to the native gate alphabet of the quantum computer used.

\subsubsection{IBM's quantum computers}\label{sctIBM}

Here, we consider the 5-qubit IBMQX4 and the 16-qubit IBMQX5. For these quantum computers, the native gate set is 
\begin{equation} \label{eqn:used-gate-alphabet-for-ibm}
\begin{aligned}
&\mathcal{A}_{\text{IBM}} = \{ R_x(\pi / 2), R_z(\theta), \text{CNOT}_{ij} \} %\subseteq \mathcal{A}_{\textsf{IBM}}
\end{aligned}
\end{equation}
where the single-qubit gates $R_x(\pi/2)$ and $R_z(\theta)$ can be performed on any qubit and the two-qubit CNOT gate can be performed between any two qubits allowed in the topology; see \cite{IBMQ5} for the topology of IBMQX4 and \cite{IBMQ16} for the topology of IBMQX5.

To compile a given unitary $U$, we use the general procedure outlined in Sec. \ref{sctsmall}. Specifically, our initial gate structure, given by $V_{\vec{k}}(\vec{\alpha})$, is selected at random from the gate alphabet in \eqref{eqn:used-gate-alphabet-for-ibm}. We then calculate the cost $C_{\text{HST}}(U,V_{\vec{k}}(\vec{\alpha}))$ by executing the HST shown in Fig.~\ref{fig:hilbert-schmidt-inner-product-circuit}(a) on the quantum computer. To perform the continuous parameter optimization over the angles $\theta$ of the $R_z$ gates, we make use of Algorithm~\ref{algo:GfQC_bisect} outlined in Appendix \ref{app:bisection}. This method is designed to limit the number of objective function calls to the quantum computer, which is an important consideration when using queue-based quantum computers like IBMQX4 and IBMQX5 since these can entail a significant amount of idle time in the queue.

In essence, our method in Algorithm \ref{algo:GfQC_bisect} discretizes the continuous parameter space of angles $\theta$ to perform the continuous optimization. These angles are selected uniformly over the unit circle and the grid spacing between them decreases in the number of iterations. See Appendix \ref{app:bisection} for full details. If the cost of the new sequence is less than the cost of the previous sequence, then we accept the change. Otherwise, we accept the change with a probability that decreases exponentially in the magnitude of the difference in cost. This change in cost defines one iteration.

\begin{figure}
\centering 
\includegraphics[width=\columnwidth]{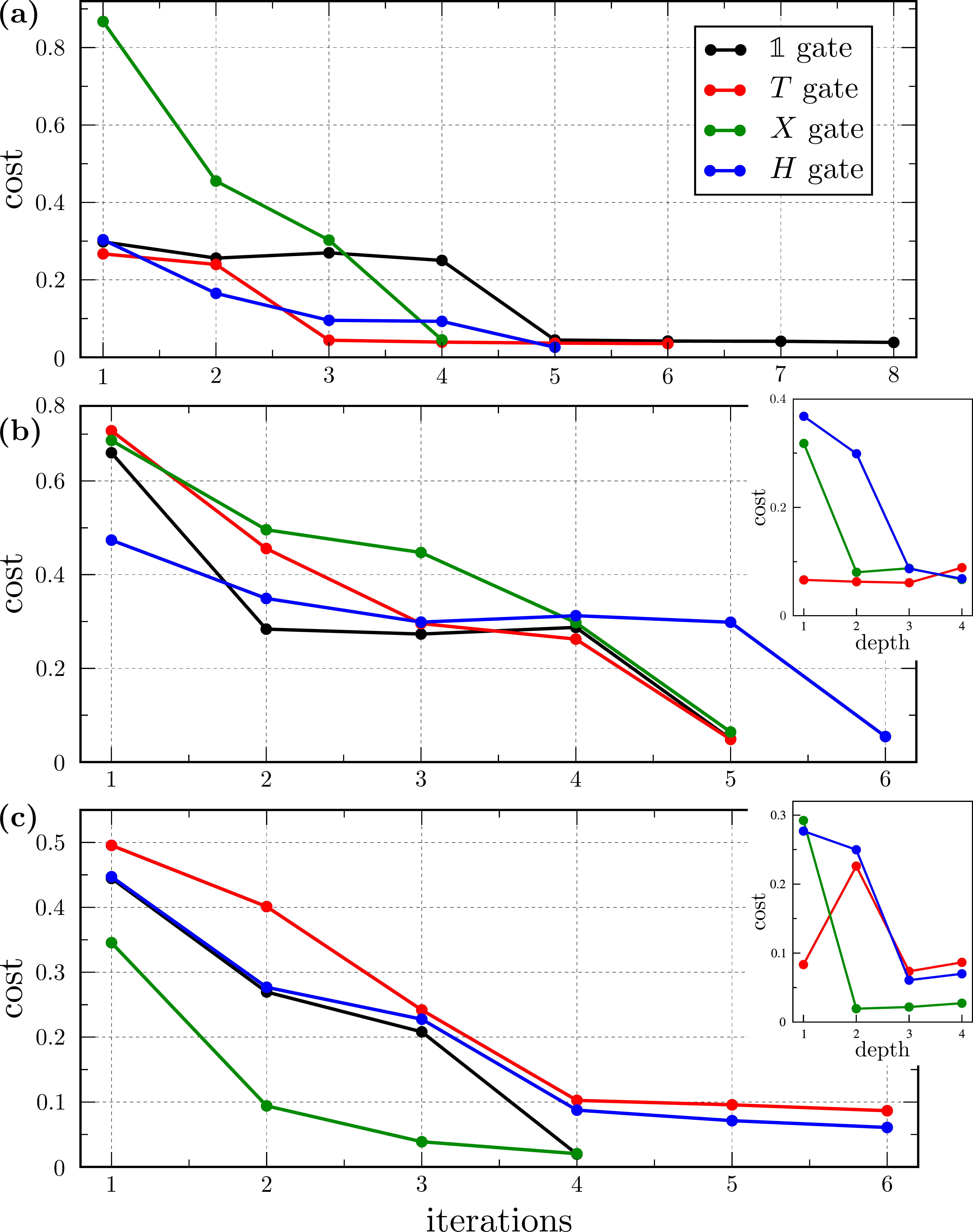}
\caption{Compiling the one-qubit gates $\id$, $X$, $H$, and $T$ using the gradient-free optimization technique described in Appendix \ref{sctGF}. The plots show the cost $C_{\text{HST}}$ as a function of the number of iterations, where an iteration is defined by an accepted update to the gate structure; see Sec. \ref{sctsmall} for a description of the procedure. The insets display the minimum cost achieved by optimizing over gate sequences with a fixed depth, where the depth is defined relative to the native gate alphabet of the quantum computer used.  {\bf (a)} Compiling on the IBMQX4 quantum computer, in which we took $8,000$ samples to evaluate the cost for each run of the Hilbert-Schmidt Test.
%Note that the compiled circuit for the $X$ gate on depth two is $R_x(\pi / 2) R_x (\pi / 2)$, which is an exact decomposition of $X$, yet the cost is not zero. This is most likely due to gate errors and readout errors on the actual device.
{\bf (b)} Compiling on the IBMQX5 quantum computer, in which we again took $8,000$ samples to evaluate the cost for each run of the Hilbert-Schmidt Test. {\bf (c)} Compiling on Rigetti's 8Q-Agave quantum computer. In the plot, each iteration uses 50 cost function evaluations to perform the continuous optimization. For each run of the Hilbert-Schmidt Test to evaluate the cost, we took $10,000$ samples (calls to the quantum computer).
}
\label{fig:ibm_rigetti-one-qubit-gates}
\end{figure}

In Fig.~\ref{fig:ibm_rigetti-one-qubit-gates}(a), we show results for compiling single-qubit gates on IBMQX4. All gates ($\id$, $T$, $X$, and $H$) converge to a cost below 0.1, but no gate achieves a cost below our tolerance of $10^{-2}$. As elaborated upon in Sec. \ref{sct-discussion}, this is due to a combination of finite sampling, gate fidelity, decoherence, and readout error on the device. The single-qubit gates compile to the following gate sequences:
\begin{enumerate}
    \item $\id$ gate: $R_z(\theta)$, with $\theta\approx 0.01 \pi$.  % Newly-collected data, std = 0.02 \pi
    \item $T$ gate: $R_z(\theta)$, with $\theta\approx 0.30 \pi $. % new data, std = 0.05 \pi
    \item $X$ gate: $R_x(\pi/2)R_x(\pi/2)$.
    %\item $X$ gate: $R_x(-\pi/2)R_z(\theta)R_x(-\pi/2)$, with $\theta\approx 0$.
    \item $H$ gate: $R_z(\theta_1) R_x(\pi/2) R_z(\theta_2)$, with $\theta_1 = \theta_2 = 0.50 \pi$.
\end{enumerate}
%In particular, note that $R_x(\pi / 2) R_x (\pi / 2)$ is a textbook decomposition of the $X$ gate, yet the cost is not zero.

Figure~\ref{fig:ibm_rigetti-one-qubit-gates}(b) shows results for compiling the same single-qubit gates as above on IBMQX5. The gate sequences have the same structure as listed above for \mbox{IBMQX4}. The optimal angles achieved are $\theta = - 0.03 \pi$ % std = 0.03 \pi
for the $\id$ gate and $\theta =  0.23  \pi$ % std = 0.03 \pi
for the $T$ gate. The $X$ gate compiles to $R_x(\pi / 2) R_x(\pi / 2)$, and the Hadamard gate $H$ compiles to $R_x(\pi / 2) R_z(\pi / 2) R_x(\pi / 2)$.

In our data collection, we performed on the order of 10 independent optimization runs for each target gate above. The standard deviations of the angles $\theta$ were on the order of $0.05\pi$, and this can be viewed as the error bars on the average values quoted above.

\subsubsection{Rigetti's quantum computer}\label{sctRigetti}

The native gate set of Rigetti's 8Q-Agave 8-qubit quantum computer is
\begin{equation}
    \mathcal{A}_{\text{Rigetti}}=\{R_x(\pm\pi/2),R_z(\theta),\text{CZ}_{ij}\}
\end{equation}
where the single-qubit gates $R_x(\pm\pi/2)$ and $R_z(\theta)$ can be performed on any qubit and the two-qubit CZ gate can be performed between any two qubits allowed in the topology; see \cite{rigettiQPU} for the topology of the 8Q-Agave quantum computer.

\begin{figure}
    \centering
    \includegraphics[width=\columnwidth]{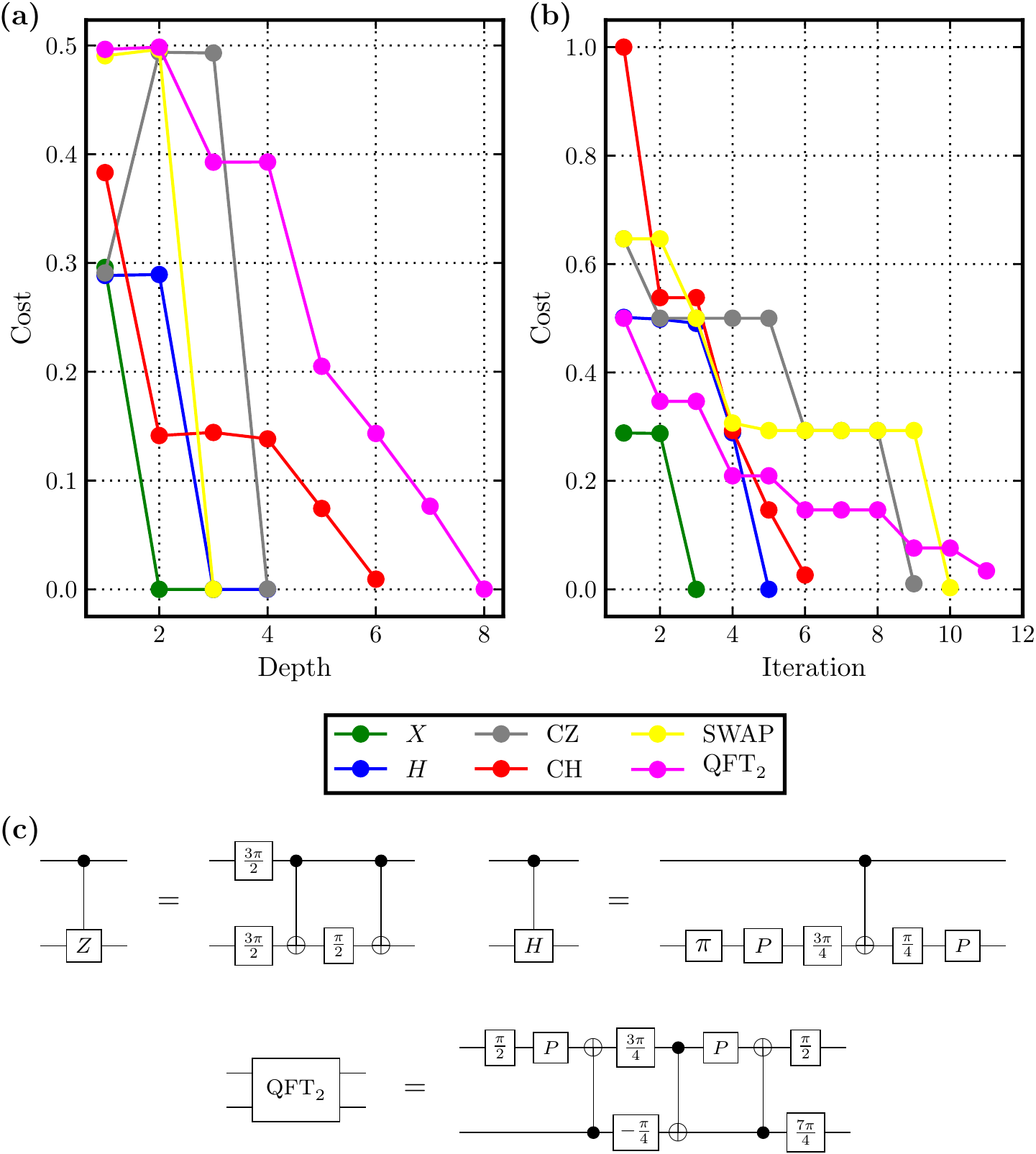}
    \caption{%\textcolor{red}{\bf [Please fix the caption and make it shorter. This Fig was merged from three]}
    Compiling one- and two-qubit gates on Rigetti's quantum virtual machine with the gate alphabet in \eqref{eqn:used-gate-alphabet-for-ibm_full} using the gradient-free optimization technique described in Algorithm \ref{algo:GfQC} in Appendix \ref{sctGF}.
    {\bf (a)} The minimum cost achieved by optimizing over gate sequences with a fixed depth. {\bf (b)} The cost as a function of the number of iterations of the full gate structure and continuous parameter optimization; see Sec.~\ref{sctsmall} for a description of the procedure. Note that each iteration uses 50 cost function evaluations, and each cost function evaluation uses $10,000$ samples (calls to the quantum computer).
    {\bf (c)} Shortest-depth decompositions of the two-qubit controlled-$Z$, controlled-Hadamard, and quantum Fourier transform gates as determined by the compilation procedure. The equalities indicated are true up to a global phase factor. Here, {\protect\includegraphics[height=0.35cm]{theta.pdf}} denotes the rotation gate $R_z(\theta)$, while  {\protect\includegraphics[height=0.35cm]{P.pdf}} represents the rotation gate $R_x(\pi/2)$.
    }
    \label{fig:2qubitgates_sim}
\end{figure}

As with the implementation on IBM's quantum computers, for the implementation on Rigetti's quantum computer we make use of the general procedure outlined in Sec. \ref{sctsmall}. Specifically, we perform random updates to the gate structure followed by continuous optimization over the parameters $\theta$ of the $R_z$ gates using the gradient-free stochastic optimization technique described in Algorithm~\ref{algo:GfQC} in Appendix \ref{sctGF}. In this optimization algorithm, we use fifty cost function evaluations to perform the continuous optimization over parameters. (That is, each iteration in Fig.~\ref{fig:ibm_rigetti-one-qubit-gates}(c) and Fig.~\ref{fig:2qubitgates_sim} uses fifty cost function evaluations, and each cost function evaluation uses $10,000$ calls to the quantum computer for finite sampling.) We take the cost error tolerance (the parameter $\varepsilon'$ in Algorithm~\ref{algo:GfQC}) to be $10^{-2}$, and for each run of the Hilbert-Schmidt Test, we take $10,000$ samples in order to estimate the cost. Our results are shown in Fig. \ref{fig:ibm_rigetti-one-qubit-gates}(c). As described in Algorithm~\ref{algo:GfQC}, we define an iteration to be one accepted update in gate structure followed by a continuous optimization over the internal gate parameters. 

The gates compiled in Fig.~\ref{fig:ibm_rigetti-one-qubit-gates}(c) have the following optimal decompositions. The same decompositions also achieve the lowest cost in the cost vs. depth plot in the inset.
\begin{enumerate}
    \item $\id$ gate: $R_z(\theta)$, with $\theta\approx 0$.  % Newly-collected data
    \item $T$ gate: $R_z(\theta)$, with $\theta\approx 0.342\pi$.
    \item $X$ gate: $R_x(-\pi/2)R_x(-\pi/2)$.
    %\item $X$ gate: $R_x(-\pi/2)R_z(\theta)R_x(-\pi/2)$, with $\theta\approx 0$.
    \item $H$ gate: $R_z(\theta_1)R_x(\pi/2)R_z(\theta_2)$, with $\theta_1\approx 0.50\pi$ and $\theta_2\approx 0.49\pi$.
\end{enumerate}
As with the results on IBM's quantum computers, none of the gates achieve a cost less than $10^{-2}$, due to factors such as finite sampling, gate fidelity, decoherence, and readout error. In addition, similar to the IBM results, the standard deviations of the angles $\theta$ here were on the order of $0.05\pi$, which can be viewed as the error bars on the average values (over 10 independent runs) quoted above.

\subsection{Quantum simulator}\label{sctsim}

We now present our results on executing QAQC for single-qubit and two-qubit gates using a simulator. We use the gate alphabet 
\begin{equation}\label{eqn:used-gate-alphabet-for-ibm_full}
    \mathcal{A}=\{R_x(\pi/2),R_z(\theta),\text{CNOT}_{ij}\},
\end{equation}
which is the gate alphabet defined in Eq.~\eqref{eqn:used-gate-alphabet-for-ibm} except with full connectivity between the qubits. We again use the gradient-free optimization method outlined in Appendix~\ref{sctGF} to perform the continuous parameter optimization. The simulations are performed assuming perfect connectivity between the qubits, no gate errors, and no decoherence.

Using Rigetti's quantum virtual machine \cite{rigetti}, we compile the controlled-Hadamard (CH) gate, the CZ gate, the SWAP gate, and the two-qubit quantum Fourier transform $\text{QFT}_2$ by adopting the gradient-free continuous optimization procedure in Algorithm \ref{algo:GfQC}. We also compile the single-qubit gates $X$ and $H$. For each run of the Hilbert-Schmidt Test to determine the cost, we took $20,000$ samples. Our results are shown in Fig. \ref{fig:2qubitgates_sim}.
For the SWAP gate, we find that circuits of depth one and two cannot achieve zero cost, but there exists a circuit with depth three for which the cost vanishes. The circuit achieving this zero cost is the well-known decomposition of the SWAP gate into three CNOT gates.
While our compilation procedure reproduces the known decomposition of the SWAP gate, it discovers a decomposition of both the CZ and the $\text{QFT}_2$ gates that differs from their conventional ``textbook'' decompositions, as shown in Fig. \ref{fig:2qubitgates_sim}(c). In particular, these decompositions have shorter depths than the conventional decompositions when written in terms of the gate alphabet in \eqref{eqn:used-gate-alphabet-for-ibm_full}.

In Appendix \ref{sctGB}, we likewise implement QAQC for one- and two-qubit gates on a simulator, but instead using a gradient-based continuous parameter optimization method outlined therein.

\section{Larger-scale implementations}\label{sctLargeImplementations}

While in the previous section we considered one- and two-qubit unitaries, in this section we explore larger unitaries, up to nine qubits. The purpose of this section is to see how QAQC scales, and in particular, to study the performance of our $C_{\text{HST}}$ and $C_{\text{LHST}}$ cost functions as the problem size increases. We consider two different examples. 

\begin{example}\label{ex-large_scale_1}
In the first example, we let $U$ be a tensor product of one-qubit unitaries. Namely we consider
\begin{equation}
    U = \bigotimes_{j=1}^n R_z(\theta_j)
\end{equation}
where the $\theta_j$ are randomly chosen, and $R_z(\theta)$ is a rotation about the $z$-axis of the Bloch sphere by angle $\theta$. Similarly, our ansatz for $V$ is of the same form,
\begin{equation}
    V = \bigotimes_{j=1}^n R_z(\phi_j)
\end{equation}
where the initial values of the angles $\phi_j$ are randomnly chosen. 
\end{example}

\begin{example}\label{ex-large_scale_2}
In the second example, we go beyond the tensor-product situation and explore a unitary that entangles all the qubits. The target unitary has the form $U = U_4(\vec{\theta}') U_3 U_2 U_1(\vec{\theta})$, with
\begin{align}
    U_1(\vec{\theta})&=  \bigotimes_{j=1}^n R_z(\theta_j), \quad U_2=...\text{CNOT}_{34} \text{CNOT}_{12}\\
     U_3 &=...\text{CNOT}_{45} \text{CNOT}_{23}, \quad U_4(\vec{\theta}')=  \bigotimes_{j=1}^n R_z(\theta_{j}') \,.
\end{align}
Here, $\text{CNOT}_{kl}$ denotes a CNOT with qubit $k$ the control and qubit $l$ the target, while $\vec{\theta} = \{\theta_j\}$ and $\vec{\theta}' = \{\theta'_j\}$ are $n$-dimensional vectors of angles. Hence $U_2$ and $U_3$ are layers of CNOTs where the CNOTs in $U_3$ are shifted down by one qubit relative to those in $U_2$. Our ansatz for the trainable unitary $V$ has the same form as $U$ but with different angles, i.e., $V = U_4(\vec{\phi}') U_3 U_2 U_1(\vec{\phi})$ where $\vec{\phi}$ and $\vec{\phi}'$ are randomly initialized.
\end{example}

In what follows we discuss our implementations of QAQC for these two examples. We first discuss the implementation on a simulator without noise, and then we move onto the implementation on a simulator with a noise model.

\subsection{Noiseless implementations}\label{sct-noiseless}

\begin{figure}
    \centering
    \includegraphics[width=\columnwidth]{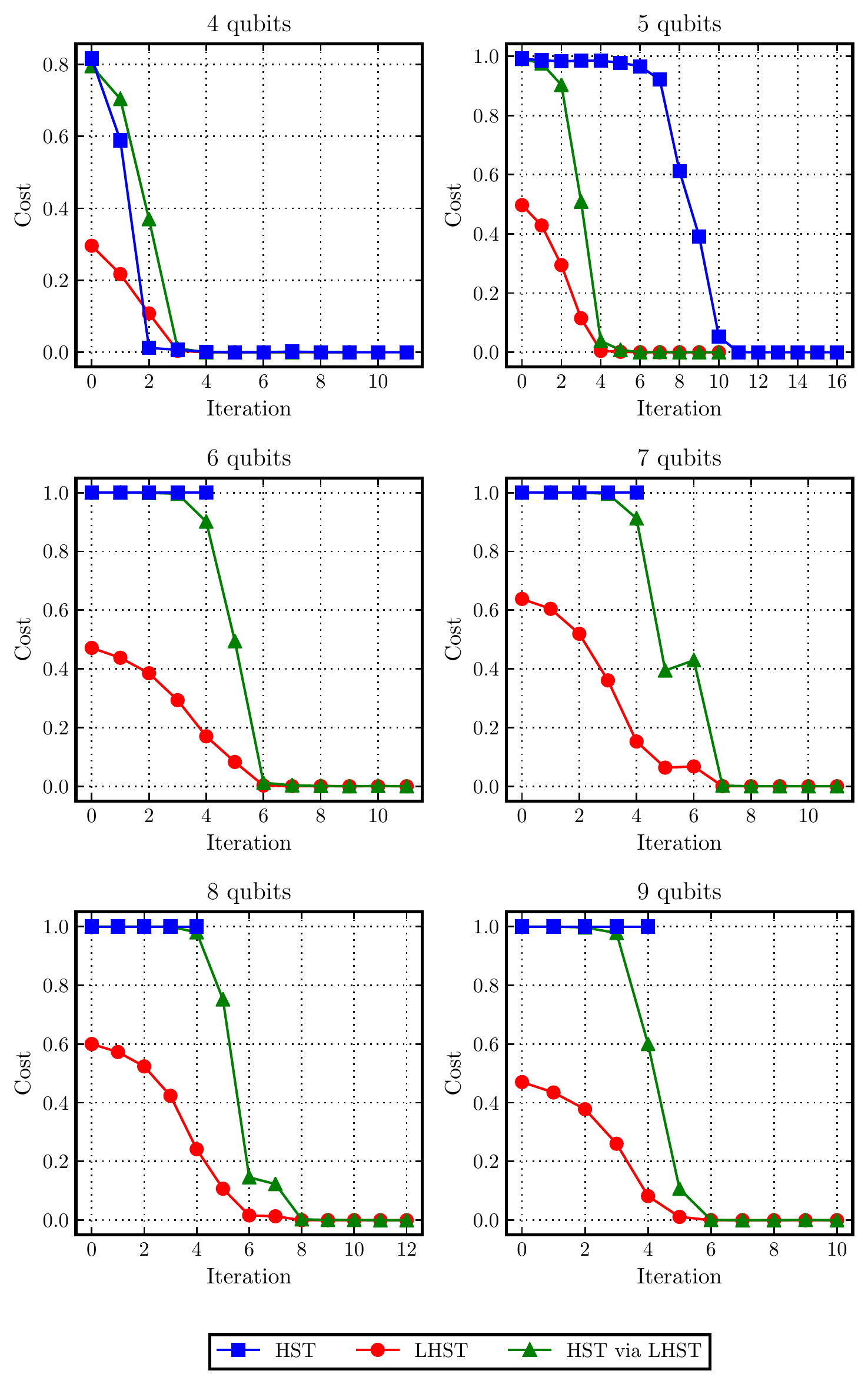}
    \caption{Results of performing continuous parameter optimization using the HST and the LHST for the scenario described in Example \ref{ex-large_scale_1}. We make use of the gradient-based optimization algorithm given by Algorithm~\ref{algo:GbQC_2} in Appendix~\ref{sctGB}. The curves ``HST via LHST'' are given by evaluating $C_{\text{HST}}$ using the angles obtained during the optimization iterations of $C_{\text{LHST}}$. For each run of the HST and LHST, we use 1000 samples to estimate the cost function.}
    \label{fig:HSTvsLHST_tensorproduct_noiseless}
\end{figure}

\begin{figure}
    \centering
    \includegraphics[width=\columnwidth]{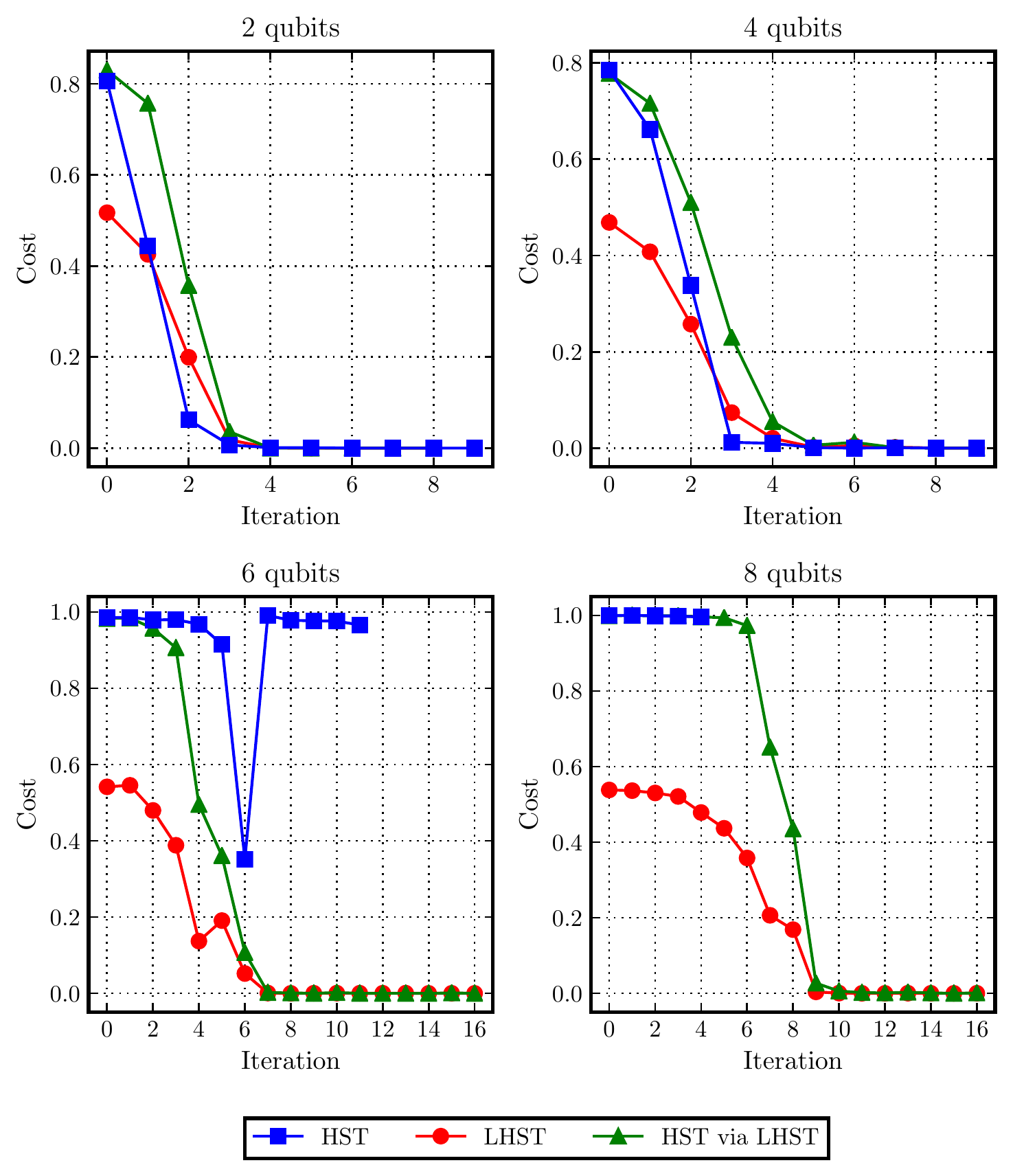}
    \caption{Results of performing continuous parameter optimization using the HST and the LHST for the scenario described in Example \ref{ex-large_scale_2}. We make use of the gradient-based optimization algorithm given by Algorithm~\ref{algo:GbQC_2} in Appendix \ref{sctGB}, in which each iteration can involve several calls to the quantum computer. The curves ``HST via LHST'' are given by evaluating $C_{\text{HST}}$ using the angles obtained during the optimization iterations of $C_{\text{LHST}}$. For each run of the HST and LHST, we use 1000 samples to estimate the cost function.}
    \label{fig:HSTvsLHST_entangling_noiseless}
\end{figure}

We implemented Examples \ref{ex-large_scale_1} and \ref{ex-large_scale_2} on a noiseless simulator. In each case, starting with the ansatz for $V$ at a randomly chosen set of angles, we performed the continuous parameter optimization over the angles using a gradient-based approach. We made use of Algorithm \ref{algo:GbQC_2} in Appendix \ref{sctHST_grad}, which is a gradient descent algorithm that explicitly evaluates the gradient using the formulas provided in Appendix \ref{sctHST_grad}. For each run of the HST and LHST, we took 1000 samples in order to estimate the value of the cost function. The results of this implementation are shown in Figs. \ref{fig:HSTvsLHST_tensorproduct_noiseless} and \ref{fig:HSTvsLHST_entangling_noiseless}. 

In the case of Example \ref{ex-large_scale_1} (Fig.~\ref{fig:HSTvsLHST_tensorproduct_noiseless}), both the $C_{\text{HST}}$ and $C_{\text{LHST}}$ cost functions converge to the desired global minimum up to 5 qubits. However, for $n=$ 6, 7, 8, and 9 qubits, we find cases in which the $C_{\text{HST}}$ cost function does not converge to the global minimum but the $C_{\text{LHST}}$ cost function does. Specifically, the cost $C_{\text{HST}}$ stays very close to one, with a gradient value smaller than the pre-set threshold of $10^{-3}$ for four consecutive iterations, causing the gradient descent algorithm to declare convergence. Interestingly, even in the cases that the $C_{\text{HST}}$ cost does not converge to the global minimum, training with the $C_{\text{LHST}}$ cost allows us to fully minimize the $C_{\text{HST}}$ cost. (See the green curves labelled ``HST via LHST'' in Fig.~\ref{fig:HSTvsLHST_tensorproduct_noiseless}, in which we evaluate the $C_{\text{HST}}$ cost at the angles obtained during the optimization of the $C_{\text{LHST}}$ cost.) This fascinating feature implies that, for $n\geq 6$ qubits in Example~\ref{ex-large_scale_1}, training our $C_{\text{LHST}}$ cost is better at minimizing the $C_{\text{HST}}$ cost than is directly attempting to train the $C_{\text{HST}}$ cost.

%In other words, we find not only that the local cost function outperforms the global cost function for a large number of qubits, but that the local cost function can be used to optimize the global cost function.

We find very similar behavior for Example \ref{ex-large_scale_2} (Fig.~\ref{fig:HSTvsLHST_entangling_noiseless}). In particular, for $n\geq 6$ qubits, we were unable to directly train the $C_{\text{HST}}$ cost. However, the $C_{\text{LHST}}$ cost converges to the global minimum for $n=$ 6 and 8 qubits. Furthermore, as with Example \ref{ex-large_scale_1}, we find that minimizing the $C_{\text{LHST}}$ cost also minimizes the $C_{\text{HST}}$ cost.

\subsection{Noisy implementations}\label{sct-noisy}

\begin{figure}
    \centering
    \includegraphics[width=\columnwidth]{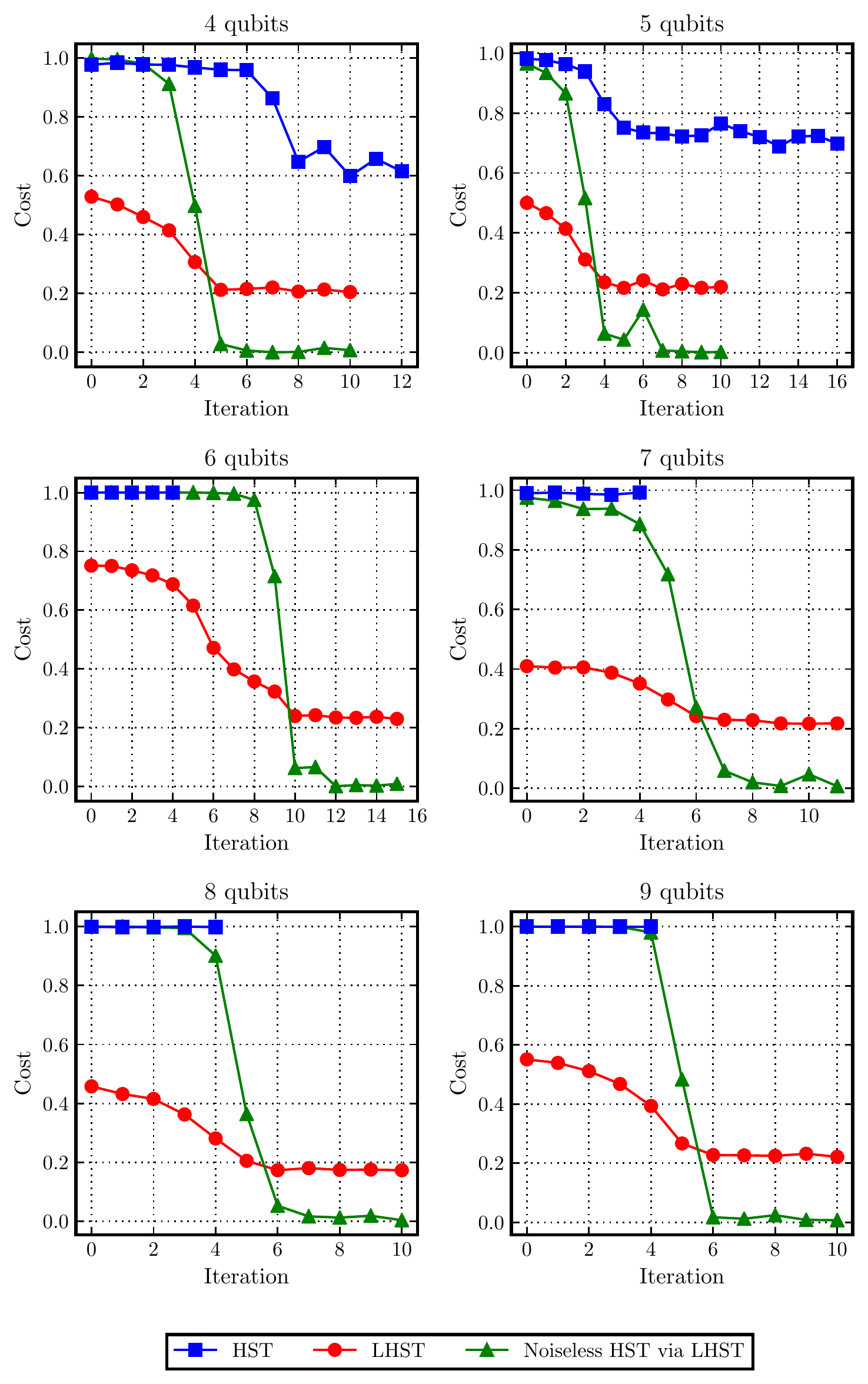}
    \caption{Results of performing continuous parameter optimization using the HST and the LHST, in the presence of noise, for the scenario described in Example \ref{ex-large_scale_1}. The noise model used matches that of the IBMQX5 quantum computer. We make use of the gradient-based optimization algorithm given by Algorithm~\ref{algo:GbQC_2} in Appendix \ref{sctGB}. The curves ``Noiseless HST via LHST'' are given by evaluating $C_{\text{HST}}$ (without noise) using the angles obtained during the optimization iterations of $C_{\text{LHST}}$. For each run of the HST and LHST, we use 1000 samples to estimate the cost function.}
    \label{fig:HSTvsLHST_tensorproduct_noisy}
\end{figure}

\begin{figure}
    \centering
    \includegraphics[width=\columnwidth]{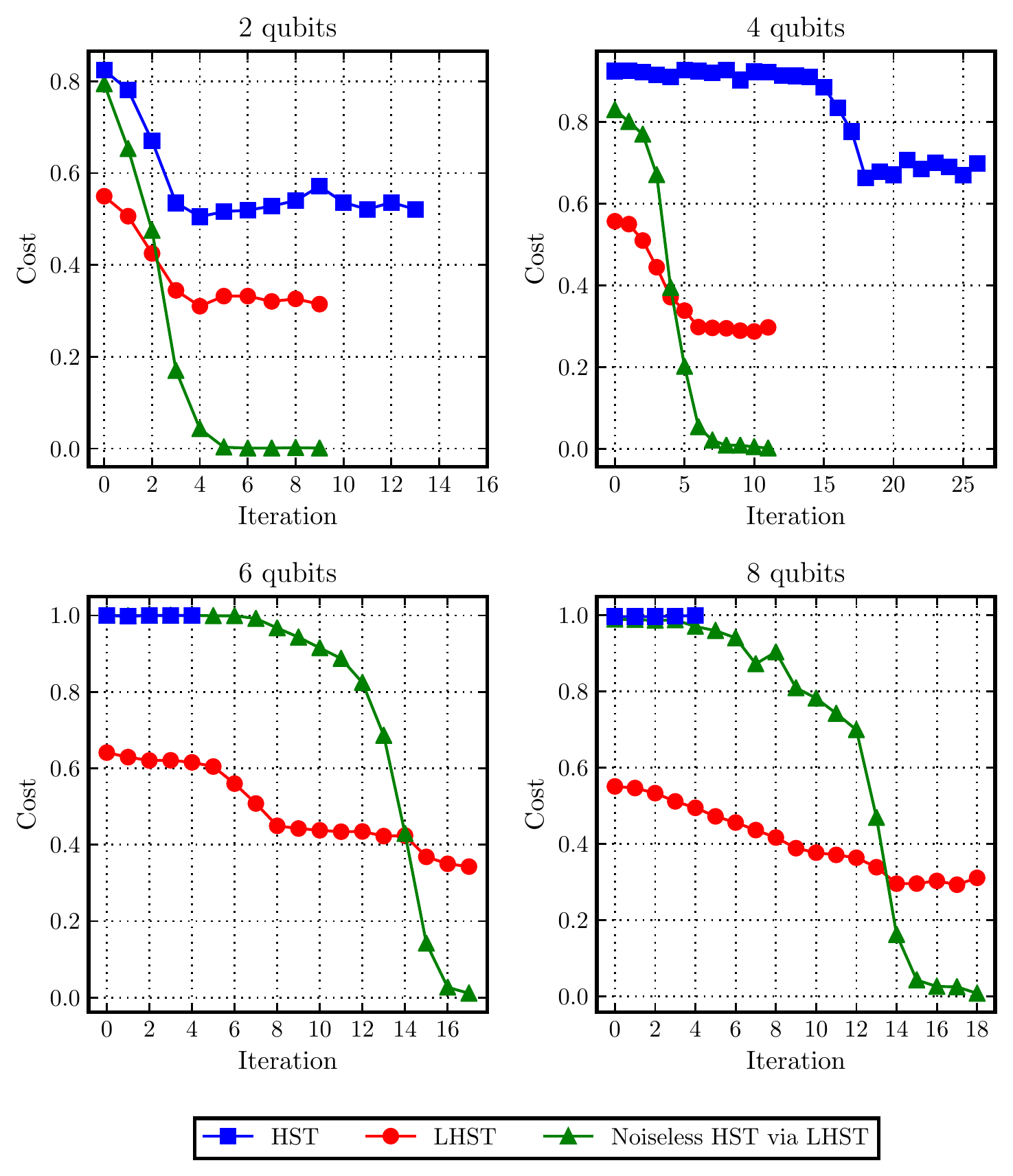}
    \caption{Results of performing continuous parameter optimization using the HST and the LHST, in the presence of noise, for the scenario described in Example \ref{ex-large_scale_2}. The noise model used matches that of the IBMQX5 quantum computer. We make use of the gradient-based optimization algorithm given by Algorithm~\ref{algo:GbQC_2} in Appendix~\ref{sctGB}. The curves ``Noiseless HST via LHST'' are given by evaluating $C_{\text{HST}}$ (without noise) using the angles obtained during the optimization iterations of $C_{\text{LHST}}$. For each run of the HST and LHST, we use 1000 samples to estimate the cost function.}
    \label{fig:HSTvsLHST_entangling_noisy}
\end{figure}

We implemented Examples \ref{ex-large_scale_1} and \ref{ex-large_scale_2} on IBM's noisy simulator, where the noise model matches that of the 16-qubit IBMQX5 quantum computer. This noise model accounts for $T_1$ noise, $T_2$ noise, gate errors, and measurement errors. We emphasize that these are realistic noise parameters since they simulate the noise on currently available quantum hardware. (Note that when our implementations required more than 16 qubits, we applied similar noise parameters to the additional qubits as those for the 16 qubits of the IBMQX5.) We used the same training algorithm as the one we used in the noiseless case above. The results of these implementations are shown in Figs. \ref{fig:HSTvsLHST_tensorproduct_noisy} and \ref{fig:HSTvsLHST_entangling_noisy}. 

Similar to the noiseless case, for Example \ref{ex-large_scale_1} (Fig.~\ref{fig:HSTvsLHST_tensorproduct_noisy}) and for Example \ref{ex-large_scale_2} (Fig.~\ref{fig:HSTvsLHST_entangling_noisy}), we find that both the $C_{\text{HST}}$ and $C_{\text{LHST}}$ cost functions converge up to a problem size of 5 qubits. Due to the noise, as expected, both cost functions converge to a value greater than zero. For $n\geq 6$ qubits, however, we find that the $C_{\text{HST}}$ cost function does not converge to a local minimum. Specifically, this cost stays very close to one with a gradient value smaller than the pre-set threshold of $10^{-3}$ for four consecutive iterations, causing the gradient descent algorithm to declare convergence. The local cost, on the other hand, converges to a local minimum in every case.

Remarkably, despite the noise in the simulation, we find that the angles obtained during the iterations of the $C_{\text{LHST}}$ optimization correspond to the optimal angles in the noiseless case. This result is indicated by the green curves labeled ``Noiseless HST via LHST''. One can see that the green curves go to zero for the local minima found by training the noisy $C_{\text{LHST}}$ cost function. Hence, in these examples, training the noisy $C_{\text{LHST}}$ cost function can be used to minimize the noiseless $C_{\text{HST}}$ cost function to the global minimum. This intriguing behavior suggests that the noise has not affected the location (i.e., the value for the angles) of the global minimum. We thus find evidence of the robustness of QAQC to the kind of noise present in actual devices. We elaborate on this point in the next section.

\section{Discussion}\label{sct-discussion}

On both IBM's and Rigetti's quantum hardware, we were able to successfully compile one-qubit gates with no \textit{a priori} assumptions about gate structure or gate parameters. We also successfully implemented QAQC for simple 9-qubit gates on both a noiseless and noisy simulator. These implementations highlighted two important issues, (1) barren plateuas in cost landscape and (2) the effect of hardware noise, which we discuss further now.

\subsection{Barren Plateaus}

Recent results~\cite{mcclean2018barren,DayEtAl2018} on gradient-based optimization with random quantum circuits suggest that the probability of observing non-zero gradients tends to become exponentially small as a function of the number of qubits. That work showed that a hardware-efficient ansatz leads to vanishing gradients as the ansatz's depth becomes deeper (and hence begins to look more like a random unitary). This is an important issue for many variational hybrid algorithms, including QAQC, and motivates the need to avoid a deep, random ansatz. Strategies to address this ``barren plateau'' issue for QAQC include restricting to a short-depth ansatz, or alternatively employing an application-specific ansatz that takes into account some information about the target unitary $U$. We intend to explore application-specific ansatze in future work to address this issue. There may be other strategies based on the fact that similar issues have been identified in classical deep learning \cite{GlorotBengio2010}. For instance, recent work \cite{BenedettiEtAl2018} shows that gradient descent with momentum (GDM) using an adaptive (multiplicative) integration step update, called resilient backpropagation (rProp), can help with convergence. But, this is an active research area and will likely be important to the success of variational hybrid algorithms.

Interestingly, in this work, we identified another barren plateau issue that is completely independent and distinct from the issue raised in Refs.~\cite{mcclean2018barren,DayEtAl2018}. Namely, we found that our operationally meaningful cost function, $C_{\text{HST}}$, can have barren plateaus even when the ansatz is a depth-one circuit. The gradient of $C_{\text{HST}}$ can vanish exponentially in $n$ even when the ansatz has only a single parameter. This issue became apparent in our implementations (see Figs.~\ref{fig:HSTvsLHST_tensorproduct_noiseless} through \ref{fig:HSTvsLHST_entangling_noisy}), where we were unable to directly train the $C_{\text{HST}}$ cost for $n\geq 6$ qubits. Fortunately, we fixed this issue by introducing the $C_{\text{LHST}}$ cost, which successfully trained in all cases we attemped (we attempted up to $n=9$ qubits). Although $C_{\text{LHST}}$ is not directly operationally meaningful, it is indirectly related to $C_{\text{HST}}$ via Eqs.~\eqref{eqnDirectBound} and \eqref{eqnConverseBound}. Hence it can be used to indirectly train $C_{\text{HST}}$, as shown in Figs.~\ref{fig:HSTvsLHST_tensorproduct_noiseless} through \ref{fig:HSTvsLHST_entangling_noisy}. We believe this barren plateau issue will show up in other variational hybrid algorithms. For example, we encountered the same issue in a recently introduced variational algorithm for state diagonalization~\cite{larose2018variational}.

%While this is not an issue for the small circuits considered in this paper, as the number of qubits increases this becomes an increasingly important consideration. Our introduction of the local cost function $C_{\text{LHST}}$ in Sec. \ref{sctlarge} (see also Sec. \ref{sctlocalHST}) for large problem sizes is meant to address this issue for QAQC. The success in practice of optimization via the local cost function will ultimately be determined by its implementation on large-scale quantum computers once they become available, and we will explore this in future work.

\subsection{Effect of Hardware Noise}

%Noise affects HST and LHST circuits
%Empirical evidence of noise resilience in our numerics (will investigate in future work)

%XXXX Rewrite this subsection after we've done our noisy implementations XXXX

The impact of hardware noise, such as decoherence, gate infidelity, and readout error, is important to consider. This is especially true since QAQC is aimed at being a useful algorithm in the era of NISQ computers, although we remark that QAQC may also be useful for fault-tolerant quantum computing.

%In principle, our approach extends to compiling larger unitaries $U$ with an exponential speedup over classical methods. In practice, limitations arise in this approach due to decoherence, gate infidelity, and readout error on NISQ computers.

On the one hand, we intuitively expect noise to significantly affect the HST and LHST cost evaluation circuits. On the other hand, we see empirical evidence of noise resilience in Figs.~\ref{fig:HSTvsLHST_tensorproduct_noisy} and \ref{fig:HSTvsLHST_entangling_noisy}. Let us elaborate on both our intuition and our empirical observations now.

%Our gradient-free optimization method employs a short-depth circuit (the HST in Fig. \ref{fig:hilbert-schmidt-inner-product-circuit}(a)), and hence we are able to implement it on NISQ hardware. Nevertheless, the imperfections of NISQ hardware do affect the accuracy of estimating the cost.

A qualitative noise analysis of the HST circuit in Fig.~\ref{fig:hilbert-schmidt-inner-product-circuit}(a) is as follows. To compile a unitary $U$ acting on $n$ qubits, a circuit with $2 n $ qubits is needed. Preparing the maximally-entangled state $| \Phi^+\rangle$ in the first portion of the circuit requires $n$ \text{CNOT} gates, which are significantly noisier than one-qubit gates and propagate errors to other qubits through entanglement. In principle, all Hadamard and \text{CNOT} gates can be implemented in parallel, but on near-term devices this may not be the case. Additionally, due to limited connectivity of NISQ devices, it is generally not possible to directly implement CNOTs between arbitrary qubits. Instead, the CNOTs need to be ``chained'' between qubits that are connected, a procedure that can significantly increase the depth of the circuit.

The next level of the circuit involves implementing $U$ in the top $n$-qubit register and $V^*$ in the bottom $n$-qubit register. Here, the noise of the computer on $V^*$ is not necessarily undesirable since it could allow us to compile noise-tailored algorithms that counteract the noise of the specific computer, as described in Sec. \ref{sctapps}. Nevertheless, the depth of $V^*$ and/or of $U$ essentially determines the overall circuit depth as noted in \eqref{eqn6}, and quantum coherence decays exponentially with the circuit depth. Hence, compiling larger gate sequences involves additional loss of coherence on NISQ computers.

The final level of the HST circuit involves making a Bell measurement on all qubits and is the reverse of the first part of the circuit. As such, the same noise analysis of the first portion of the circuit applies here. Readout errors can be significant on NISQ devices~\cite{KTC18}, and our HST circuit involves a number of measurements that scales linearly in the number of qubits. Hence, compiling larger unitaries can increase overall readout error.

A similar qualitative noise analysis holds for the LHST circuit in Fig.~\ref{fig:hilbert-schmidt-inner-product-circuit}(b), except we note that to calculate the functions $C_{\text{LHST}}^{(j)}$ in \eqref{eq-LHST_2} we require only one CNOT gate in the last portion of the LHST circuit before the measurement. Furthermore, we measure only two qubits regardless of the total number of qubits.

With that said, we observed a (somewhat surprising) noise resilience in Figs.~\ref{fig:HSTvsLHST_tensorproduct_noisy} and \ref{fig:HSTvsLHST_entangling_noisy}. In these implementations, we imported the noise model of the IBMQX5 quantum computer, which is a currently available cloud quantum computer. Hence, we considered realistic noise parameters for decoherence, gate infedility, and readout error. This noise affected all circuit elements of the LHST circuit in Fig.~\ref{fig:hilbert-schmidt-inner-product-circuit}(b). Yet we still obtained the correct unitary $V$ via QAQC, as shown by the green curves going to zero in Figs.~\ref{fig:HSTvsLHST_tensorproduct_noisy} and \ref{fig:HSTvsLHST_entangling_noisy}.  

Naturally, we plan to investigate this noise resilience in full detail in future work. But it is worth emphasizing the following point here. \textit{The value of the cost could be significantly affected by noise without shifting the location of the global minimum in parameter space.} In fact, one can see in Figs.~\ref{fig:HSTvsLHST_tensorproduct_noisy} and \ref{fig:HSTvsLHST_entangling_noisy} that the value of the $C_{\text{LHST}}$ cost is significantly affected by noise. Namely, note that the red curves in these plots do not go to zero for larger iterations. However, the green curves do go to zero, which means that QAQC found the correct parameters for $V$ despite the noisy cost values.

We could speculate reasons for why the global minimum appears not shift in parameter space with noise. For example, it could be due to the nature of our cost functions. These cost functions can be thought of as entanglement fidelities and hence are related to Hilbert-space averages of input-output fidelities, see Eq.~\eqref{eq:HST_Fbar}. By averaging the input-output fidelity over the whole Hilbert space, the effect of noise could essentially be averaged away. This is just speculation at this point, and we will perform a detailed analysis of the effect of noise in future work. Regardless, our preliminary results in Figs.~\ref{fig:HSTvsLHST_tensorproduct_noisy} and \ref{fig:HSTvsLHST_entangling_noisy} suggest that QAQC may indeed be useful in the NISQ era.

\section{Conclusions}\label{sctconclusion}

%XXXX Rewrite this section after we've done our noisy implementations XXXX

Quantum compiling is crucial in the era of NISQ devices, where constraints on NISQ computers (such as limited connectivity, limited circuit depth, etc.) place severe restrictions on the quantum algorithms that can be implemented in practice. In this work, we presented a methodology for quantum compilation called quantum-assisted quantum compiling (QAQC), whereby a quantum computer provides an exponential speedup in evaluating the cost of a gate sequence, i.e., how well the gate sequence matches the target. In principle, QAQC should allow for the compiling of larger algorithms than standard classical methods for quantum compiling due to this exponential speedup. As a proof-of-principle, we implemented QAQC on IBM's and Rigetti's quantum computers to compile various one-qubit gates to their native gate alphabets. To our knowledge, this is the first time NISQ hardware has been used to compile a target unitary. In addition, we successfully implemented QAQC on a noiseless and noisy simulator for simple 9-qubit unitaries.

%Current noise levels in the hardware prevented us from compiling multi-qubit gates, although we are optimistic that future noise reduction could enable larger scale QAQC.

Our main technical results were the following. First, we carefully chose a cost function (which involved global and local overlaps between a target unitary $U$ and a trainable unitary $V$) and proved that it satisfied four criteria: it is faithful, it is efficient to compute on a quantum computer, it has an operational meaning, and it scales well with the size of the problem. Second, we presented short-depth circuits (see Sections~\ref{sctHST} and \ref{sctlocalHST}) for computing our cost function. Third, we proved that evaluating our cost function is $\DQC$-hard, and hence no classical algorithm can efficiently evaluate our cost function, under reasonable complexity assumptions. This established a rigorous proof for the difficulty of classically simulating QAQC. We also remark that, in the Appendix, we detailed our gradient-free and gradient-based methods for optimizing our cost function. This included a circuit for gradient computation that generalizes the famous Power of One Qubit \cite{knill1998POOQ} and hence is likely of interest to a broader community.

As elaborated in the Discussion section, our noisy implementations of QAQC showed a surprising resilience to noise. While simulating realistic noise parameters based on a currently available cloud quantum computer (IBMQX5), we were able to run QAQC on a 9-qubit unitary and obtain the correct parameters for $V$. We plan to investigate this intriguing noise resilience in future work.

QAQC is a novel variational hybrid algorithm, similar to other well-known variational hybrid algorithms such as VQE \cite{peruzzo2014VQE} and QAOA \cite{farhi2014QAOA}. Variational hybrid algorithms are likely to provide some of the first real applications of quantum computers in the NISQ era. In the case of QAQC, it is an algorithm that makes other algorithms more efficient to implement, via algorithm depth compression. We note that the ability to compress algorithm depth will also be useful (to reduce the run-time of quantum circuits) in the era of fault-tolerant quantum computing. The central application of QAQC is thus to make quantum computers more useful.

\begin{acknowledgements}

We thank IBM and Rigetti for providing  access to their quantum computers. The views expressed in this paper are those of the authors and do not reflect those of IBM or Rigetti. SK, RL, and AP acknowledge support from the U.S. Department of Energy through a quantum computing program sponsored by the LANL Information Science \& Technology Institute. RL acknowledges support from an Engineering Distinguished Fellowship through Michigan State University. 
AP is partially supported by AFOSR YIP award number FA9550-16-1-0495 and the Institute for Quantum Information and Matter, an NSF Physics Frontiers Center (NSF Grant PHY-1733907) and the Kortschak Scholars program. LC was supported by the U.S. Department of Energy through the J. Robert Oppenheimer fellowship. ATS and PJC were supported by the LANL ASC Beyond Moore's Law project. LC, ATS, and PJC were also supported by the LDRD program at LANL. We thank Alexandru Gheorghiu and Thomas Vidick for useful discussions.

\end{acknowledgements}

% ============
% bibliography
% ============
%\bibliographystyle{apsrev4-1mod}
%\bibliography{qc_qc}

%merlin.mbs apsrev4-1.bst 2010-07-25 4.21a (PWD, AO, DPC) hacked
%Control: key (0)
%Control: author (72) initials jnrlst
%Control: editor formatted (1) identically to author
%Control: production of article title (-1) disabled
%Control: page (0) single
%Control: year (1) truncated
%Control: production of eprint (0) enabled
\begin{thebibliography}{63}%
\makeatletter
\providecommand \@ifxundefined [1]{%
 \@ifx{#1\undefined}
}%
\providecommand \@ifnum [1]{%
 \ifnum #1\expandafter \@firstoftwo
 \else \expandafter \@secondoftwo
 \fi
}%
\providecommand \@ifx [1]{%
 \ifx #1\expandafter \@firstoftwo
 \else \expandafter \@secondoftwo
 \fi
}%
\providecommand \natexlab [1]{#1}%
\providecommand \enquote  [1]{``#1''}%
\providecommand \bibnamefont  [1]{#1}%
\providecommand \bibfnamefont [1]{#1}%
\providecommand \citenamefont [1]{#1}%
\providecommand \href@noop [0]{\@secondoftwo}%
\providecommand \href [0]{\begingroup \@sanitize@url \@href}%
\providecommand \@href[1]{\@@startlink{#1}\@@href}%
\providecommand \@@href[1]{\endgroup#1\@@endlink}%
\providecommand \@sanitize@url [0]{\catcode `\\12\catcode `\$12\catcode
  `\&12\catcode `\#12\catcode `\^12\catcode `\_12\catcode `\%12\relax}%
\providecommand \@@startlink[1]{}%
\providecommand \@@endlink[0]{}%
\providecommand \url  [0]{\begingroup\@sanitize@url \@url }%
\providecommand \@url [1]{\endgroup\@href {#1}{\urlprefix }}%
\providecommand \urlprefix  [0]{URL }%
\providecommand \Eprint [0]{\href }%
\providecommand \doibasemod [0]{http://dx.doi.org/}%
\providecommand \selectlanguage [0]{\@gobble}%
\providecommand \bibinfo  [0]{\@secondoftwo}%
\providecommand \bibfield  [0]{\@secondoftwo}%
\providecommand \translation [1]{[#1]}%
\providecommand \BibitemOpen [0]{}%
\providecommand \bibitemStop [0]{}%
\providecommand \bibitemNoStop [0]{.\EOS\space}%
\providecommand \EOS [0]{\spacefactor3000\relax}%
\providecommand \BibitemShut  [1]{\csname bibitem#1\endcsname}%
\let\auto@bib@innerbib\@empty
%</preamble>
\bibitem [{\citenamefont {Shor}(1997)}]{shor1997factoring}%
  \BibitemOpen
  \bibfield  {author} {\bibinfo {author} {\bibfnamefont {P.}~\bibnamefont
  {Shor}},\ }\bibfield  {title} {\emph {\bibinfo {title} {Polynomial-time
  algorithms for prime factorization and discrete logarithms on a quantum
  computer},\ }}\href {\doibasemod 10.1137/S0097539795293172} {\bibfield
  {journal} {\bibinfo  {journal} {SIAM Journal on Computing}\ }\textbf
  {\bibinfo {volume} {26}},\ \bibinfo {pages} {1484} (\bibinfo {year}
  {1997})}\BibitemShut {NoStop}%
\bibitem [{\citenamefont {Farhi}\ \emph {et~al.}(2014)\citenamefont {Farhi},
  \citenamefont {Goldstone},\ and\ \citenamefont {Gutmann}}]{farhi2014QAOA}%
  \BibitemOpen
  \bibfield  {author} {\bibinfo {author} {\bibfnamefont {E.}~\bibnamefont
  {Farhi}}, \bibinfo {author} {\bibfnamefont {J.}~\bibnamefont {Goldstone}}, \
  and\ \bibinfo {author} {\bibfnamefont {S.}~\bibnamefont {Gutmann}},\
  }\bibfield  {title} {\emph {\bibinfo {title} {A quantum approximate
  optimization algorithm},\ }}\href {https://arxiv.org/abs/1411.4028}
  {\bibfield  {journal} {\bibinfo  {journal} {arXiv:1411.4028}\ } (\bibinfo
  {year} {2014})}\BibitemShut {NoStop}%
\bibitem [{\citenamefont {Feynman}(1982)}]{feynman1982simulating}%
  \BibitemOpen
  \bibfield  {author} {\bibinfo {author} {\bibfnamefont {R.~P.}\ \bibnamefont
  {Feynman}},\ }\bibfield  {title} {\emph {\bibinfo {title} {Simulating physics
  with computers},\ }}\href {\doibasemod 10.1007/BF02650179} {\bibfield
  {journal} {\bibinfo  {journal} {International Journal of Theoretical
  Physics}\ }\textbf {\bibinfo {volume} {21}},\ \bibinfo {pages} {467}
  (\bibinfo {year} {1982})}\BibitemShut {NoStop}%
\bibitem [{\citenamefont {Preskill}(2018)}]{preskill2018quantum}%
  \BibitemOpen
  \bibfield  {author} {\bibinfo {author} {\bibfnamefont {J.}~\bibnamefont
  {Preskill}},\ }\bibfield  {title} {\emph {\bibinfo {title} {Quantum computing
  in the {NISQ} era and beyond},\ }}\href {\doibasemod
  10.22331/q-2018-08-06-79} {\bibfield  {journal} {\bibinfo  {journal}
  {Quantum}\ }\textbf {\bibinfo {volume} {2}},\ \bibinfo {pages} {79} (\bibinfo
  {year} {2018})}\BibitemShut {NoStop}%
\bibitem [{\citenamefont {Preskill}(2012)}]{preskill2012quantum}%
  \BibitemOpen
  \bibfield  {author} {\bibinfo {author} {\bibfnamefont {J.}~\bibnamefont
  {Preskill}},\ }\bibfield  {title} {\emph {\bibinfo {title} {Quantum computing
  and the entanglement frontier},\ }}\href {https://arxiv.org/abs/1203.5813}
  {\bibfield  {journal} {\bibinfo  {journal} {arXiv:1203.5813}\ } (\bibinfo
  {year} {2012})}\BibitemShut {NoStop}%
\bibitem [{\citenamefont {Neill}\ \emph {et~al.}(2018)\citenamefont {Neill},
  \citenamefont {Roushan}, \citenamefont {Kechedzhi}, \citenamefont {Boixo},
  \citenamefont {Isakov}, \citenamefont {Smelyanskiy} \emph
  {et~al.}}]{neill2017blueprint}%
  \BibitemOpen
  \bibfield  {author} {\bibinfo {author} {\bibfnamefont {C.}~\bibnamefont
  {Neill}}, \bibinfo {author} {\bibfnamefont {P.}~\bibnamefont {Roushan}},
  \bibinfo {author} {\bibfnamefont {K.}~\bibnamefont {Kechedzhi}}, \bibinfo
  {author} {\bibfnamefont {S.}~\bibnamefont {Boixo}}, \bibinfo {author}
  {\bibfnamefont {S.~V.}\ \bibnamefont {Isakov}}, \bibinfo {author}
  {\bibfnamefont {V.}~\bibnamefont {Smelyanskiy}},  \emph {et~al.},\ }\bibfield
   {title} {\emph {\bibinfo {title} {A blueprint for demonstrating quantum
  supremacy with superconducting qubits},\ }}\href {\doibasemod
  10.1126/science.aao4309} {\bibfield  {journal} {\bibinfo  {journal}
  {Science}\ }\textbf {\bibinfo {volume} {360}},\ \bibinfo {pages} {195}
  (\bibinfo {year} {2018})}\BibitemShut {NoStop}%
\bibitem [{\citenamefont {Venturelli}\ \emph {et~al.}(2018)\citenamefont
  {Venturelli}, \citenamefont {Do}, \citenamefont {Rieffel},\ and\
  \citenamefont {Frank}}]{venturelli2018compiling}%
  \BibitemOpen
  \bibfield  {author} {\bibinfo {author} {\bibfnamefont {D.}~\bibnamefont
  {Venturelli}}, \bibinfo {author} {\bibfnamefont {M.}~\bibnamefont {Do}},
  \bibinfo {author} {\bibfnamefont {E.}~\bibnamefont {Rieffel}}, \ and\
  \bibinfo {author} {\bibfnamefont {J.}~\bibnamefont {Frank}},\ }\bibfield
  {title} {\emph {\bibinfo {title} {Compiling quantum circuits to realistic
  hardware architectures using temporal planners},\ }}\href {\doibasemod
  10.1088/2058-9565/aaa331} {\bibfield  {journal} {\bibinfo  {journal} {Quantum
  Science and Technology}\ }\textbf {\bibinfo {volume} {3}},\ \bibinfo {pages}
  {025004} (\bibinfo {year} {2018})}\BibitemShut {NoStop}%
\bibitem [{\citenamefont {Booth}\ \emph {et~al.}(2018)\citenamefont {Booth},
  \citenamefont {Do}, \citenamefont {Beck}, \citenamefont {Rieffel},
  \citenamefont {Venturelli},\ and\ \citenamefont
  {Frank}}]{booth2018comparing}%
  \BibitemOpen
  \bibfield  {author} {\bibinfo {author} {\bibfnamefont {K.~E.~C.}\
  \bibnamefont {Booth}}, \bibinfo {author} {\bibfnamefont {M.}~\bibnamefont
  {Do}}, \bibinfo {author} {\bibfnamefont {J.~C.}\ \bibnamefont {Beck}},
  \bibinfo {author} {\bibfnamefont {E.}~\bibnamefont {Rieffel}}, \bibinfo
  {author} {\bibfnamefont {D.}~\bibnamefont {Venturelli}}, \ and\ \bibinfo
  {author} {\bibfnamefont {J.}~\bibnamefont {Frank}},\ }\bibfield  {title}
  {\emph {\bibinfo {title} {Comparing and integrating constraint programming
  and temporal planning for quantum circuit compilation},\ }}\href
  {https://arxiv.org/abs/1803.06775} {\bibfield  {journal} {\bibinfo  {journal}
  {arXiv:1803.06775}\ } (\bibinfo {year} {2018})}\BibitemShut {NoStop}%
\bibitem [{\citenamefont {Cincio}\ \emph {et~al.}(2018)\citenamefont {Cincio},
  \citenamefont {Suba{\c{s}}{\i}}, \citenamefont {Sornborger},\ and\
  \citenamefont {Coles}}]{cincio2018learning}%
  \BibitemOpen
  \bibfield  {author} {\bibinfo {author} {\bibfnamefont {L.}~\bibnamefont
  {Cincio}}, \bibinfo {author} {\bibfnamefont {Y.}~\bibnamefont
  {Suba{\c{s}}{\i}}}, \bibinfo {author} {\bibfnamefont {A.~T.}\ \bibnamefont
  {Sornborger}}, \ and\ \bibinfo {author} {\bibfnamefont {P.~J.}\ \bibnamefont
  {Coles}},\ }\bibfield  {title} {\emph {\bibinfo {title} {Learning the quantum
  algorithm for state overlap},\ }}\href {\doibasemod 10.1088/1367-2630/aae94a}
  {\bibfield  {journal} {\bibinfo  {journal} {New Journal of Physics}\ }\textbf
  {\bibinfo {volume} {20}},\ \bibinfo {eid} {113022} (\bibinfo {year}
  {2018})}\BibitemShut {NoStop}%
\bibitem [{\citenamefont {Maslov}\ \emph {et~al.}(2008)\citenamefont {Maslov},
  \citenamefont {Dueck}, \citenamefont {Miller},\ and\ \citenamefont
  {Negrevergne}}]{maslov2008quantum}%
  \BibitemOpen
  \bibfield  {author} {\bibinfo {author} {\bibfnamefont {D.}~\bibnamefont
  {Maslov}}, \bibinfo {author} {\bibfnamefont {G.~W.}\ \bibnamefont {Dueck}},
  \bibinfo {author} {\bibfnamefont {D.~M.}\ \bibnamefont {Miller}}, \ and\
  \bibinfo {author} {\bibfnamefont {C.}~\bibnamefont {Negrevergne}},\
  }\bibfield  {title} {\emph {\bibinfo {title} {Quantum circuit simplification
  and level compaction},\ }}\href {\doibasemod 10.1109/TCAD.2007.911334}
  {\bibfield  {journal} {\bibinfo  {journal} {IEEE Transactions on
  Computer-Aided Design of Integrated Circuits and Systems}\ }\textbf {\bibinfo
  {volume} {27}},\ \bibinfo {pages} {436} (\bibinfo {year} {2008})}\BibitemShut
  {NoStop}%
\bibitem [{\citenamefont {Fowler}(2011)}]{fowler2011}%
  \BibitemOpen
  \bibfield  {author} {\bibinfo {author} {\bibfnamefont {A.~G.}\ \bibnamefont
  {Fowler}},\ }\bibfield  {title} {\emph {\bibinfo {title} {Constructing
  arbitrary {S}teane code single logical qubit fault-tolerant gates},\ }}\href
  {http://dl.acm.org/citation.cfm?id=2230936.2230946} {\bibfield  {journal}
  {\bibinfo  {journal} {Quantum Information and Computation}\ }\textbf
  {\bibinfo {volume} {11}},\ \bibinfo {pages} {867} (\bibinfo {year}
  {2011})}\BibitemShut {NoStop}%
\bibitem [{\citenamefont {Booth~Jr}(2012)}]{booth2012quantum}%
  \BibitemOpen
  \bibfield  {author} {\bibinfo {author} {\bibfnamefont {J.}~\bibnamefont
  {Booth~Jr}},\ }\bibfield  {title} {\emph {\bibinfo {title} {Quantum compiler
  optimizations},\ }}\href {https://arxiv.org/abs/1206.3348} {\bibfield
  {journal} {\bibinfo  {journal} {arXiv:1206.3348}\ } (\bibinfo {year}
  {2012})}\BibitemShut {NoStop}%
\bibitem [{\citenamefont {Nam}\ \emph {et~al.}(2018)\citenamefont {Nam},
  \citenamefont {Ross}, \citenamefont {Su}, \citenamefont {Childs},\ and\
  \citenamefont {Maslov}}]{nam2017automated}%
  \BibitemOpen
  \bibfield  {author} {\bibinfo {author} {\bibfnamefont {Y.}~\bibnamefont
  {Nam}}, \bibinfo {author} {\bibfnamefont {N.~J.}\ \bibnamefont {Ross}},
  \bibinfo {author} {\bibfnamefont {Y.}~\bibnamefont {Su}}, \bibinfo {author}
  {\bibfnamefont {A.~M.}\ \bibnamefont {Childs}}, \ and\ \bibinfo {author}
  {\bibfnamefont {D.}~\bibnamefont {Maslov}},\ }\bibfield  {title} {\emph
  {\bibinfo {title} {Automated optimization of large quantum circuits with
  continuous parameters},\ }}\href {\doibasemod 10.1038/s41534-018-0072-4}
  {\bibfield  {journal} {\bibinfo  {journal} {npj Quantum Information}\
  }\textbf {\bibinfo {volume} {4}},\ \bibinfo {pages} {23} (\bibinfo {year}
  {2018})}\BibitemShut {NoStop}%
\bibitem [{\citenamefont {Chong}\ \emph {et~al.}(2017)\citenamefont {Chong},
  \citenamefont {Franklin},\ and\ \citenamefont
  {Martonosi}}]{chong2017programming}%
  \BibitemOpen
  \bibfield  {author} {\bibinfo {author} {\bibfnamefont {F.~T.}\ \bibnamefont
  {Chong}}, \bibinfo {author} {\bibfnamefont {D.}~\bibnamefont {Franklin}}, \
  and\ \bibinfo {author} {\bibfnamefont {M.}~\bibnamefont {Martonosi}},\
  }\bibfield  {title} {\emph {\bibinfo {title} {Programming languages and
  compiler design for realistic quantum hardware},\ }}\href {\doibasemod
  10.1038/nature23459} {\bibfield  {journal} {\bibinfo  {journal} {Nature}\
  }\textbf {\bibinfo {volume} {549}},\ \bibinfo {pages} {180} (\bibinfo {year}
  {2017})}\BibitemShut {NoStop}%
\bibitem [{\citenamefont {Heyfron}\ and\ \citenamefont
  {Campbell}(2018)}]{heyfron2017efficient}%
  \BibitemOpen
  \bibfield  {author} {\bibinfo {author} {\bibfnamefont {L.~E.}\ \bibnamefont
  {Heyfron}}\ and\ \bibinfo {author} {\bibfnamefont {E.~T.}\ \bibnamefont
  {Campbell}},\ }\bibfield  {title} {\emph {\bibinfo {title} {An efficient
  quantum compiler that reduces {T} count},\ }}\href {\doibasemod
  10.1088/2058-9565/aad604} {\bibfield  {journal} {\bibinfo  {journal} {Quantum
  Science and Technology}\ }\textbf {\bibinfo {volume} {4}},\ \bibinfo {pages}
  {015004} (\bibinfo {year} {2018})}\BibitemShut {NoStop}%
\bibitem [{\citenamefont {H{\"a}ner}\ \emph {et~al.}(2018)\citenamefont
  {H{\"a}ner}, \citenamefont {Steiger}, \citenamefont {Svore},\ and\
  \citenamefont {Troyer}}]{haner2018software}%
  \BibitemOpen
  \bibfield  {author} {\bibinfo {author} {\bibfnamefont {T.}~\bibnamefont
  {H{\"a}ner}}, \bibinfo {author} {\bibfnamefont {D.~S.}\ \bibnamefont
  {Steiger}}, \bibinfo {author} {\bibfnamefont {K.}~\bibnamefont {Svore}}, \
  and\ \bibinfo {author} {\bibfnamefont {M.}~\bibnamefont {Troyer}},\
  }\bibfield  {title} {\emph {\bibinfo {title} {A~software methodology for
  compiling quantum programs},\ }}\href {\doibasemod 10.1088/2058-9565/aaa5cc}
  {\bibfield  {journal} {\bibinfo  {journal} {Quantum Science and Technology}\
  }\textbf {\bibinfo {volume} {3}},\ \bibinfo {pages} {020501} (\bibinfo {year}
  {2018})}\BibitemShut {NoStop}%
\bibitem [{\citenamefont {Oddi}\ and\ \citenamefont
  {Rasconi}(2018)}]{oddi2018greedy}%
  \BibitemOpen
  \bibfield  {author} {\bibinfo {author} {\bibfnamefont {A.}~\bibnamefont
  {Oddi}}\ and\ \bibinfo {author} {\bibfnamefont {R.}~\bibnamefont {Rasconi}},\
  }in\ \href {https://link.springer.com/chapter/10.1007/978-3-319-93031-2_32}
  {\emph {\bibinfo {booktitle} {International Conference on the Integration of
  Constraint Programming, Artificial Intelligence, and Operations Research}}}\
  (\bibinfo {organization} {Springer},\ \bibinfo {year} {2018})\ pp.\ \bibinfo
  {pages} {446--461}\BibitemShut {NoStop}%
\bibitem [{\citenamefont {Peruzzo}\ \emph {et~al.}(2014)\citenamefont
  {Peruzzo}, \citenamefont {McClean}, \citenamefont {Shadbolt}, \citenamefont
  {Yung}, \citenamefont {Zhou}, \citenamefont {Love}, \citenamefont
  {Aspuru-Guzik},\ and\ \citenamefont {O'Brien}}]{peruzzo2014VQE}%
  \BibitemOpen
  \bibfield  {author} {\bibinfo {author} {\bibfnamefont {A.}~\bibnamefont
  {Peruzzo}}, \bibinfo {author} {\bibfnamefont {J.}~\bibnamefont {McClean}},
  \bibinfo {author} {\bibfnamefont {P.}~\bibnamefont {Shadbolt}}, \bibinfo
  {author} {\bibfnamefont {M.-H.}\ \bibnamefont {Yung}}, \bibinfo {author}
  {\bibfnamefont {X.-Q.}\ \bibnamefont {Zhou}}, \bibinfo {author}
  {\bibfnamefont {P.~J.}\ \bibnamefont {Love}}, \bibinfo {author}
  {\bibfnamefont {A.}~\bibnamefont {Aspuru-Guzik}}, \ and\ \bibinfo {author}
  {\bibfnamefont {J.~L.}\ \bibnamefont {O'Brien}},\ }\bibfield  {title} {\emph
  {\bibinfo {title} {A variational eigenvalue solver on a photonic quantum
  processor},\ }}\href {\doibasemod 10.1038/ncomms5213} {\bibfield  {journal}
  {\bibinfo  {journal} {Nature Communications}\ }\textbf {\bibinfo {volume}
  {5}},\ \bibinfo {pages} {4213} (\bibinfo {year} {2014})}\BibitemShut
  {NoStop}%
\bibitem [{\citenamefont {Johnson}\ \emph {et~al.}(2017)\citenamefont
  {Johnson}, \citenamefont {Romero}, \citenamefont {Olson}, \citenamefont
  {Cao},\ and\ \citenamefont {Aspuru-Guzik}}]{johnson2017qvector}%
  \BibitemOpen
  \bibfield  {author} {\bibinfo {author} {\bibfnamefont {P.~D.}\ \bibnamefont
  {Johnson}}, \bibinfo {author} {\bibfnamefont {J.}~\bibnamefont {Romero}},
  \bibinfo {author} {\bibfnamefont {J.}~\bibnamefont {Olson}}, \bibinfo
  {author} {\bibfnamefont {Y.}~\bibnamefont {Cao}}, \ and\ \bibinfo {author}
  {\bibfnamefont {A.}~\bibnamefont {Aspuru-Guzik}},\ }\bibfield  {title} {\emph
  {\bibinfo {title} {{QVECTOR}: an algorithm for device-tailored quantum error
  correction},\ }}\href {https://arxiv.org/abs/1711.02249} {\bibfield
  {journal} {\bibinfo  {journal} {arXiv:1711.02249}\ } (\bibinfo {year}
  {2017})}\BibitemShut {NoStop}%
\bibitem [{\citenamefont {{Benedetti}}\ \emph
  {et~al.}(2018{\natexlab{a}})\citenamefont {{Benedetti}}, \citenamefont
  {{Garcia-Pintos}}, \citenamefont {{Perdomo}}, \citenamefont
  {{Leyton-Ortega}}, \citenamefont {{Nam}},\ and\ \citenamefont
  {{Perdomo-Ortiz}}}]{benedetti2018generative}%
  \BibitemOpen
  \bibfield  {author} {\bibinfo {author} {\bibfnamefont {M.}~\bibnamefont
  {{Benedetti}}}, \bibinfo {author} {\bibfnamefont {D.}~\bibnamefont
  {{Garcia-Pintos}}}, \bibinfo {author} {\bibfnamefont {O.}~\bibnamefont
  {{Perdomo}}}, \bibinfo {author} {\bibfnamefont {V.}~\bibnamefont
  {{Leyton-Ortega}}}, \bibinfo {author} {\bibfnamefont {Y.}~\bibnamefont
  {{Nam}}}, \ and\ \bibinfo {author} {\bibfnamefont {A.}~\bibnamefont
  {{Perdomo-Ortiz}}},\ }\bibfield  {title} {\emph {\bibinfo {title} {A
  generative modeling approach for benchmarking and training shallow quantum
  circuits},\ }}\href {https://arxiv.org/abs/1801.07686} {\bibfield  {journal}
  {\bibinfo  {journal} {arXiv:1801.07686}\ } (\bibinfo {year}
  {2018}{\natexlab{a}})}\BibitemShut {NoStop}%
\bibitem [{\citenamefont {Mitarai}\ \emph {et~al.}(2018)\citenamefont
  {Mitarai}, \citenamefont {Negoro}, \citenamefont {Kitagawa},\ and\
  \citenamefont {Fujii}}]{mitarai2018quantum}%
  \BibitemOpen
  \bibfield  {author} {\bibinfo {author} {\bibfnamefont {K.}~\bibnamefont
  {Mitarai}}, \bibinfo {author} {\bibfnamefont {M.}~\bibnamefont {Negoro}},
  \bibinfo {author} {\bibfnamefont {M.}~\bibnamefont {Kitagawa}}, \ and\
  \bibinfo {author} {\bibfnamefont {K.}~\bibnamefont {Fujii}},\ }\bibfield
  {title} {\emph {\bibinfo {title} {Quantum circuit learning},\ }}\href
  {\doibasemod 10.1103/PhysRevA.98.032309} {\bibfield  {journal} {\bibinfo
  {journal} {Physical Review A}\ }\textbf {\bibinfo {volume} {98}},\ \bibinfo
  {eid} {032309} (\bibinfo {year} {2018})}\BibitemShut {NoStop}%
\bibitem [{\citenamefont {{Verdon}}\ \emph {et~al.}(2018)\citenamefont
  {{Verdon}}, \citenamefont {{Pye}},\ and\ \citenamefont
  {{Broughton}}}]{verdon2018universal}%
  \BibitemOpen
  \bibfield  {author} {\bibinfo {author} {\bibfnamefont {G.}~\bibnamefont
  {{Verdon}}}, \bibinfo {author} {\bibfnamefont {J.}~\bibnamefont {{Pye}}}, \
  and\ \bibinfo {author} {\bibfnamefont {M.}~\bibnamefont {{Broughton}}},\
  }\bibfield  {title} {\emph {\bibinfo {title} {{A Universal Training Algorithm
  for Quantum Deep Learning}},\ }}\href {https://arxiv.org/abs/1806.09729}
  {\bibfield  {journal} {\bibinfo  {journal} {arXiv:1806.09729}\ } (\bibinfo
  {year} {2018})}\BibitemShut {NoStop}%
\bibitem [{\citenamefont {{Romero}}\ \emph {et~al.}(2017)\citenamefont
  {{Romero}}, \citenamefont {{Olson}},\ and\ \citenamefont
  {{Aspuru-Guzik}}}]{Romero17}%
  \BibitemOpen
  \bibfield  {author} {\bibinfo {author} {\bibfnamefont {J.}~\bibnamefont
  {{Romero}}}, \bibinfo {author} {\bibfnamefont {J.~P.}\ \bibnamefont
  {{Olson}}}, \ and\ \bibinfo {author} {\bibfnamefont {A.}~\bibnamefont
  {{Aspuru-Guzik}}},\ }\bibfield  {title} {\emph {\bibinfo {title} {{Quantum
  autoencoders for efficient compression of quantum data}},\ }}\href
  {\doibasemod 10.1088/2058-9565/aa8072} {\bibfield  {journal} {\bibinfo
  {journal} {Quantum Science and Technology}\ }\textbf {\bibinfo {volume}
  {2}},\ \bibinfo {pages} {045001} (\bibinfo {year} {2017})}\BibitemShut
  {NoStop}%
\bibitem [{\citenamefont {{Romero}}\ \emph {et~al.}(2018)\citenamefont
  {{Romero}}, \citenamefont {{Olson}},\ and\ \citenamefont
  {{Aspuru-Guzik}}}]{Romero18}%
  \BibitemOpen
  \bibfield  {author} {\bibinfo {author} {\bibfnamefont {J.}~\bibnamefont
  {{Romero}}}, \bibinfo {author} {\bibfnamefont {J.~P.}\ \bibnamefont
  {{Olson}}}, \ and\ \bibinfo {author} {\bibfnamefont {A.}~\bibnamefont
  {{Aspuru-Guzik}}},\ }\bibfield  {title} {\emph {\bibinfo {title} {{Quantum
  autoencoders for short depth quantum circuit synthesis}},\ }}\href
  {https://github.com/zapatacomputing/cusp_cirq_demo/blob/master/cusp_protocol.pdf}
  {\bibfield  {journal} {\bibinfo  {journal} {GitHub article}\ } (\bibinfo
  {year} {2018})}\BibitemShut {NoStop}%
\bibitem [{\citenamefont {{Dive}}\ \emph {et~al.}(2018)\citenamefont {{Dive}},
  \citenamefont {{Pitchford}}, \citenamefont {{Mintert}},\ and\ \citenamefont
  {{Burgarth}}}]{Dive17}%
  \BibitemOpen
  \bibfield  {author} {\bibinfo {author} {\bibfnamefont {B.}~\bibnamefont
  {{Dive}}}, \bibinfo {author} {\bibfnamefont {A.}~\bibnamefont {{Pitchford}}},
  \bibinfo {author} {\bibfnamefont {F.}~\bibnamefont {{Mintert}}}, \ and\
  \bibinfo {author} {\bibfnamefont {D.}~\bibnamefont {{Burgarth}}},\ }\bibfield
   {title} {\emph {\bibinfo {title} {{In situ upgrade of quantum simulators to
  universal computers}},\ }}\href {\doibasemod 10.22331/q-2018-08-08-80}
  {\bibfield  {journal} {\bibinfo  {journal} {{Quantum}}\ }\textbf {\bibinfo
  {volume} {2}},\ \bibinfo {pages} {80} (\bibinfo {year} {2018})}\BibitemShut
  {NoStop}%
\bibitem [{\citenamefont {Knill}\ and\ \citenamefont
  {Laflamme}(1998)}]{knill1998POOQ}%
  \BibitemOpen
  \bibfield  {author} {\bibinfo {author} {\bibfnamefont {E.}~\bibnamefont
  {Knill}}\ and\ \bibinfo {author} {\bibfnamefont {R.}~\bibnamefont
  {Laflamme}},\ }\bibfield  {title} {\emph {\bibinfo {title} {Power of one bit
  of quantum information},\ }}\href {\doibasemod 10.1103/PhysRevLett.81.5672}
  {\bibfield  {journal} {\bibinfo  {journal} {Physical Review Letters}\
  }\textbf {\bibinfo {volume} {81}},\ \bibinfo {pages} {5672} (\bibinfo {year}
  {1998})}\BibitemShut {NoStop}%
\bibitem [{\citenamefont {Fujii}\ \emph {et~al.}(2018)\citenamefont {Fujii},
  \citenamefont {Kobayashi}, \citenamefont {Morimae}, \citenamefont
  {Nishimura}, \citenamefont {Tamate},\ and\ \citenamefont
  {Tani}}]{DBLP:journals/corr/FujiiKMNTT14}%
  \BibitemOpen
  \bibfield  {author} {\bibinfo {author} {\bibfnamefont {K.}~\bibnamefont
  {Fujii}}, \bibinfo {author} {\bibfnamefont {H.}~\bibnamefont {Kobayashi}},
  \bibinfo {author} {\bibfnamefont {T.}~\bibnamefont {Morimae}}, \bibinfo
  {author} {\bibfnamefont {H.}~\bibnamefont {Nishimura}}, \bibinfo {author}
  {\bibfnamefont {S.}~\bibnamefont {Tamate}}, \ and\ \bibinfo {author}
  {\bibfnamefont {S.}~\bibnamefont {Tani}},\ }\bibfield  {title} {\emph
  {\bibinfo {title} {{Impossibility of Classically Simulating One-Clean-Qubit
  Model with Multiplicative Error}},\ }}\href {\doibasemod
  10.1103/PhysRevLett.120.200502} {\bibfield  {journal} {\bibinfo  {journal}
  {Physical Review Letters}\ }\textbf {\bibinfo {volume} {120}},\ \bibinfo
  {eid} {200502} (\bibinfo {year} {2018})}\BibitemShut {NoStop}%
\bibitem [{\citenamefont {Rosgen}\ and\ \citenamefont
  {Watrous}(2005)}]{DBLP:journals/corr/cs-CC-0407056}%
  \BibitemOpen
  \bibfield  {author} {\bibinfo {author} {\bibfnamefont {B.}~\bibnamefont
  {Rosgen}}\ and\ \bibinfo {author} {\bibfnamefont {J.}~\bibnamefont
  {Watrous}},\ }in\ \href {\doibasemod 10.1109/CCC.2005.21} {\emph {\bibinfo
  {booktitle} {20th Annual IEEE Conference on Computational Complexity
  (CCC'05)}}}\ (\bibinfo {year} {2005})\ pp.\ \bibinfo {pages}
  {344--354}\BibitemShut {NoStop}%
\bibitem [{\citenamefont {Smith}\ \emph {et~al.}(2016)\citenamefont {Smith},
  \citenamefont {Curtis},\ and\ \citenamefont {Zeng}}]{rigetti}%
  \BibitemOpen
  \bibfield  {author} {\bibinfo {author} {\bibfnamefont {R.~S.}\ \bibnamefont
  {Smith}}, \bibinfo {author} {\bibfnamefont {M.~J.}\ \bibnamefont {Curtis}}, \
  and\ \bibinfo {author} {\bibfnamefont {W.~J.}\ \bibnamefont {Zeng}},\
  }\bibfield  {title} {\emph {\bibinfo {title} {A practical quantum instruction
  set architecture},\ }}\href {https://arxiv.org/abs/1608.03355} {\bibfield
  {journal} {\bibinfo  {journal} {arXiv:1608.03355}\ } (\bibinfo {year}
  {2016})}\BibitemShut {NoStop}%
\bibitem [{\citenamefont {{Cross}}\ \emph {et~al.}(2017)\citenamefont
  {{Cross}}, \citenamefont {{Bishop}}, \citenamefont {{Smolin}},\ and\
  \citenamefont {{Gambetta}}}]{cross17ibm}%
  \BibitemOpen
  \bibfield  {author} {\bibinfo {author} {\bibfnamefont {A.~W.}\ \bibnamefont
  {{Cross}}}, \bibinfo {author} {\bibfnamefont {L.~S.}\ \bibnamefont
  {{Bishop}}}, \bibinfo {author} {\bibfnamefont {J.~A.}\ \bibnamefont
  {{Smolin}}}, \ and\ \bibinfo {author} {\bibfnamefont {J.~M.}\ \bibnamefont
  {{Gambetta}}},\ }\bibfield  {title} {\emph {\bibinfo {title} {{Open Quantum
  Assembly Language}},\ }}\href {https://arxiv.org/abs/1707.03429} {\bibfield
  {journal} {\bibinfo  {journal} {arXiv:1707.03429}\ } (\bibinfo {year}
  {2017})}\BibitemShut {NoStop}%
\bibitem [{\citenamefont {Nielsen}\ and\ \citenamefont {Chuang}(2000)}]{NC00}%
  \BibitemOpen
  \bibfield  {author} {\bibinfo {author} {\bibfnamefont {M.~A.}\ \bibnamefont
  {Nielsen}}\ and\ \bibinfo {author} {\bibfnamefont {I.~L.}\ \bibnamefont
  {Chuang}},\ }\href {\doibasemod 10.1017/CBO9780511976667} {\emph {\bibinfo
  {title} {Quantum {C}omputation and Quantum {I}nformation}}}\ (\bibinfo
  {publisher} {Cambridge University Press},\ \bibinfo {year}
  {2000})\BibitemShut {NoStop}%
\bibitem [{\citenamefont {Kitaev}(1997)}]{K97}%
  \BibitemOpen
  \bibfield  {author} {\bibinfo {author} {\bibfnamefont {A.}~\bibnamefont
  {Kitaev}},\ }\bibfield  {title} {\emph {\bibinfo {title} {Quantum
  computations: algorithms and error correction},\ }}\href {\doibasemod
  10.1070/RM1997v052n06ABEH002155} {\bibfield  {journal} {\bibinfo  {journal}
  {Russian Mathematical Surveys}\ }\textbf {\bibinfo {volume} {52}},\ \bibinfo
  {pages} {1191} (\bibinfo {year} {1997})}\BibitemShut {NoStop}%
\bibitem [{\citenamefont {Dawson}\ and\ \citenamefont {Nielsen}(2006)}]{DN06}%
  \BibitemOpen
  \bibfield  {author} {\bibinfo {author} {\bibfnamefont {C.~M.}\ \bibnamefont
  {Dawson}}\ and\ \bibinfo {author} {\bibfnamefont {M.~A.}\ \bibnamefont
  {Nielsen}},\ }\bibfield  {title} {\emph {\bibinfo {title} {The
  {S}olovay-{K}itaev algorithm},\ }}\href
  {http://dl.acm.org/citation.cfm?id=2011679.2011685} {\bibfield  {journal}
  {\bibinfo  {journal} {Quantum Information and Compututation}\ }\textbf
  {\bibinfo {volume} {6}},\ \bibinfo {pages} {81} (\bibinfo {year}
  {2006})}\BibitemShut {NoStop}%
\bibitem [{\citenamefont {Pham}\ \emph {et~al.}(2013)\citenamefont {Pham},
  \citenamefont {Van~Meter},\ and\ \citenamefont {Horsman}}]{PVH13}%
  \BibitemOpen
  \bibfield  {author} {\bibinfo {author} {\bibfnamefont {T.~T.}\ \bibnamefont
  {Pham}}, \bibinfo {author} {\bibfnamefont {R.}~\bibnamefont {Van~Meter}}, \
  and\ \bibinfo {author} {\bibfnamefont {C.}~\bibnamefont {Horsman}},\
  }\bibfield  {title} {\emph {\bibinfo {title} {Optimization of the
  {S}olovay-{K}itaev algorithm},\ }}\href {\doibasemod
  10.1103/PhysRevA.87.052332} {\bibfield  {journal} {\bibinfo  {journal}
  {Physical Review A}\ }\textbf {\bibinfo {volume} {87}},\ \bibinfo {pages}
  {052332} (\bibinfo {year} {2013})}\BibitemShut {NoStop}%
\bibitem [{\citenamefont {Kliuchnikov}\ \emph {et~al.}(2013)\citenamefont
  {Kliuchnikov}, \citenamefont {Maslov},\ and\ \citenamefont {Mosca}}]{KMM13}%
  \BibitemOpen
  \bibfield  {author} {\bibinfo {author} {\bibfnamefont {V.}~\bibnamefont
  {Kliuchnikov}}, \bibinfo {author} {\bibfnamefont {D.}~\bibnamefont {Maslov}},
  \ and\ \bibinfo {author} {\bibfnamefont {M.}~\bibnamefont {Mosca}},\
  }\bibfield  {title} {\emph {\bibinfo {title} {Asymptotically optimal
  approximation of single qubit unitaries by {C}lifford and {T} circuits using
  a constant number of ancillary qubits},\ }}\href {\doibasemod
  10.1103/PhysRevLett.110.190502} {\bibfield  {journal} {\bibinfo  {journal}
  {Physical Review Letters}\ }\textbf {\bibinfo {volume} {110}},\ \bibinfo
  {pages} {190502} (\bibinfo {year} {2013})}\BibitemShut {NoStop}%
\bibitem [{\citenamefont {Kliuchnikov}\ \emph {et~al.}(2014)\citenamefont
  {Kliuchnikov}, \citenamefont {Bocharov},\ and\ \citenamefont
  {Svore}}]{KBS14}%
  \BibitemOpen
  \bibfield  {author} {\bibinfo {author} {\bibfnamefont {V.}~\bibnamefont
  {Kliuchnikov}}, \bibinfo {author} {\bibfnamefont {A.}~\bibnamefont
  {Bocharov}}, \ and\ \bibinfo {author} {\bibfnamefont {K.~M.}\ \bibnamefont
  {Svore}},\ }\bibfield  {title} {\emph {\bibinfo {title} {Asymptotically
  optimal topological quantum compiling},\ }}\href {\doibasemod
  10.1103/PhysRevLett.112.140504} {\bibfield  {journal} {\bibinfo  {journal}
  {Physical Review Letters}\ }\textbf {\bibinfo {volume} {112}},\ \bibinfo
  {pages} {140504} (\bibinfo {year} {2014})}\BibitemShut {NoStop}%
\bibitem [{\citenamefont {Zhiyenbayev}\ \emph {et~al.}(2018)\citenamefont
  {Zhiyenbayev}, \citenamefont {Akulin},\ and\ \citenamefont
  {Mandilara}}]{ZAM18}%
  \BibitemOpen
  \bibfield  {author} {\bibinfo {author} {\bibfnamefont {Y.}~\bibnamefont
  {Zhiyenbayev}}, \bibinfo {author} {\bibfnamefont {V.~M.}\ \bibnamefont
  {Akulin}}, \ and\ \bibinfo {author} {\bibfnamefont {A.}~\bibnamefont
  {Mandilara}},\ }\bibfield  {title} {\emph {\bibinfo {title} {Quantum
  compiling with diffusive sets of gates},\ }}\href {\doibasemod
  10.1103/PhysRevA.98.012325} {\bibfield  {journal} {\bibinfo  {journal}
  {Physical Review A}\ }\textbf {\bibinfo {volume} {98}},\ \bibinfo {pages}
  {012325} (\bibinfo {year} {2018})}\BibitemShut {NoStop}%
\bibitem [{\citenamefont {Horodecki}\ \emph {et~al.}(1999)\citenamefont
  {Horodecki}, \citenamefont {Horodecki},\ and\ \citenamefont
  {Horodecki}}]{HHH99}%
  \BibitemOpen
  \bibfield  {author} {\bibinfo {author} {\bibfnamefont {M.}~\bibnamefont
  {Horodecki}}, \bibinfo {author} {\bibfnamefont {P.}~\bibnamefont
  {Horodecki}}, \ and\ \bibinfo {author} {\bibfnamefont {R.}~\bibnamefont
  {Horodecki}},\ }\bibfield  {title} {\emph {\bibinfo {title} {General
  teleportation channel, singlet fraction, and quasidistillation},\ }}\href
  {\doibasemod 10.1103/PhysRevA.60.1888} {\bibfield  {journal} {\bibinfo
  {journal} {Physical Review A}\ }\textbf {\bibinfo {volume} {60}},\ \bibinfo
  {pages} {1888} (\bibinfo {year} {1999})}\BibitemShut {NoStop}%
\bibitem [{\citenamefont {Nielsen}(2002)}]{nielsen02}%
  \BibitemOpen
  \bibfield  {author} {\bibinfo {author} {\bibfnamefont {M.~A.}\ \bibnamefont
  {Nielsen}},\ }\bibfield  {title} {\emph {\bibinfo {title} {A simple formula
  for the average gate fidelity of a quantum dynamical operation},\ }}\href
  {\doibasemod 10.1016/S0375-9601(02)01272-0} {\bibfield  {journal} {\bibinfo
  {journal} {Physics Letters A}\ }\textbf {\bibinfo {volume} {303}},\ \bibinfo
  {pages} {249} (\bibinfo {year} {2002})}\BibitemShut {NoStop}%
\bibitem [{\citenamefont {Gepp}\ and\ \citenamefont {Stocks}(2009)}]{Gepp2009}%
  \BibitemOpen
  \bibfield  {author} {\bibinfo {author} {\bibfnamefont {A.}~\bibnamefont
  {Gepp}}\ and\ \bibinfo {author} {\bibfnamefont {P.}~\bibnamefont {Stocks}},\
  }\bibfield  {title} {\emph {\bibinfo {title} {A review of procedures to
  evolve quantum algorithms},\ }}\href {\doibasemod 10.1007/s10710-009-9080-7}
  {\bibfield  {journal} {\bibinfo  {journal} {Genetic Programming and Evolvable
  Machines}\ }\textbf {\bibinfo {volume} {10}},\ \bibinfo {pages} {181}
  (\bibinfo {year} {2009})}\BibitemShut {NoStop}%
\bibitem [{\citenamefont {Suzuki}(1990)}]{SUZUKI1990319}%
  \BibitemOpen
  \bibfield  {author} {\bibinfo {author} {\bibfnamefont {M.}~\bibnamefont
  {Suzuki}},\ }\bibfield  {title} {\emph {\bibinfo {title} {Fractal
  decomposition of exponential operators with applications to many-body
  theories and monte carlo simulations},\ }}\href {\doibasemod
  10.1016/0375-9601(90)90962-N} {\bibfield  {journal} {\bibinfo  {journal}
  {Physics Letters A}\ }\textbf {\bibinfo {volume} {146}},\ \bibinfo {pages}
  {319 } (\bibinfo {year} {1990})}\BibitemShut {NoStop}%
\bibitem [{\citenamefont {Jones}\ and\ \citenamefont
  {Benjamin}(2018)}]{jones2018quantum}%
  \BibitemOpen
  \bibfield  {author} {\bibinfo {author} {\bibfnamefont {T.}~\bibnamefont
  {Jones}}\ and\ \bibinfo {author} {\bibfnamefont {S.~C.}\ \bibnamefont
  {Benjamin}},\ }\bibfield  {title} {\emph {\bibinfo {title} {Quantum
  compilation and circuit optimisation via energy dissipation},\ }}\href
  {https://arxiv.org/abs/1811.03147} {\bibfield  {journal} {\bibinfo  {journal}
  {arXiv:1811.03147}\ } (\bibinfo {year} {2018})}\BibitemShut {NoStop}%
\bibitem [{\citenamefont {Garcia-Escartin}\ and\ \citenamefont
  {Chamorro-Posada}(2013)}]{garcia2013swap}%
  \BibitemOpen
  \bibfield  {author} {\bibinfo {author} {\bibfnamefont {J.~C.}\ \bibnamefont
  {Garcia-Escartin}}\ and\ \bibinfo {author} {\bibfnamefont {P.}~\bibnamefont
  {Chamorro-Posada}},\ }\bibfield  {title} {\emph {\bibinfo {title} {Swap test
  and {H}ong-{O}u-{M}andel effect are equivalent},\ }}\href {\doibasemod
  10.1103/PhysRevA.87.052330} {\bibfield  {journal} {\bibinfo  {journal}
  {Physical Review A}\ }\textbf {\bibinfo {volume} {87}},\ \bibinfo {pages}
  {052330} (\bibinfo {year} {2013})}\BibitemShut {NoStop}%
\bibitem [{\citenamefont {Shor}\ and\ \citenamefont
  {Jordan}(2008)}]{DBLP:journals/qic/ShorJ08}%
  \BibitemOpen
  \bibfield  {author} {\bibinfo {author} {\bibfnamefont {P.~W.}\ \bibnamefont
  {Shor}}\ and\ \bibinfo {author} {\bibfnamefont {S.~P.}\ \bibnamefont
  {Jordan}},\ }\bibfield  {title} {\emph {\bibinfo {title} {Estimating jones
  polynomials is a complete problem for one clean qubit},\ }}\href
  {http://www.rintonpress.com/xxqic8/qic-8-89/0681-0714.pdf} {\bibfield
  {journal} {\bibinfo  {journal} {Quantum Information {\&} Computation}\
  }\textbf {\bibinfo {volume} {8}},\ \bibinfo {pages} {681} (\bibinfo {year}
  {2008})}\BibitemShut {NoStop}%
\bibitem [{IBM(2018{\natexlab{a}})}]{IBMQ5}%
  \BibitemOpen
  \href
  {https://github.com/QISKit/qiskit-backend-information/tree/master/backends/tenerife/V1}
  {\bibinfo {title} {{IBM} {Q} 5 {T}enerife backend specification},\ }
  (\bibinfo {year} {2018}{\natexlab{a}})\BibitemShut {NoStop}%
\bibitem [{IBM(2018{\natexlab{b}})}]{IBMQ16}%
  \BibitemOpen
  \href
  {https://github.com/Qiskit/qiskit-backend-information/tree/master/backends/rueschlikon/V1}
  {\bibinfo {title} {{IBM} {Q} 16 {R}ueschlikon backend specification},\ }
  (\bibinfo {year} {2018}{\natexlab{b}})\BibitemShut {NoStop}%
\bibitem [{rig(2018)}]{rigettiQPU}%
  \BibitemOpen
  \href {http://docs.rigetti.com/en/latest/qpu.html} {\bibinfo {title}
  {{Rigetti} {8Q-Agave} specification v.2.0.0.dev0},\ } (\bibinfo {year}
  {2018})\BibitemShut {NoStop}%
\bibitem [{\citenamefont {McClean}\ \emph {et~al.}(2018)\citenamefont
  {McClean}, \citenamefont {Boixo}, \citenamefont {Smelyanskiy}, \citenamefont
  {Babbush},\ and\ \citenamefont {Neven}}]{mcclean2018barren}%
  \BibitemOpen
  \bibfield  {author} {\bibinfo {author} {\bibfnamefont {J.~R.}\ \bibnamefont
  {McClean}}, \bibinfo {author} {\bibfnamefont {S.}~\bibnamefont {Boixo}},
  \bibinfo {author} {\bibfnamefont {V.~N.}\ \bibnamefont {Smelyanskiy}},
  \bibinfo {author} {\bibfnamefont {R.}~\bibnamefont {Babbush}}, \ and\
  \bibinfo {author} {\bibfnamefont {H.}~\bibnamefont {Neven}},\ }\bibfield
  {title} {\emph {\bibinfo {title} {Barren plateaus in quantum neural network
  training landscapes},\ }}\href {\doibasemod 10.1038/s41467-018-07090-4}
  {\bibfield  {journal} {\bibinfo  {journal} {Nature Communications}\ }\textbf
  {\bibinfo {volume} {9}},\ \bibinfo {eid} {4812} (\bibinfo {year}
  {2018})}\BibitemShut {NoStop}%
\bibitem [{\citenamefont {Day}\ \emph {et~al.}(2019)\citenamefont {Day},
  \citenamefont {Bukov}, \citenamefont {Weinberg}, \citenamefont {Mehta},\ and\
  \citenamefont {Sels}}]{DayEtAl2018}%
  \BibitemOpen
  \bibfield  {author} {\bibinfo {author} {\bibfnamefont {A.~G.~R.}\
  \bibnamefont {Day}}, \bibinfo {author} {\bibfnamefont {M.}~\bibnamefont
  {Bukov}}, \bibinfo {author} {\bibfnamefont {P.}~\bibnamefont {Weinberg}},
  \bibinfo {author} {\bibfnamefont {P.}~\bibnamefont {Mehta}}, \ and\ \bibinfo
  {author} {\bibfnamefont {D.}~\bibnamefont {Sels}},\ }\bibfield  {title}
  {\emph {\bibinfo {title} {Glassy phase of optimal quantum control},\ }}\href
  {\doibasemod 10.1103/PhysRevLett.122.020601} {\bibfield  {journal} {\bibinfo
  {journal} {Physical Review Letters}\ }\textbf {\bibinfo {volume} {122}},\
  \bibinfo {pages} {020601} (\bibinfo {year} {2019})}\BibitemShut {NoStop}%
\bibitem [{\citenamefont {Glorot}\ and\ \citenamefont
  {Bengio}(2010)}]{GlorotBengio2010}%
  \BibitemOpen
  \bibfield  {author} {\bibinfo {author} {\bibfnamefont {X.}~\bibnamefont
  {Glorot}}\ and\ \bibinfo {author} {\bibfnamefont {Y.}~\bibnamefont
  {Bengio}},\ }in\ \href
  {http://proceedings.mlr.press/v9/glorot10a/glorot10a.pdf?hc_location=ufi}
  {\emph {\bibinfo {booktitle} {In Proceedings of the International Conference
  on Artificial Intelligence and Statistics}}}\ (\bibinfo {year} {2010})\ pp.\
  \bibinfo {pages} {249--256}\BibitemShut {NoStop}%
\bibitem [{\citenamefont {{Benedetti}}\ \emph
  {et~al.}(2018{\natexlab{b}})\citenamefont {{Benedetti}}, \citenamefont
  {{Garcia-Pintos}}, \citenamefont {{Perdomo}}, \citenamefont
  {{Leyton-Ortega}}, \citenamefont {{Nam}},\ and\ \citenamefont
  {{Perdomo-Ortiz}}}]{BenedettiEtAl2018}%
  \BibitemOpen
  \bibfield  {author} {\bibinfo {author} {\bibfnamefont {M.}~\bibnamefont
  {{Benedetti}}}, \bibinfo {author} {\bibfnamefont {D.}~\bibnamefont
  {{Garcia-Pintos}}}, \bibinfo {author} {\bibfnamefont {O.}~\bibnamefont
  {{Perdomo}}}, \bibinfo {author} {\bibfnamefont {V.}~\bibnamefont
  {{Leyton-Ortega}}}, \bibinfo {author} {\bibfnamefont {Y.}~\bibnamefont
  {{Nam}}}, \ and\ \bibinfo {author} {\bibfnamefont {A.}~\bibnamefont
  {{Perdomo-Ortiz}}},\ }\bibfield  {title} {\emph {\bibinfo {title} {{A
  generative modeling approach for benchmarking and training shallow quantum
  circuits}},\ }}\href {https://arxiv.org/abs/1801.07686} {\bibfield  {journal}
  {\bibinfo  {journal} {arXiv:1801.07686}\ } (\bibinfo {year}
  {2018}{\natexlab{b}})}\BibitemShut {NoStop}%
\bibitem [{\citenamefont {LaRose}\ \emph {et~al.}(2018)\citenamefont {LaRose},
  \citenamefont {Tikku}, \citenamefont {O'Neel-Judy}, \citenamefont {Cincio},\
  and\ \citenamefont {Coles}}]{larose2018variational}%
  \BibitemOpen
  \bibfield  {author} {\bibinfo {author} {\bibfnamefont {R.}~\bibnamefont
  {LaRose}}, \bibinfo {author} {\bibfnamefont {A.}~\bibnamefont {Tikku}},
  \bibinfo {author} {\bibfnamefont {{\'E}.}~\bibnamefont {O'Neel-Judy}},
  \bibinfo {author} {\bibfnamefont {L.}~\bibnamefont {Cincio}}, \ and\ \bibinfo
  {author} {\bibfnamefont {P.~J.}\ \bibnamefont {Coles}},\ }\bibfield  {title}
  {\emph {\bibinfo {title} {Variational quantum state diagonalization},\
  }}\href {https://arxiv.org/abs/1810.10506} {\bibfield  {journal} {\bibinfo
  {journal} {arXiv:1810.10506}\ } (\bibinfo {year} {2018})}\BibitemShut
  {NoStop}%
\bibitem [{\citenamefont {Kandala}\ \emph {et~al.}(2018)\citenamefont
  {Kandala}, \citenamefont {Temme}, \citenamefont {Corcoles}, \citenamefont
  {Mezzacapo}, \citenamefont {Chow},\ and\ \citenamefont {Gambetta}}]{KTC18}%
  \BibitemOpen
  \bibfield  {author} {\bibinfo {author} {\bibfnamefont {A.}~\bibnamefont
  {Kandala}}, \bibinfo {author} {\bibfnamefont {K.}~\bibnamefont {Temme}},
  \bibinfo {author} {\bibfnamefont {A.~D.}\ \bibnamefont {Corcoles}}, \bibinfo
  {author} {\bibfnamefont {A.}~\bibnamefont {Mezzacapo}}, \bibinfo {author}
  {\bibfnamefont {J.~M.}\ \bibnamefont {Chow}}, \ and\ \bibinfo {author}
  {\bibfnamefont {J.~M.}\ \bibnamefont {Gambetta}},\ }\bibfield  {title} {\emph
  {\bibinfo {title} {Extending the computational reach of a noisy
  superconducting quantum processor},\ }}\href {\doibasemod
  10.1038/s41586-019-1040-7} {\bibfield  {journal} {\bibinfo  {journal}
  {Nature}\ }\textbf {\bibinfo {volume} {567}},\ \bibinfo {pages} {491}
  (\bibinfo {year} {2018})}\BibitemShut {NoStop}%
\bibitem [{sci(2018{\natexlab{a}})}]{scikit-opt}%
  \BibitemOpen
  \href {https://github.com/scikit-optimize/scikit-optimize} {\bibinfo {title}
  {Scikit-optimize},\ } (\bibinfo {year} {2018}{\natexlab{a}})\BibitemShut
  {NoStop}%
\bibitem [{\citenamefont {Mo{\v{c}}kus}(1975)}]{mockus1975BayesOpt}%
  \BibitemOpen
  \bibfield  {author} {\bibinfo {author} {\bibfnamefont {J.}~\bibnamefont
  {Mo{\v{c}}kus}},\ }in\ \href {\doibasemod 10.1007/978-3-662-38527-2_55}
  {\emph {\bibinfo {booktitle} {Optimization Techniques IFIP Technical
  Conference Novosibirsk, July 1--7, 1974}}}\ (\bibinfo  {publisher} {Springer
  Berlin Heidelberg},\ \bibinfo {address} {Berlin, Heidelberg},\ \bibinfo
  {year} {1975})\ pp.\ \bibinfo {pages} {400--404}\BibitemShut {NoStop}%
\bibitem [{\citenamefont {Osborne}\ \emph {et~al.}(2009)\citenamefont
  {Osborne}, \citenamefont {Garnett},\ and\ \citenamefont
  {Roberts}}]{osborne2009Gaussian}%
  \BibitemOpen
  \bibfield  {author} {\bibinfo {author} {\bibfnamefont {M.~A.}\ \bibnamefont
  {Osborne}}, \bibinfo {author} {\bibfnamefont {R.}~\bibnamefont {Garnett}}, \
  and\ \bibinfo {author} {\bibfnamefont {S.~J.}\ \bibnamefont {Roberts}},\ }in\
  \href
  {https://www.cse.wustl.edu/~garnett/files/papers/osborne_et_al_lion_2009.pdf}
  {\emph {\bibinfo {booktitle} {3rd International Conference on Learning and
  Intelligent Optimization (LION3) 2009}}}\ (\bibinfo {year}
  {2009})\BibitemShut {NoStop}%
\bibitem [{\citenamefont {Rebentrost}\ \emph {et~al.}(2016)\citenamefont
  {Rebentrost}, \citenamefont {Schuld}, \citenamefont {Wossnig}, \citenamefont
  {Petruccione},\ and\ \citenamefont {Lloyd}}]{RS16}%
  \BibitemOpen
  \bibfield  {author} {\bibinfo {author} {\bibfnamefont {P.}~\bibnamefont
  {Rebentrost}}, \bibinfo {author} {\bibfnamefont {M.}~\bibnamefont {Schuld}},
  \bibinfo {author} {\bibfnamefont {L.}~\bibnamefont {Wossnig}}, \bibinfo
  {author} {\bibfnamefont {F.}~\bibnamefont {Petruccione}}, \ and\ \bibinfo
  {author} {\bibfnamefont {S.}~\bibnamefont {Lloyd}},\ }\bibfield  {title}
  {\emph {\bibinfo {title} {Quantum gradient descent and {N}ewton's method for
  constrained polynomial optimization},\ }}\href
  {https://arxiv.org/abs/1612.01789} {\bibfield  {journal} {\bibinfo  {journal}
  {arXiv:1612.01789}\ } (\bibinfo {year} {2016})}\BibitemShut {NoStop}%
\bibitem [{\citenamefont {Kerenidis}\ and\ \citenamefont
  {Prakash}(2017)}]{KP17}%
  \BibitemOpen
  \bibfield  {author} {\bibinfo {author} {\bibfnamefont {I.}~\bibnamefont
  {Kerenidis}}\ and\ \bibinfo {author} {\bibfnamefont {A.}~\bibnamefont
  {Prakash}},\ }\bibfield  {title} {\emph {\bibinfo {title} {Quantum gradient
  descent for linear systems and least squares},\ }}\href
  {https://arxiv.org/abs/1704.04992} {\bibfield  {journal} {\bibinfo  {journal}
  {arXiv:1704.04992}\ } (\bibinfo {year} {2017})}\BibitemShut {NoStop}%
\bibitem [{\citenamefont {Gily{\'e}n}\ \emph {et~al.}()\citenamefont
  {Gily{\'e}n}, \citenamefont {Arunachalam},\ and\ \citenamefont
  {Wiebe}}]{GA17}%
  \BibitemOpen
  \bibfield  {author} {\bibinfo {author} {\bibfnamefont {A.}~\bibnamefont
  {Gily{\'e}n}}, \bibinfo {author} {\bibfnamefont {S.}~\bibnamefont
  {Arunachalam}}, \ and\ \bibinfo {author} {\bibfnamefont {N.}~\bibnamefont
  {Wiebe}},\ }\bibinfo {title} {Optimizing quantum optimization algorithms via
  faster quantum gradient computation},\ in\ \href {\doibasemod
  10.1137/1.9781611975482.87} {\emph {\bibinfo {booktitle} {Proceedings of the
  Thirtieth Annual ACM-SIAM Symposium on Discrete Algorithms}}},\ pp.\ \bibinfo
  {pages} {1425--1444}\BibitemShut {NoStop}%
\bibitem [{\citenamefont {Sousa}\ and\ \citenamefont {Ramos}(2007)}]{SR07}%
  \BibitemOpen
  \bibfield  {author} {\bibinfo {author} {\bibfnamefont {P.~B.~M.}\
  \bibnamefont {Sousa}}\ and\ \bibinfo {author} {\bibfnamefont {R.~V.}\
  \bibnamefont {Ramos}},\ }\bibfield  {title} {\emph {\bibinfo {title}
  {Universal quantum circuit for $n$-qubit quantum gate: A programmable quantum
  gate},\ }}\href {http://dl.acm.org/citation.cfm?id=2011717.2011721}
  {\bibfield  {journal} {\bibinfo  {journal} {Quantum Information and
  Computation}\ }\textbf {\bibinfo {volume} {7}},\ \bibinfo {pages} {228}
  (\bibinfo {year} {2007})}\BibitemShut {NoStop}%
\bibitem [{\citenamefont {Vatan}\ and\ \citenamefont {Williams}(2004)}]{VW04}%
  \BibitemOpen
  \bibfield  {author} {\bibinfo {author} {\bibfnamefont {F.}~\bibnamefont
  {Vatan}}\ and\ \bibinfo {author} {\bibfnamefont {C.}~\bibnamefont
  {Williams}},\ }\bibfield  {title} {\emph {\bibinfo {title} {Optimal quantum
  circuits for general two-qubit gates},\ }}\href {\doibasemod
  10.1103/PhysRevA.69.032315} {\bibfield  {journal} {\bibinfo  {journal}
  {Physical Review A}\ }\textbf {\bibinfo {volume} {69}},\ \bibinfo {pages}
  {032315} (\bibinfo {year} {2004})}\BibitemShut {NoStop}%
\bibitem [{sci(2018{\natexlab{b}})}]{scipy-opt}%
  \BibitemOpen
  \href {https://docs.scipy.org/doc/scipy/reference/optimize.html} {\bibinfo
  {title} {Scipy optimization and root finding},\ } (\bibinfo {year}
  {2018}{\natexlab{b}})\BibitemShut {NoStop}%
\bibitem [{\citenamefont {Zhou}\ \emph {et~al.}(2011)\citenamefont {Zhou},
  \citenamefont {Ralph}, \citenamefont {Kalasuwan}, \citenamefont {Zhang},
  \citenamefont {Peruzzo}, \citenamefont {Lanyon},\ and\ \citenamefont
  {{O'Brien}}}]{ZR11}%
  \BibitemOpen
  \bibfield  {author} {\bibinfo {author} {\bibfnamefont {X.-Q.}\ \bibnamefont
  {Zhou}}, \bibinfo {author} {\bibfnamefont {T.~C.}\ \bibnamefont {Ralph}},
  \bibinfo {author} {\bibfnamefont {P.}~\bibnamefont {Kalasuwan}}, \bibinfo
  {author} {\bibfnamefont {M.}~\bibnamefont {Zhang}}, \bibinfo {author}
  {\bibfnamefont {A.}~\bibnamefont {Peruzzo}}, \bibinfo {author} {\bibfnamefont
  {B.~P.}\ \bibnamefont {Lanyon}}, \ and\ \bibinfo {author} {\bibfnamefont
  {J.~L.}\ \bibnamefont {{O'Brien}}},\ }\bibfield  {title} {\emph {\bibinfo
  {title} {Adding control to arbitrary unknown quantum operations},\ }}\href
  {\doibasemod 10.1038/ncomms1392} {\bibfield  {journal} {\bibinfo  {journal}
  {Nature Communications}\ }\textbf {\bibinfo {volume} {2}},\ \bibinfo {eid}
  {413} (\bibinfo {year} {2011})}\BibitemShut {NoStop}%
\end{thebibliography}%
%merlin.mbs apsrev4-1.bst 2010-07-25 4.21a (PWD, AO, DPC) hacked
%Control: key (0)
%Control: author (72) initials jnrlst
%Control: editor formatted (1) identically to author
%Control: production of article title (-1) disabled
%Control: page (0) single
%Control: year (1) truncated
%Control: production of eprint (0) enabled
%

\appendix

\section{Remark on implementation of $V^*$}\label{app:Vstar}

As mentioned in Sec.~\ref{sctcircuits}, a subtle point about evaluating the cost functions $C_{\HST}(U,V_{\vec{k}}(\vec{\alpha}))$ and $C_{\LHST}(U,V_{\vec{k}}(\vec{\alpha}))$ is that the complex conjugate $V_{\vec{k}}(\vec{\alpha})^*$ must be executed on the quantum computer, not $V_{\vec{k}}(\vec{\alpha})$ itself. The complex conjugate of a unitary corresponding to a gate sequence can be obtained by taking the complex conjugate of each unitary in the gate sequence. However, if each gate in the sequence comes from a gate alphabet $\mathcal{A}$, it is possible that the complex conjugate of a gate in the sequence is not contained in the alphabet; for example, if $\mathcal{A}=\{R_x(\pi/2),R_z(\theta)\}$, then the complex conjugate of $R_x(\pi/2)$, which is $R_x(-\pi/2)$, is not contained in $\mathcal{A}$. But the unitary $R_z(\pi)R_x(\pi/2)R_z(\pi)$ is equal (up to a global phase) to $R_x(-\pi/2)$. There are thus two ways to proceed when performing the compilation procedure: during the optimization over the continuous parameters, directly run the gate sequence corresponding to $V_{\vec{k}}(\vec{\alpha})$, expressing it in terms of the native gate alphabet of the quantum computer, then at the end establish the complex conjugate of the optimal unitary as the unitary to which $U$ has been compiled. This would involve translating the complex conjugate of each gate in the optimal sequence into the native gate alphabet of the quantum computer. An alternative is to first take the complex conjugate $V_{\vec{k}}(\vec{\alpha})^*$ by translating the complex conjugate of each gate in the sequence into the native gate alphabet, then execute the resulting sequence on the quantum computer. In each case, we allow for a small-scale classical compiler that can perform the simple translation of the complex conjugate of a gate sequence into the native gate alphabet of the quantum computer. Note that this small-scale classical compiler does not come with exponential overhead because it is only compiling one- and two-qubit gates. 

Also, observe that if a gate alphabet is not closed under complex conjugation, then the depth of a gate sequence from that alphabet can increase by taking its complex conjugate. This is true for the example given above, in which the complex conjugate $R_x(-\pi/2)$ of $R_x(\pi/2)$ has a depth of three under the alphabet $\mathcal{A}=\{R_x(\pi/2),R_z(\theta)\}$, while the original gate has a depth of only one. However, in general, note that the final depth increases by at most a constant factor relative to the original depth.

\section{Faithfulness of LHST cost function}\label{app:LHST_faithful}

\PropositionLHSTfaithfulness*

\begin{specialproof}
First, we note that since $0\leq C_{\text{LHST}}^{(j)}(U,V)\leq 1$ for all $j\in\{1,2,\dotsc,n\}$, we get that $C_{\text{LHST}}(U,V)=0$ if and only if $C_{\text{LHST}}^{(j)}(U,V)=0$, i.e., $F_e^{(j)}=1$, for all $j\in\{1,2,\dotsc,n\}$. Next, since $F_e^{(j)}$ is by definition the entanglement fidelity of the channel $\mathcal{E}_j$, we have that $F_e^{(j)}=1$ if and only if $\mathcal{E}_j$ is the identity channel $\mathcal{I}$. Finally, the condition $U=V$ is equivalent to $W\coloneqq UV^\dagger=\id$. Therefore, it suffices to prove that $W=\id$ if and only if $\mathcal{E}_j$ is the identity channel for all $j\in\{1,2,\dotsc,n\}$. The implication $W=\id\Rightarrow \mathcal{E}_j=\mathcal{I}$ for all $j\in\{1,2,\dotsc,n\}$ is immediate. We now prove the converse.

Let $j=1$, and suppose that $W$ has the following operator Schmidt decomposition under the bipartite cut $A_1|A_2\dotsb A_n$:
\begin{equation}
    W=\sum_{i=1}^r \sqrt{\sigma_i}X_i^{A_1}\otimes Y_i^{A_2\dotsb A_n},
\end{equation}
where $\{X_i\}_{i=1}^r$ and $\{Y_i\}_{i=1}^r$ are orthonormal sets of operators, $\sigma_i>0$ are the Schmidt coefficients of $W$, and $r$ is the Schmidt rank of $W$. Since $W$ is unitary, we have
\begin{equation}
    W^\dagger W=\sum_{i,i'=1}^r\sqrt{\sigma_i\sigma_{i'}} X_i^\dagger X_{i'}\otimes Y_i^\dagger Y_{i'}=\id_{A_1\dotsb A_n},
\end{equation}
which implies that
\begin{equation}\label{eq-LHST_faithful_pf1}
    \Tr_{A_2\dotsb A_n}(W^\dagger W)=\sum_{i=1}^r\sigma_i X_i^\dagger X_i=2^{n-1}\id_{A_1}.
\end{equation}
Plugging in the Schmidt decomposition of $W$ into the definition of $\mathcal{E}_1$ in \eqref{eq-HST_local_channel}, we get
\begin{align}
    \mathcal{E}_1(\rho)&=\sum_{i=1}^r\frac{1}{2^{n-1}}\sigma_i X_i\rho X_i^\dagger.
\end{align}
The operators $K_i:=\sqrt{\frac{\sigma_i}{2^{n-1}}}X_i$ can therefore be regarded as Kraus operators for $\mathcal{E}_1$. Indeed, they satisfy the following condition for trace preservation:
\begin{align}
    \sum_{i=1}^r K_i^\dagger K_i&=\frac{1}{2^{n-1}}\sum_{i=1}^r\sigma_i X_i^\dagger X_i\\
    &=\frac{1}{2^{n-1}}\Tr_{A_2\dotsb A_n}(W^\dagger W)\\
    &=\id_{A_1},
\end{align}
where to obtain the second equality we used \eqref{eq-LHST_faithful_pf1}.

Now, we assume that $\mathcal{E}_1$ is the identity channel, meaning that $\mathcal{E}_1(\rho)=\sum_{i=1}^r\frac{\sigma_i}{2^{n-1}}X_i\rho X_i^\dagger=\rho$ for all states $\rho$. By the non-uniqueness of Kraus representations of quantum channels, there exists an isometry $V$ relating the Kraus operators $\{K_i\}_{i=1}^r$ to another set $\{N_j\}_{j=1}^s$ of Kraus operators according to $K_i=\sum_{j=1}^s V_{i,j}N_j$. Since one Kraus representation of the identity channel is the one consisting of only the identity operator $\id$, we let the set $\{N_j\}_{j=1}^s$ consist of only the identity operator. The isometry $V$ is then a $r\times 1$ matrix, so that $V_{i,1}=\alpha_i\in\mathbb{C}$ for all $i\in\{1,2,\dotsc,r\}$. This implies that $K_i=\sqrt{\frac{\sigma_i}{2^{n-1}}}X_i=\alpha_i\id_{A_1}$ for all $i\in\{1,2,\dotsc,r\}$. Therefore,
\begin{align}
    W_{A_1\dotsb A_n}&=\sum_{i=1}^r\sqrt{\sigma_i}X_i^{A_1}\otimes Y_i^{A_2\dotsb A_n}\\
    &=\sum_{i=1}^r\sqrt{\sigma_i}\left(\sqrt{\frac{2^{n-1}}{\sigma_i}}\alpha_i\id_{A_1}\right)\otimes Y_i^{A_2\dotsb A_n}\\
    &=\id_{A_1}\otimes\sqrt{2^{n-1}}\sum_{i=1}^r\alpha_i Y_i^{A_2\dotsb A_n}\\
    &=:\id_{A_1}\otimes W'_{A_2\dotsb A_n},\label{eq-local_HST_pf}
\end{align}
where in the last line we have defined the unitary $W'_{A_2\dotsb A_n}=\sqrt{2^{n-1}}\sum_{i=1}^r\alpha_i Y_i^{A_2\dotsb A_n}$.

Now, given the assumption that $\mathcal{E}_1=\mathcal{I}$, so that $W$ has the form in \eqref{eq-local_HST_pf}, we get that
\begin{equation}
    \mathcal{E}_2(\rho)=\Tr_{A_3\dotsb A_n}\left(W'\left(\rho\otimes \frac{\id_{A_3\dotsb A_n}}{2^{n-2}}\right)(W')^\dagger\right).
\end{equation}
Therefore, applying the procedure above for $j=2$ by taking the bipartite cut in the operator Schmidt decomposition of $W'$ to be $A_2|A_3\dotsb A_n$, we get that if $\mathcal{E}_2$ is the identity channel, then $W=\id_{A_1}\otimes\id_{A_2}\otimes W''$ for some unitary $W''$ acting on $A_3\dotsb A_n$. Continuing in this manner for all $j$ up to $j=n$, assuming in each case that $\mathcal{E}_j$ is the identity channel, we ultimately obtain $W=\id_{A_1}\otimes\id_{A_2}\otimes\dotsb\otimes\id_{A_n}$, which implies that $U=V$, as required.
\end{specialproof}

\section{Relation between $C_{\text{LHST}}$ and $C_{\text{HST}}$}\label{sctConverseBound}

\TheoremConverseBound*

\begin{specialproof}
%We prove Eq.~\eqref{eqnDirectBound} and Eq.~\eqref{eqnConverseBound}. Let us start by recalling the definitions of the global cost function $C_{\text{HST}}$ and the local cost function $C_{\text{LHST}}$: we have
First we rewrite the global cost function:
%\begin{equation}
%	C_{\text{HST}}(U,V)=1-\frac{1}{d^2}\%left|\Tr[V^\dagger U]\right|^2,
%\end{equation}
%which we can write as
\begin{equation}
    \begin{aligned}
	C_{\text{HST}}(U,V) &=1-\frac{1}{d^2}\left|\Tr[V^\dagger U]\right|^2\\
	&=1-\Tr[\ket{\Phi^+}\bra{\Phi^+}_{AB}\\
	&\quad\times(W\otimes\id_B)\ket{\Phi^+}\bra{\Phi^+}_{AB}(W^\dagger\otimes\id_B)],
	\end{aligned}
\end{equation}
where $W=UV^\dagger$. Also, for the local cost function, we have
\begin{equation}
	C_{\text{LHST}}(U,V)\coloneqq\frac{1}{n}\sum_{j=1}^n C_{\text{LHST}}^{(j)}(U,V),
\end{equation}
where
\begin{equation}
	\begin{aligned}
	&C_{\text{LHST}}^{(j)}(U,V)\\
	&= 1-\Tr[\Pi_j(W\otimes\id_B)\ket{\Phi^+}\bra{\Phi^+}_{AB}(W^\dagger\otimes\id)]
	\end{aligned}
\end{equation}
and we have defined
\begin{equation}
	\Pi_j\coloneqq\id_{A_1B_1}\otimes\dotsb\otimes\ket{\Phi^+}\bra{\Phi^+}_{A_jB_j}\otimes\dotsb\otimes\id_{A_nB_n},
\end{equation}
which are projectors that all mutually commute. Let
\begin{equation}
	\rho\coloneqq (W\otimes\id_B)\ket{\Phi^+}\bra{\Phi^+}_{AB}(W^\dagger\otimes\id_B).
\end{equation}
Then, we can write $C_{\text{HST}}(U,V)$ as
\begin{equation}
	C_{\text{HST}}(U,V)=1-\Tr[\Pi_n\dotsb \Pi_1\rho],
\end{equation}
and we can write $C_{\text{LHST}}^{(j)}(U,V)$ as
\begin{equation}
	C_{\text{LHST}}^{(j)}(U,V)=1-\Tr[\Pi_j\rho]
\end{equation}
for all $1\leq j\leq n$. If we associate the events $E_j$ with the projectors $\Pi_j$, so that $\Pr[E_j]=\Tr[\Pi_j\rho]$, then, $\Tr[\Pi_n\dotsb\Pi_1\rho]=\Pr\left[\bigcap_{i=1}^n E_i\right]$.

To prove \eqref{eqnDirectBound}, namely $C_{\text{LHST}}(U,V)\leq C_{\text{HST}}(U,V)$, we recall a basic inequality in probability theory. For any set $\{A_1,A_2,\dotsc,A_n\}$ of events, it holds that
\begin{equation}\label{eq-LHST_HST_bounds_lem1}
    \Pr\left[\bigcup_{i=1}^n A_i\right]\geq\frac{1}{n}\sum_{i=1}^n\Pr[A_i].
\end{equation}
Let us take $A_i=\overline{E_i}$ in \eqref{eq-LHST_HST_bounds_lem1}. Then,
\begin{align}
    \Pr\left[\bigcup_{i=1}^n\overline{E_i}\right]&\geq\frac{1}{n}\sum_{i=1}^n\Pr[\overline{E_i}]\\
    \Rightarrow 1-\Pr\left[\bigcap_{i=1}^n E_i\right]&\geq\frac{1}{n}\sum_{i=1}^n(1-\Pr[E_i]).
\end{align}
By definition of the events $E_i$, the last equality is precisely $C_{\text{HST}}(U,V)\geq C_{\text{LHST}}(U,V)$, as required.

To prove \eqref{eqnConverseBound}, we make use of the union bound:
\begin{align}
    \Pr\left[\bigcup_{i=1}^n \overline{E_i}\right]&\leq\sum_{i=1}^n\Pr[\overline{E_i}]\\
    \Rightarrow 1-\Pr\left[\bigcap_{i=1}^n E_i\right]&\leq \sum_{i=1}^n(1-\Pr[E_i])\\
    &=nC_{\text{LHST}}(U,V).
\end{align}
Given that the left-hand side of the above inequality is precisely $C_{\text{HST}}(U,V)$, we have that $C_{\text{HST}}(U,V)\leq nC_{\text{LHST}}(U,V)$, as required.
\end{specialproof}

\section{Proofs of complexity theorems}\label{sctComplexityProofs}

\TheoremDQCHST*

\begin{specialproof}
% First, we show that the problem of approximating the cost $C_\HST(U,V)$ lies in \DQC. We are given as input two $2^n$-dimensional unitaries specifying $\poly(n)$-sized quantum circuits $U$ and $V$ on $n$ qubits, and the task is to approximate $C_\HST(U,V)$ up to inverse polynomial precision in a one-clean-qubit model of quantum computation. This essentially follows from \cite{DBLP:journals/qic/ShorJ08}, although we spell out the argument here for completeness. Using the well known power-of-one-qubit (POOQ) circuit \cite{knill1998POOQ} in Fig. \ref{fig:POOQ_circuit}, we can compute the normalized trace of any $2^n$-dimensional unitary that specifies a $\poly(n)$-sized quantum circuit $Q$ by separately estimating its real and imaginary parts. On the input state $\rho=\dya{0} \ot (\id /2)^{\ot n}$, the circuit applies a Hadamard gate on the first register, followed by a controlled-$Q$ operation onto the remaining registers. Then, depending on whether we seek to estimate the real or imaginary part, the circuit implements a gate $R$ that represents either $H$, in which case the circuit computes $\Re[\Tr(Q)]$, or the $S$ gate followed by $H$, in which case the circuit computes $\Im[\Tr(Q)]$. 
% Thus, if we let $Q = V^\dagger U$ and run the POOQ circuit over an appropriate choice of $O(1/\epsilon^2)$ many samples, we can obtain approximations for the real and imaginary parts of $\Tr(V^\dagger U)$ respectively. Therefore, we can approximate $C_\HST(U,V)$ up to $\epsilon$-precision, where $\epsilon$ is at most $O(1/\poly(n))$.
We show that the problem of approximating the cost $C_\HST(U,V)$ is hard for \DQC. In other words, we have to show that any problem in \DQC reduces to an instance of approximating $C_\HST(U,V)$ for some $\epsilon = O(1/\poly(n))$. Recall that, given as input a $\poly(n)$-sized unitary $Q$ on $n$-qubits, any problem in \DQC requires us to estimate the acceptance probability $p_\text{acc}$ when measuring the outcome ``$0$'' on input $\rho = \ketbra{0}{0} \otimes \id/{2^{n}}$, i.e.
\begin{align}\label{dqc1-comp}
p_\text{acc} =  \Tr[ (\ketbra{0}{0} \otimes \id) Q \rho Q^\dagger].
\end{align}
Note that, since the above equation describes a probability via the positive semi-definite operator $\ketbra{0}{0} \otimes \id$, the trace will result in a non-negative real number.
Let us re-write Eq.~\eqref{dqc1-comp} as follows:
\begin{align}\label{DQC1-normalized}
p_\text{acc} = \frac{1}{2^n} \big|\Tr[ (\ketbra{0}{0} \otimes \id) Q  (\ketbra{0}{0} \otimes \id) Q^\dagger]\big|.
\end{align}
\begin{figure}
    \centering
    \includegraphics[width=0.8\columnwidth]{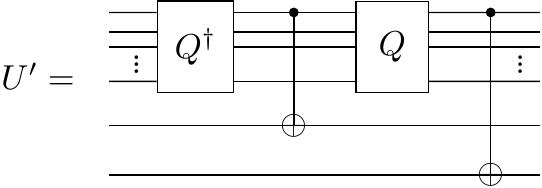}
    \caption{The trace of the unitary $U'$ defined by the circuit above is equal to the trace of the non-unitary operator $(\ketbra{0}{0} \otimes \id) Q  (\ketbra{0}{0} \otimes \id)Q^\dagger$ up to a factor of $4$ \cite{DBLP:journals/qic/ShorJ08}.}
    \label{unitary_fix}
\end{figure}
When letting $U'$ as in Fig.~\ref{unitary_fix}, we can also write
\begin{align}
\label{eqntraceUprime}
\Tr[ (\ketbra{0}{0} \otimes \id) Q  (\ketbra{0}{0} \otimes \id) Q^\dagger] = \Tr[U']/4,
\end{align}
hence the problem is equivalent to approximating the absolute value of the trace of a unitary $U'$.
In fact, given our choice of $U'$ and when taking $V$ to be the identity, the problem reduces to an instance of approximating the cost $C_\HST(U',\id)$ up to some precision $\epsilon = O(1/\poly(n))$ via a simple reduction. 
% Hence, over an appropriate choice of $O(1/\epsilon^2)$ many samples, we can obtain $\epsilon$-approximations for the absolute value of the normalized trace using the Hilbert-Schmidt test circuit. 
Therefore, we have shown that the problem of approximating $C_\HST(U,V)$ up to $\epsilon$-precision is \DQC-hard.
\end{specialproof}

% \begin{figure}
%         \centering
%         \includegraphics[width=0.7\columnwidth]{POOQ.pdf}
%         \caption{The Power of One Qubit (POOQ) \cite{knill1998POOQ}, which can be used to compute the trace of a unitary $U$ acting on a $d$-dimensional space. The $R$ gate represents either $H$, in which case the circuit computes $\Re[\Tr(U)]$, or the $S$ gate followed by $H$, in which case the circuit computes $\Im[\Tr(U)]$.}
%         \label{fig:POOQ_circuit}
% \end{figure}

\TheoremDQCLHST*

\begin{specialproof}
We show that any problem in \DQC reduces to an instance of approximating $C_\LHST(U,V)$ via a reduction. We are given as input a $\poly(n)$-sized unitary $Q$ on $n$-qubits, and the task is to estimate the acceptance probability of outputting $``0"$. Our proof strategy is to show that one can efficiently extract $\Tr(U')$, the trace of an $n$-qubit unitary $U'$, from two distinct evaluations of $C_\LHST$ and elementary post-processing. This implies that computing $C_\LHST$ is hard for \DQC, since all problems in \DQC can be seen as estimating the real part of $\Tr(U')$ via Eq.~\eqref{DQC1-normalized} and Eq.~\eqref{eqntraceUprime}.

The two cost function evaluations that we consider are $C_\LHST(U_1,\id)$ and $C_\LHST(U_2,\id)$, where
\begin{align}
    U_1 &= U' \\
    U_2 &= C_{U'}\,.
\end{align}
Here, $C_{U'}$ denotes controlled-$U'$ operation.

First, consider $U_2$ and let the $j=n+1$ qubit correspond to the control qubit for the $C_{U'}$ controlled unitary. Then one can show that
\begin{align}
    C_\LHST^{(j)}(U_2,\id) &=  \frac{1}{2}C_\LHST^{(j)}(U_1,\id)\quad \forall j\in \{1,...,n\}\,,\\
      C_\LHST^{(n+1)}(U_2,\id) &= \frac{1}{2} - \frac{1}{2^{n+1}}\Re(\Tr(U'))\,.
\end{align}
This gives
\begin{align}
\label{eqnCLHSTU2}
    C_\LHST(U_2,\id) = \frac{1}{2(n+1)}&\bigg(1+nC_\LHST(U_1,\id)\nonumber\\
    &-\frac{\Re(\Tr(U'))}{2^n}\bigg)\,.
\end{align}

For notational simplicity, let $B(U):= 1- C_\LHST(U,\id)$. Then, we can rewrite Eqs.~\eqref{eqnCLHSTU2} as
\begin{align}
    B(U_2) &= \frac{1}{2}\left( 1+\frac{2^{-n}}{n+1}\Re(\Tr(U'))+\frac{n}{n+1}B(U_1)\right)\,.
\end{align}
Hence, we have that
\begin{align}
    \Re(\Tr(U'))&= 2^n\left((n+1)(2B(U_2)-1)-nB(U_1)\right)\,.
\end{align}
By choosing $U'$ according to Fig.~\ref{unitary_fix}, one can see from Eq.~\eqref{DQC1-normalized} and Eq.~\eqref{eqntraceUprime} that the problem is equivalent to $\epsilon$-approximating our local cost function for some $\epsilon = O(1/\poly(n))$. Hence, any \DQC problem can be efficiently solved for by computing a simple linear combination of two instances of $C_\LHST$. Therefore, we have shown that the problem of approximating the cost $C_\LHST(U,V)$ is hard for \DQC.
\end{specialproof}

\section{Gradient-free optimization method}\label{sctGF}

%\section{Our gradient-free optimization method}\label{sctGF}

We now outline our approach to gradient-free optimization of over the continuous gate parameters $\vec{\alpha}$ in the trainable unitary $V_{\vec{k}}(\vec{\alpha})$. This approach was used to obtain the results in Sec. \ref{sctimplementations}. Given that this is an implementation for small problem size, we employ the cost function $C_{\text{HST}}(U,V_{\vec{k}}(\vec{\alpha}))$. However, we note that one can replace $C_{\text{HST}}$ with our general cost function $C_q$ for larger problem sizes.

\begin{algorithm}
\DontPrintSemicolon % Some LaTeX compilers require you to use \dontprintsemicolon instead
\KwIn{
    Unitary $U$ to be compiled; trainable unitary $V_{\vec{k}}(\vec \alpha)$ of a given structure; error tolerance $\varepsilon'\in (0,1)$; maximum number of starting points $N$; maximum number of iterations $N_{\text{iter}}$ for \texttt{gp\_minimize}; sample precision $\delta > 0$.}
\KwOut{
    Parameters $\vec{\alpha}_{\text{opt}}$ such that at best $C_{\text{HST}}(U,V_{\vec{k}}(\vec{\alpha}_{\text{opt}}))\leq \varepsilon'$. }
\kwInit{$\vec{\alpha}_{\text{opt}} \gets 0; \texttt{cost} \gets 1$}
    \Repeat{\normalfont{$\texttt{cost}\leq\varepsilon'$, at most $N$ times.}}{
    choose an initial parameter $\vec \alpha^{(0)}$ at random;\;
    run \texttt{gp\_minimize} with $\vec \alpha^{(0)}$ and $N_{\text{iter}}$ as input and $\vec \alpha_{\min}$ as output; whenever the cost is called upon for some $\vec \alpha$, run the \text{HST} on $V_{\vec{k}}(\vec \alpha)^*$ and $U$ approximately $1/\delta^2$ times to estimate the cost $C_{\text{HST}}(U,V_{\vec{k}}(\vec \alpha))$;\;
      \If{$\normalfont{\texttt{cost}} \geq C_{\text{HST}}(U,V_{\vec k}(\vec{\alpha}_{\min}))$} 
    {$\normalfont{\texttt{cost}} \gets C_{\text{HST}}(U,V_{\vec k}(\vec{\alpha}_{\min}))$; $\vec \alpha_{\text{opt}} \gets  \vec \alpha_{\min}$}
    }
    
\Return{$\vec{\alpha}_{\textnormal{opt}},\normalfont{\texttt{cost}}$}\;
\caption{\sc Gradient-free Continuous\newline Optimization for QAQC via the HST}
\label{algo:GfQC}
\end{algorithm}
        
Recall that we compute the cost function $C_{\text{HST}}(U,V_{\vec{k}}(\vec{\alpha}))$ using the Hilbert-Schmidt Test (HST), as described in Sec. \ref{sctHST} and illustrated in Fig. \ref{fig:hilbert-schmidt-inner-product-circuit}(a). For a given set of gate structure parameters $\vec{k}$, the calculation of the cost on a quantum computer (as well as on a simulator) is affected by the fact that, due to finite sampling, the HST allows us to obtain only an estimate of the magnitude of the Hilbert-Schmidt inner product. Noise within the quantum computer itself also affects the calculation of the cost. Therefore, in order to perform gradient-free optimization over the continuous gate parameters $\vec{\alpha}$, we make use of stochastic optimization techniques that are designed to optimize noisy functions. Specifically, we make use of the \texttt{gp\_minimize} routine in the scikit-optimize Python library \cite{scikit-opt}, which is a gradient-free optimization routine that performs Bayesian optimization using Gaussian processes \cite{mockus1975BayesOpt,osborne2009Gaussian}. See Algorithm \ref{algo:GfQC} for a general overview of the optimization procedure. Note that with this algorithm, we obtain an $\varepsilon$-approximate compilation of $U$, with \begin{equation}
\label{eqnEpsilonCHST}
    \varepsilon=\left(\frac{d}{d+1}\right)\varepsilon'\,.
\end{equation}
In the small-scale quantum computer implementations of Fig.~\ref{fig:ibm_rigetti-one-qubit-gates}(c) and Fig.~\ref{fig:2qubitgates_sim}, we use 50 objective function evaluations in \texttt{gp\_minimize} per iteration. Note that evaluating the objective function involves running the quantum circuit many times in order to sample from the output distribution of the circuit.

For large problem sizes, as described in Sec. \ref{sctlarge}, we propose using the cost function $C_q=qC_{\text{HST}}+(1-q)C_{\text{LHST}}$. The gradient-free continuous parameter optimization algorithm for $C_q$ is similar to the one for $C_{\text{HST}}$ in Algorithm \ref{algo:GfQC}, except that in addition to running the HST we run the LHST for every qubit $j\in\{1,2,\dotsc,n\}$ in order to compute the local cost $C_{\text{LHST}}$. In this case, the algorithm provides an $\varepsilon$-approximate compilation of $U$, with \begin{equation}
\label{eqnEpsilonCq}
    \varepsilon=\left(\frac{n}{1-q+nq}\right)\left(\frac{d}{d+1}\right)\varepsilon'\,.
\end{equation}

We emphasize that our approach to gradient-free optimization avoids the exponential overhead of evaluating the cost function classically, yet at the same time makes use of fast and efficient classical heuristics for optimization. In fact, using the HST, Algorithm \ref{algo:GfQC} requires only $ O(1/\delta^2)$ calls to the quantum computer in order to evaluate the cost, where $\delta = 1/\sqrt{n_\text{shots}}$ is the sample precision, which is related to the number of samples $n_\text{shots}$ taken from the device.

\subsection{Alternative method for gradient-free optimization}\label{app:bisection}

Here we propose an alternative algorithm for gradient-free optimization that, on average, significantly reduces the number of times the objective function is evaluated. As a result, it is more suitable for cloud computing under a queue submission system (e.g., IBM's Quantum Experience). This algorithm performs a ``multi-scale bisection'' of the parameter space based on simulated annealing. We implement this method in Sec. \ref{sctIBM} specifically for the hardware of IBM because the queue submission system can require a significant amount of time to make many calls to the quantum computer.

This alternative approach to performing gradient-free continuous parameter optimization is outlined in Algorithm \ref{algo:GfQC_bisect}. We start with four angles spread uniformly in the interval $[0, 2 \pi)$---namely $0, \pi / 2, \pi,$ and $ 3 \pi / 2$. This significantly reduces the size of the search space and allows us to get close to, or find exactly, an optimal gate sequence. Once the optimal structure is reached from this step, we then bisect the angles for each gate $R_z(\alpha)$ by evaluating the cost with a new circuit containing $R_z( \alpha \pm \pi / 2^{t+1})$, where $t = 1, 2,\dots,t_{\max}$ is determined by the iteration in the procedure. Although we do not explore all angles in the interval, the runtime is logarithmically faster than a continuous search due to the bisection procedure. An additional advantage of this approach is that many gates have angles that are simple fractions of $\pi$, e.g., $T = R_z(\pi / 4)$ and $H = R_z( \pi / 2) R_x ( \pi / 2) R_z ( \pi / 2)$.
In a noiseless environment, the two steps above are sufficient. On actual devices, we implement a third step of stochastic optimization by evaluating the cost for the new circuit with each gate $R_z(\alpha)$ replaced by $R_z(\alpha \pm \Delta(t))$ for some small value $\Delta(t) \ll 1$ decreasing monotonically with the iteration $t$. This allows us to compile for a given device by accounting for noise and gate errors. This can be thought of as a ``fine-grained'' angular optimization in contrast to the previous ``coarse-grained'' angular optimization.\\

\begin{algorithm}
\DontPrintSemicolon % Some LaTeX compilers require you to use \dontprintsemicolon instead
\KwIn{
    Unitary $U$ to be compiled; trainable unitary $V_{\vec k}(\vec{\alpha})$ of a given structure and gate alphabet $\mathcal{A}$; error tolerance $\varepsilon' \in (0,1)$; maximum number of iterations $N$; maximum number of bisections $t_{\max}$ of the unit circle; sample precision $\delta > 0$.}
\KwOut{
    Parameters $\vec{\alpha}_{\text{opt}}$ such that at best $C_{\text{HST}}(U,V_{\vec{k}}(\vec{\alpha}_{\text{opt}}))\leq \varepsilon'$. }
\kwInit{Restrict all gates in $\mathcal A$ with continuous parameters to discrete angles in the set $\Omega_0=\{0, \pi / 2, \pi, 3 \pi / 2\}$;
$\alpha_{\text{opt}}\gets 0; \texttt{cost} \gets 1$}
\For{$t = 1,2,\dots,t_{\max}$}{    
\Repeat{\normalfont{$\texttt{cost}\leq\varepsilon'$ at most $N$ times.}}{
 anneal over all possible bisected angles in the set
 $\Omega_t :=\{\alpha \pm \pi/2^{t+1} \, | \, \text{for } \alpha \in \Omega_0 \} \cup \Omega_{t-1}$;\\
  whenever the cost is called upon for some $\alpha \in \Omega_t$, run the \text{HST} on $V_{\vec{k}}(\vec{\alpha})^*$ and $U$ approximately $1/\delta^2$ times to estimate the cost $C_{\text{HST}}(U,V_{\vec{k}}(\vec{\alpha}))$;\;
  \If{$\normalfont{\texttt{cost}} \geq C_{\text{HST}}(U,V_{\vec{k}}(\vec{\alpha}))$} 
{$\normalfont{\texttt{cost}} \gets C_{\text{HST}}(U,V_{\vec{k}}(\vec{\alpha}))$;}}
 
 \Repeat{\normalfont{$\texttt{cost}\leq\varepsilon'$ at most $N$ times.}}{
minimize the cost over all small continuous increments $\Delta(t) \ll 1$ within the set of bisected angles $\Omega_t$;
whenever the cost is called upon for some $\alpha + \Delta(t)$, with $\alpha \in \Omega_t$, run the \text{HST} on $V_{\vec{k}}(\alpha+\Delta(t))^*$ and $U$ approximately $1/\delta^2$ times to estimate the cost $C_{\text{HST}}(U,V_{\vec{k}}(\alpha+\Delta(t)))$;\;
\If{$\normalfont{\texttt{cost}} \geq C_{\text{HST}}(U,V_{\vec{k}}(\alpha+\Delta(t)))$} 
{$\normalfont{\texttt{cost}} \gets C_{\text{HST}}(U,V_{\vec{k}}(\alpha+\Delta(t)))$; $\vec{\alpha}_{\text{opt}}\gets \alpha+\Delta(t)$ }
}
 }
\Return{$\vec{\alpha}_{\textnormal{opt}},\normalfont{\texttt{cost}}$}\;
\caption{\sc Gradient-free Optimization\newline using Bisection for QAQC}
\label{algo:GfQC_bisect}
\end{algorithm}

\section{Gradient-based optimization method}\label{sctGB}

%\section{Our gradient-based optimization method}\label{sctGB}

We now describe a gradient-based approach to performing the optimization over the continuous parameters in the trainable gate sequence $V_{\vec{k}}(\vec{\alpha})$. In Sec. \ref{sctPOTQ}, we define a new cost function for this purpose, and we introduce a quantum circuit to calculate this cost function on a quantum computer. In Sec. \ref{sctPOTQopt}, we present the results of implementing this method on a quantum simulator. In Sec. \ref{sctHST_grad}, we briefly describe how the original cost functions $C_{\text{HST}}$ and $C_{\text{LHST}}$ can also be optimized using a gradient-based method.

While recent work on gradient descent continuous optimization has shown vast quantum speedups over classical variants~\cite{RS16,KP17,GA17}, the majority of proposals still appear to be out of reach for implementations on NISQ devices, mainly due to their use of certain algorithmic techniques, such as quantum random-access memory, the quantum Fourier transform, and the Grover search algorithm, which have high resource requirements. Instead, we focus on continuous optimization procedures that are feasible on current quantum computers and leave improvements to our algorithms as an open problem.

The gradient with respect to $\vec{\alpha}$ of the gate sequence $V_{\vec{k}}(\vec{\alpha})$ given by
\begin{equation}\label{eq:trainable_gate_seq}
    V_{\vec{k}}(\vec{\alpha})=G_{k_L}(\alpha_L)G_{k_{L-1}}(\alpha_{L-1})\dotsb G_{k_1}(\alpha_1),
\end{equation}
is defined by
\begin{equation}
    \nabla_{\vec{\alpha}}V_{\vec{k}}(\vec{\alpha})=\left(\frac{\partial V_{\vec{k}}(\vec{\alpha})}{\partial\alpha_1},\dotsc,\frac{\partial V_{\vec{k}}(\vec{\alpha})}{\partial \alpha_L}\right),
\end{equation}
where the $(i,j)$ matrix element of the $\ell$-th component is
\begin{equation}
    \left(\frac{\partial V_{\vec{k}}(\vec{\alpha})}{\partial\alpha_\ell}\right)_{i,j}=\frac{\partial V_{\vec{k}}(\vec{\alpha})_{i,j}}{\partial\alpha_\ell}.
\end{equation}
For example, consider the rotation gate $R_z(\alpha)=e^{-i\alpha\sigma_z/2}$, which is parametrized by the angle $\alpha$. Then, the derivative with respect to $\alpha$ can be written as
\begin{equation}\label{eq-Pauli_rot_deriv}
    \frac{\partial}{\partial\alpha} R_z(\alpha) = -\frac{i}{2}\sigma_zR_z(\alpha) \ ,
\end{equation}
which follows from the Taylor series expansion of the exponent.

\begin{figure}
\centering
\includegraphics[scale=1]{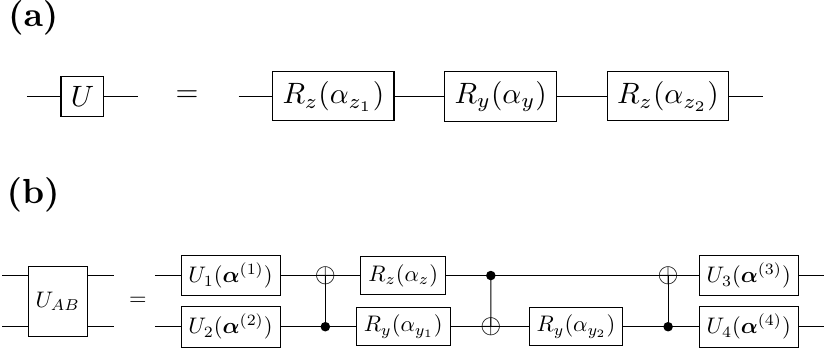}
\caption{ {\bf (a)} Any single-qubit gate $U$ can be decomposed into three elementary rotations (up to a global phase). Given appropriate parameters $\vec \alpha = (\alpha_{z_1},\alpha_y,\alpha_{z_2})$, $U$ can be written as
$V(\vec \alpha) = e^{-i \alpha_{z_2} \sigma_z/2} e^{-i \alpha_y \sigma_y/2} e^{-i \alpha_{z_1} \sigma_z/2}$. {\bf (b)} Any two-qubit gate $U_{AB}$ can be decomposed into three \text{CNOT} gates as well as $15$ elementary single-qubit gates, where each unitary $U_j(\vec \alpha^{(j)})$ can be written as in (a).
This decomposition is known to be optimal~\cite{VW04}, i.e., it uses the least number of continuous parameters and CNOT gates. General universal quantum circuits for $n$-qubit gates are discussed in~\cite{SR07}.}
\label{fig:univ_circuits}
\end{figure}

Now, evaluating the gradient on a quantum computer is possible due to the fact that for the gate alphabets we consider in this paper, only the single-qubit gates are parameterized, and these gates are simply rotation gates. In fact, any unitary gate can be decomposed into circuits in which only the single-qubit rotation gates are present. This is illustrated in Fig. \ref{fig:univ_circuits}. Furthermore, the circuits in Fig. \ref{fig:univ_circuits}(a) and Fig. \ref{fig:univ_circuits}(b) are universal for one- and two-qubit gates, respectively (see \cite{SR07}, which also contains universal circuits for $n$-qubit gates). This means that our gradient-based approach can be applied to any $n$-qubit unitary without explicitly searching over gate structures, though the compilations obtained in this manner will generally have sub-optimal depth. 
    
%We define our cost function as the normalized Hilbert-Schmidt distance,
%\begin{equation}
%\begin{aligned}
%C_{\text{GB}}(U,V_{\vec{k}}(\vec{\alpha}))&:= %\frac{1}{2d}\norm{U-V_{\vec{k}}(\vec{\alpha})}_{\HS}^2\\
%            &=1-\frac{1}{d}\Re[\Tr(V_{\vec{k}}(\vec{\alpha})^\dagger U)].
%            \end{aligned}
%\end{equation}

%We emphasize that  Due to the existence of exact universal circuits for $n$-qubit unitary operations, our gradient-based QAQC approach can, in principle, directly compile an arbitrary unitary gate at the cost of suboptimal depth. We give examples of universal circuits for single-qubit and two-qubit gates in Fig. \ref{fig:univ_circuits}.

\subsection{The Power of Two Qubits}\label{sctPOTQ}

Consider the following cost function based on the normalized Hilbert-Schmidt distance between the unitaries $U$ and $V$:
\begin{equation}
    \begin{aligned}
    C_{\text{POTQ}}(U,V)&\coloneqq\frac{1}{2d}\norm{U-V}_{\text{HS}}^2\\
    &=1-\frac{1}{d}\text{Re}\left[\Tr(V^\dagger U)\right],
    \end{aligned}
\end{equation}
where POTQ stands for ``Power of Two Qubits'' and refers to the circuit used to evaluate it, which we present below. Note that $C_{\text{POTQ}}(U,V)$ is zero if and only if $U=V$. Contrary to the cost function $C_{\text{HST}}(U,V)$, which is defined using the magnitude of the inner product $\langle V,U\rangle$, this cost function is defined using the real part of the inner product. Consequently, it does not vanish if $U$ and $V$ differ only by a global phase. Indeed, if $V=e^{i\varphi}U$, then $C_{\text{POTQ}}(U,V)=1-\cos(\varphi)$.

\begin{figure}
        \centering
        \includegraphics[scale=0.95]{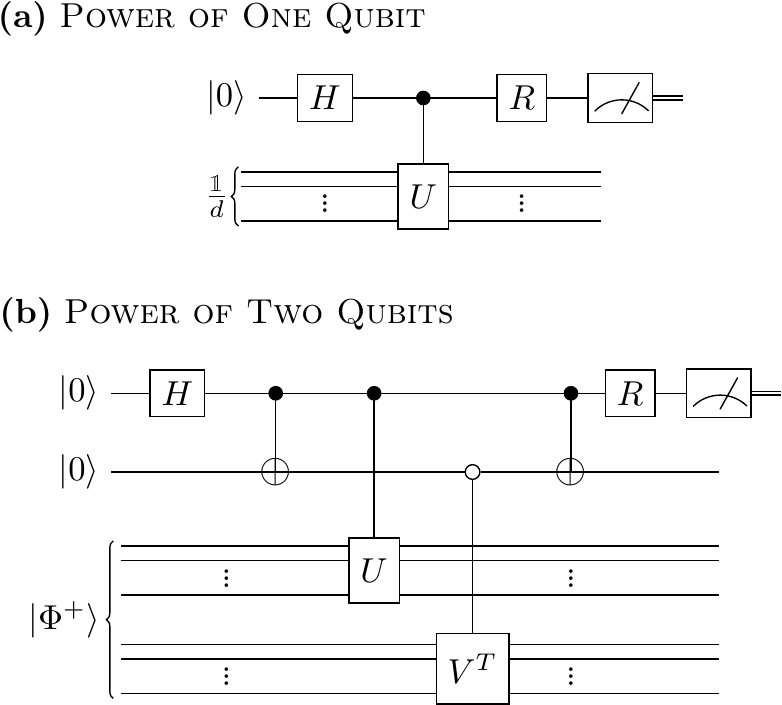}
        \caption{\textbf{(a)} The Power of One Qubit (POOQ) \cite{knill1998POOQ}. This can be used to compute the trace of a unitary $U$ acting on a $d$-dimensional space. The $R$ gate represents either $H$, in which case the circuit computes $\Re[\Tr(U)]$, or the $S$ gate followed by $H$, in which case the circuit computes $\Im[\Tr(U)]$. \textbf{(b)} The Power of Two Qubits (POTQ). This is a generalization of the POOQ, as can be seen by setting $V=\id$. The POTQ can be used to compute the Hilbert-Schmidt inner product $\Tr(V^\dagger U)$ between two unitaries $U$ and $V$ acting on a $d$-dimensional space. As with the POOQ, $R=H$ leads to $\Re[\Tr(V^\dagger U)]$, while $R=HS$ leads to $\Im[\Tr(V^\dagger U)]$. }
        \label{fig:POTQ_circuit}
\end{figure}

Before discussing the circuit used to evaluate the cost function $C_{\text{POTQ}}(U,V)$, let us review the Power of One Qubit (POOQ) \cite{knill1998POOQ}, shown in Fig.~\ref{fig:POTQ_circuit}(a), which is a circuit for computing the trace of a $d$-dimensional unitary $U$. This circuit acts on a $d$-dimensional system $A$, initially in the maximally mixed state, $\id/d$, and on a single-qubit ancilla $Q$ initially in the $\ket{0}$ state. After applying a Hadamard gate to $Q$ and a controlled-$U$ gate to $QA$ (with $Q$ the control system), the reduced density matrix $\rho_Q$ has its off-diagonal elements proportional to $\Tr(U)$. Hence, one can measure $Q$ in the $X$ and $Y$ bases, respectively, to read off the real and imaginary parts of $\Tr(U)$.

We now introduce a circuit for computing the real and imaginary parts of $\langle V,U\rangle$ that generalizes the POOQ and is called the Power of Two Qubits (POTQ), depicted in Fig.~\ref{fig:POTQ_circuit}(b). As the name suggests, the POTQ employs two single-qubit ancillas, $Q$ and $Q'$, each initially in the $\ket{0}$ state. In addition, two $d$-dimensional systems, $A$ and $B$, are initially prepared in the Bell state $\ket{\Phi^+}$ defined in Eq.~\eqref{eqn2}. (Although not shown in Fig.~\ref{fig:POTQ_circuit}(b), this Bell state is prepared with a depth-two circuit, as shown in Fig.~\ref{fig:hilbert-schmidt-inner-product-circuit}.) 

The first step in the POTQ is to prepare the two-qubit maximally entangled state $\frac{1}{\sqrt{2}}(\ket{0}\ket{0}+\ket{1}\ket{1})$ between $Q$ and $Q'$, using the Hadamard and CNOT gates as shown in Fig.~\ref{fig:POTQ_circuit}(b). The second step is to apply a controlled-$U$ gate between $Q$ and $A$ (with $Q$ the control system). In parallel with this gate, the anticontrolled-$V^T$ gate is applied to $Q'B$, with $Q'$ the control system, where anticontrolled means that the roles of the $\ket{0}$ and $\ket{1}$ states on the control system are reversed in comparison to a controlled gate. This results in the state:
\begin{align}
\label{eqn9}
&\frac{1}{\sqrt{2}}(\ket{0}_{Q}\ket{0}_{Q'}(\id_A\otimes V^T)\ket{\Phi^+} \notag\\
&\hspace{10pt}+ \ket{1}_{Q}\ket{1}_{Q'}(U\otimes \id_B)\ket{\Phi^+} )\notag\\
&=\frac{1}{\sqrt{2}}(\ket{0}_{Q}\ket{0}_{Q'}(V\otimes \id_B)\ket{\Phi^+}\notag\\
&\hspace{10pt}+ \ket{1}_{Q}\ket{1}_{Q'}(U\otimes \id_B)\ket{\Phi^+} ) ,
\end{align}
where to obtain the equality we used the ricochet property in Eq. \eqref{eq:ricochet_prop}. As in the HST, note that $V$ itself is not implemented. In this case, its transpose is implemented.

Finally, a CNOT gate is applied to $QQ'$, with $Q$ the control system. This results in the reduced state on $Q$ being
\begin{align}
        &\rho_Q = \frac{1}{2}\left(\ket{0}\bra{0}+\Tr(V^\dagger U)\ket{0}\bra{1} \right.\nonumber\\ &\qquad\quad\left.+\Tr(U^\dagger V)\ket{1}\bra{0}+\ket{1}\bra{1}\right).
\end{align}
By inspection of $\rho_Q$, one can see that measuring $Q$ in the $X$ and $Y$ bases, respectively, gives the real and imaginary parts of $\Tr(V^\dagger U)$.

Interestingly, if we set $V$ to the identity in the POTQ, then since the \text{CNOT} gate commutes with the controlled-$U$ gate and the reduced state of $\ket{\Phi^+}$ is the maximally mixed state $\id/d$, we recover the POOQ. The POTQ is therefore a generalization of the POOQ.
    
Note that while the POOQ can also be used to determine $\Tr(V^\dagger U)$, the POTQ has the advantage that the controlled gates for $U$ and $V$ can be executed in parallel, while in the POOQ they would have to be executed in series. This makes the POTQ better suited for NISQ devices, where short depth is crucial. Consider the depth of the POTQ. Denoting the controlled-$U$ and the anticontrolled-$V^T$ as $C_U$ and $\overline{C}_{V^T}$ respectively, the overall depth is
\begin{align}
\label{eqn12}
D(\text{POTQ}) = 4 + \max\{D(C_U),D(\overline{C}_{V^T})\}\end{align}
Note the similarity here to Eq.~\eqref{eqn6}. The overall depth is essentially determined by whichever controlled gate has the largest depth.

\subsection{Gradient-based optimization via the POTQ}\label{sctPOTQopt}

The gradient with respect to $\vec{\alpha}$ of $C_{\text{POTQ}}(U,V_{\vec{k}}(\vec{\alpha}))$ can be computed using the POTQ. This is due to the fact that
\begin{equation}
\frac{\partial}{\partial\alpha_{\ell}}\text{Re}\left[\Tr(V_{\vec{k}}(\vec{\alpha})^\dagger U)\right]=\frac{1}{2}\Re\left[\Tr\left(\widetilde{V}_{\vec{k}}^{(\ell)}(\vec{\alpha})^\dagger U\right)\right],
\end{equation}
where
\begin{equation}
    \begin{aligned}
    \widetilde{V}_{\vec{k}}^{(\ell)}(\vec{\alpha})&\coloneqq G_{k_L}(\alpha_L)\dotsb G_{k_{\ell+1}}(\alpha_{\ell+1})(-i\sigma_{k_\ell})\\
    &\quad\quad\times G_{k_\ell}(\alpha_\ell)G_{k_{\ell-1}}(\alpha_{\ell-1})\dotsb G_{k_1}(\alpha_1)
    \end{aligned}
\end{equation}
is the original gate sequence $V_{\vec{k}}(\vec{\alpha})$ except with an additional Pauli gate $\sigma_{k_{\ell}}$ corresponding to the variable with respect to which the derivative is taken. (Note that for the gate alphabets that we consider in this paper, only the single-qubit gates are parameterized, and these gates are simply rotation gates. The derivative of any one-qubit rotation gate is analogous to the expression in \eqref{eq-Pauli_rot_deriv} for the derivative of the rotation gate $R_z(\alpha)$.) This means that to compute the gradient of $C_{\text{POTQ}}(U,V_{\vec{k}}(\vec{\alpha}))$, we simply add the appropriate local Pauli gate to the original gate sequence and run the POTQ on this new gate sequence.

\begin{algorithm}
    \DontPrintSemicolon 
    \KwIn{
    Unitary $U$ to be compiled; a trainable unitary $V_{\vec k}(\vec \alpha)$ of a given structure,
    where $\vec \alpha$ is a continuous
    circuit parameter of dimension $L$; maximum number of iterations $N$; error tolerance $\varepsilon'\in (0,1)$; learning rate $\eta > 0$; sample precision $\delta >0$.}
    \KwOut{Parameters $\vec{\alpha}_{\text{opt}}$ such that at best $C_{\text{POTQ}}(U,V_{\vec{k}}(\vec{\alpha_{\text{opt}}}))\leq\varepsilon'$.}
    \kwInit{$\vec{\alpha}_{\text{opt}} \gets 0;\texttt{cost} \gets 1$}
    \Repeat{$\normalfont{\texttt{cost}} \leq \varepsilon'$, \normalfont{at most $N$ times}}
    {
    choose initial parameters $\vec \alpha^{(0)}$ at random\;
    \For{$\tau=1,2,\dots,T$}
    {
    \For{$i=1,2,\dots,L$}
     {
     run the POTQ on $\partial_{\alpha_i} V_{\vec k}(\vec \alpha^{(\tau-1)})^T$ and $U$ approximately $ 1/{\delta^2}$ times to estimate $\Re \left(\Tr\left[\partial_{\alpha_i} V_{\vec k}(\vec \alpha^{(\tau-1)})^\dagger U\right]\right)$
     }
     \textbf{update} $\vec \alpha^{(\tau)} \gets \vec \alpha^{(\tau-1)} - \eta \, \nabla_{\vec \alpha} C_{\text{POTQ}}(U,V_{\vec k}(\vec{\alpha}^{(\tau-1)}))$\;
    }
    run the POTQ on $V_{\vec k}(\vec \alpha^{(\tau)})^T$ and $U$ approximately $1/{\delta^2}$ times to estimate the cost $C_{\text{POTQ}}(U,V_{\vec k}(\vec{\alpha}^{(\tau)}))$\;
    \If{$\normalfont{\texttt{cost}} \geq C_{\text{POTQ}}(U,V_{\vec k}(\vec{\alpha}^{(\tau)}))$} 
    {$\normalfont{\texttt{cost}} \gets C_{\text{POTQ}}(U,V_{\vec k}(\vec{\alpha}^{(\tau)}))$; $\vec \alpha_{\text{opt}} \gets  \vec \alpha^{(\tau)}$}
    }
    \Return{$\vec\alpha_{\textnormal{opt}},\normalfont{\texttt{cost}}$}\;
    \caption{\sc Gradient-based Continuous\newline Optimization for QAQC via the POTQ}
    \label{algo:GbQC}
\end{algorithm}

Our gradient-based optimization procedure is outlined in Algorithm \ref{algo:GbQC}. Given an arbitrary unitary $U$ as input, Algorithm \ref{algo:GbQC} compiles $U$ to a unitary $V_{\vec k}(\vec {\alpha}_{\text{opt}})$ of a given structure $\vec{k}$ that minimizes the cost $C_{\text{POTQ}}$. The gradient is evaluated with the POTQ circuit as a subroutine within a classical gradient-descent algorithm.
The overall query complexity in the number of calls to the cost evaluation routine of Algorithm \ref{algo:GbQC} is
$ O(NTL/\delta^2)$, where $\delta = 1/\sqrt{n_\text{shots}}$ is the sample precision, $N$ is the maximum number of repetitions over random initial parameters $\vec \alpha^0$, $L$ is the dimension of the continuous parameter space of $\vec \alpha$, and $T$ is the number of gradient descent iterations for a suitable learning rate $\eta > 0$. In order to improve convergence, it may also be useful to supply the quantum subroutines for computing the cost function and the gradient to a more advanced minimization routine, for example as found in the Python library SciPy~\cite{scipy-opt}. We present below the results on compiling both single-qubit and two-qubit gates on a simulator.

When performing Algorithm \ref{algo:GbQC}, we rely on the ability to perform the controlled-$U$ gate. The unitary $U$ may be unknown, e.g., as in Fig.~\ref{fig:applications}(b). In general, to perform a controlled operation with respect to a target unitary $U$, one can use a method for ``remote control'' \cite{ZR11}. This method employs a local $U$ gate and controlled-SWAP operations in order to realize the controlled-$U$ gate. In practice, since any controlled unitary gate can be decomposed into native gates, the ability to compile controlled-SWAP, the Toffoli gate, and the set of controlled rotations is sufficient. In order to perform such a translation, we allow the user to have access to a small-scale classical compiler. This does not incur exponential overhead since the gates to be translated are one- and two-qubit gates (or their controlled versions). While this may cause the depth of our compiled unitary to increase, it will only be by a constant factor.

We note that decoherence, gate infidelity, and readout errors on NISQ computers are all more pronounced when attempting to execute controlled unitaries. This means that there is significant performance loss for controlled unitaries, as required in the POTQ. Consequently, we did not implement our gradient-based optimization method on current quantum devices, but we speculate that improvements to quantum hardware will enable this application.

\subsubsection{Implementation on a quantum simulator}\label{sctGBimplement}

\begin{figure}
\centering
\includegraphics[width=\columnwidth]{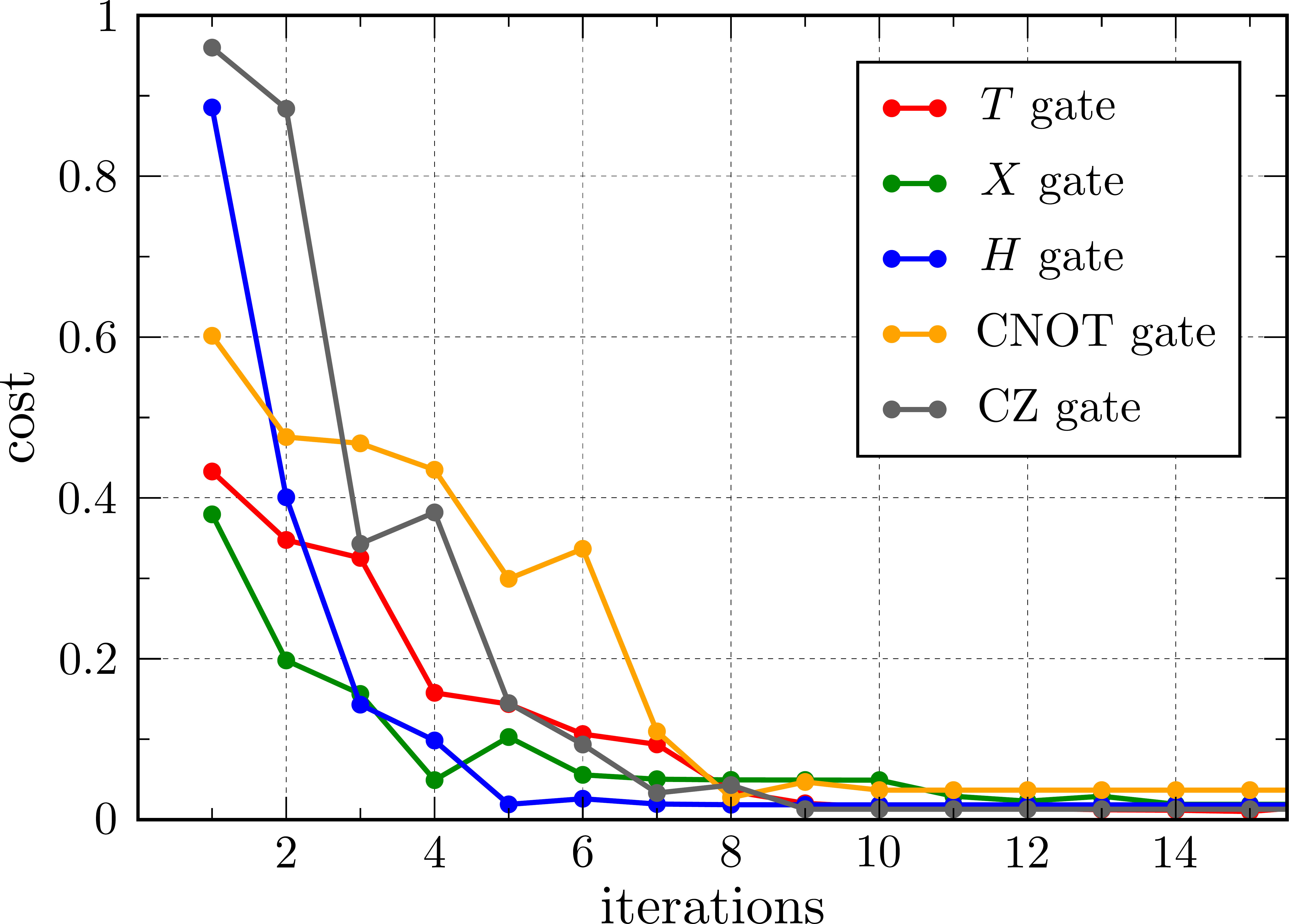}
\caption{
Compiling one- and two-qubit gates on a simulator with the gate alphabet in \eqref{eqn:used-gate-alphabet-for-ibm_full} using the gradient-based optimization technique described in Algorithm \ref{algo:GbQC}, with $n_\text{shots}=10,000$. Shown is the cost as a function of the number of gradient calls of the continuous parameter optimization using the \texttt{minimize} routine in the SciPy-optimize Python library. The gate structure for the single-qubit gates is fixed to the one shown in Fig. \ref{fig:univ_circuits}(a), while the gate structure for the two-qubit gates is fixed to the one shown in Fig. \ref{fig:univ_circuits}(b).} %Here, the cost refers to the normalized Hilbert-Schmidt distance with respect to the target gate.}
\label{fig:gradientIBM}
\end{figure}

We use IBM's simulator \cite{cross17ibm} to compile a selection of single-qubit and two-qubit gates by performing the gradient-based optimization procedure in Algorithm \ref{algo:GbQC}. In order to improve convergence, we additionally supply the gradient, as well as the cost function, to the \texttt{minimize} routine in the SciPy-optimize Python library \cite{scipy-opt}. For the single-qubit gates, we assume a fixed structure for the trainable gate sequence according to the decomposition in Fig. \ref{fig:univ_circuits}(a), while for the two-qubit gates we assume a fixed structure for the trainable gate sequence according to the decomposition in Fig. \ref{fig:univ_circuits}(b). We compile
the $T$ gate, $X$ gate, Hadamard ($H$) gate, as well as the CNOT and CZ gates, all with $n_\text{shots}=10,000$. The results are shown in Fig. \ref{fig:gradientIBM}. We note that increasing $n_{\text{shots}}$ to higher orders of magnitude significantly reduces the sampling error and results in more stable convergence at the cost of an increase in runtime.

\subsection{Gradient-based optimization via the HST and LHST}\label{sctHST_grad}

We now show that it is possible to perform gradient-based optimization of the original cost function $C_{\text{HST}}$ and its local variant $C_{\text{LHST}}$. This allows us to perform gradient-based optimization of the general cost function $C_q=qC_{\text{HST}}+(1-q)C_{\text{LHST}}$. The algorithm for gradient-based optimization of $C_{\text{HST}}$ and $C_{\text{LHST}}$ is presented in Algorithm \ref{algo:GbQC_2}. %Note that with this algorithm, we obtain an $\varepsilon$-approximate compilation of $U$ with $\varepsilon$ given by the formula in Eq.~\eqref{eqnEpsilonCq}.

%For a given unitary $U$ to be compiled and a trainable unitary $V_{\vec{k}}(\vec{\alpha})$ with a fixed structure parameter $\vec{k}$, the gradient with respect to $\vec{\alpha}$ of both $C_{\text{GF}}(U,V_{\vec{k}}(\vec{\alpha}))$ and $C_{\text{local}}(U,V_{\vec{k}}(\vec{\alpha})$ can be computed using the HST and the local HST, respectively. 

The gradient with respect to $\vec{\alpha}$ of both $C_{\text{HST}}(U,V_{\vec{k}}(\vec{\alpha}))$ and $C_{\text{LHST}}(U,V_{\vec{k}}(\vec{\alpha}))$ can be computed using the HST and the LHST, respectively. Specifically, for a gate sequence of the form in \eqref{eq:trainable_gate_seq}, in which the only parameterized gates are the single-qubit rotation gates, we have that
\begin{equation}\label{eq:HST_gradient}
    \begin{aligned}
    \frac{\partial}{\partial\alpha_{\ell}}C_{\text{HST}}(U,V_{\vec{k}}(\vec{\alpha}))&=\frac{1}{2}C_{\text{HST}}(U,\widehat{V}_{\vec{k},+}^{(\ell)}(\vec{\alpha}))\\
    &\quad -\frac{1}{2}C_{\text{HST}}(U,\widehat{V}_{\vec{k},-}^{(\ell)}(\vec{\alpha})),
    \end{aligned}
\end{equation}
and
\begin{equation}\label{eq:local_HST_gradient}
    \begin{aligned}
    \frac{\partial}{\partial\alpha_{\ell}}C_{\text{LHST}}^{(j)}(U,V_{\vec{k}}(\vec{\alpha}))&=\frac{1}{2}C_{\text{LHST}}^{(j)}(U,\widehat{V}_{\vec{k},+}^{(\ell)}(\vec{\alpha}))\\
    &\quad -\frac{1}{2}C_{\text{LHST}}^{(j)}(U,\widehat{V}_{\vec{k},-}^{(\ell)}(\vec{\alpha}))
    \end{aligned}
\end{equation}
for all $j\in\{1,2,\dotsc,n\}$. Here,
\begin{equation}
    \begin{aligned}
    \widehat{V}_{\vec{k},\pm}^{(\ell)}(\vec{\alpha})&\coloneqq G_{k_L}(\alpha_L)\dotsb G_{k_{\ell+1}}(\alpha_{\ell+1})G_{k_\ell}\left(\pm\frac{\pi}{2}\right)\\
    &\quad\quad\times G_{k_\ell}(\alpha_\ell)G_{k_{\ell-1}}(\alpha_{\ell-1})\dotsb G_{k_1}(\alpha_1)
    \end{aligned}
\end{equation}
is the original gate sequence $V_{\vec{k}}(\vec{\alpha})$ with an additional rotation gate $G_{k_\ell}\left(\pm\frac{\pi}{2}\right)$ corresponding to the variable with respect to which the derivative is taken. In other words, to compute the gradient of the cost function $C_{\text{HST}}(U,V_{\vec{k}}(\vec{\alpha}))$, we run the HST in Fig. \ref{fig:hilbert-schmidt-inner-product-circuit}(a) twice, once with the gate sequence $\widehat{V}_{\vec{k},+}^{(\ell)}(\vec{\alpha})$ and once with the gate sequence $\widehat{V}_{\vec{k},-}^{(\ell)}(\vec{\alpha})$. Similarly, to compute the gradient of the functions $C_{\text{LHST}}^{(j)}(U,V_{\vec{k}}(\vec{\alpha}))$, we run the LHST in Fig. \ref{fig:hilbert-schmidt-inner-product-circuit}(b) twice, once with the gate sequence $\widehat{V}_{\vec{k},+}^{(\ell)}(\vec{\alpha})$ and once with the gate sequence $\widehat{V}_{\vec{k},-}^{(\ell)}(\vec{\alpha})$.

The expressions for the gradient in \eqref{eq:HST_gradient} and \eqref{eq:local_HST_gradient} can be verified by recalling that only the one-qubit gates need to be parameterized and that they can always be assumed to have the form $e^{-i\alpha\sigma/2}$ for some Pauli operator $\sigma$, where $\alpha$ is the continuous parameter specifying the gate. Then, for the gate sequence $V_{\vec{k}}(\vec{\alpha})$ in \eqref{eq:trainable_gate_seq}, we get
\begin{align}
    \frac{\partial V_{\vec{k}}(\vec{\alpha})}{\partial\alpha_\ell}&=G_{k_L}(\alpha_L)\dotsb G_{k_{\ell+1}}(\alpha_{\ell+1})\frac{\partial G_{k_{\ell}}(\alpha_{\ell})}{\partial\alpha_\ell}\nonumber\\
    &\qquad\times G_{k_{\ell-1}}(\alpha_{\ell-1})\dotsb G_{k_1}(\alpha_1)\\
    &=-\frac{i}{2}G_{k_\ell}(\alpha_\ell)\dotsb G_{k_{\ell+1}}(\alpha_{\ell+1})\sigma_{k_\ell}G_{k_\ell}(\alpha_\ell)\nonumber\\
    &\qquad\times G_{k_{\ell-1}}(\alpha_{\ell-1})\dotsb G_{k_1}(\alpha_1)\label{eq-HST_grad_pf1}
\end{align}
Then, we use the identity
\begin{equation}\label{eq-HST_grad_pf2}
    \begin{aligned}
    i[\sigma_{k_\ell},\rho]&=G_{k_\ell}\left(-\frac{\pi}{2}\right)\rho G_{k_\ell}\left(-\frac{\pi}{2}\right)^\dagger \\
    &\qquad\qquad\qquad -G_{k_\ell}\left(\frac{\pi}{2}\right)\rho G_{k_\ell}\left(\frac{\pi}{2}\right)^\dagger,
    \end{aligned}
\end{equation}
which holds for any state $\rho$. We also observe that both the functions $C_{\text{HST}}(U,V_{\vec{k}}(\vec{\alpha}))$ and $C_{\text{LHST}}^{(j)}(U,V_{\vec{k}}(\vec{\alpha}))$ are of the form
\begin{equation}
    \begin{aligned}
    F(\vec{\alpha})=\Tr[H(U\otimes V_{\vec{k}}(\vec{\alpha})^*)\rho(U^\dagger\otimes V_{\vec{k}}(\vec{\alpha})^T)],
    \end{aligned}
\end{equation}
where $\rho=\ket{\Phi^+}\bra{\Phi^+}_{A_1\dotsb A_n}$ for both functions, $H=\ket{\Phi^+}\bra{\Phi^+}_{A_1\dotsb A_n}$ for $C_{\text{HST}}(U,V_{\vec{k}}(\vec{\alpha}))$, and $H=\ket{\Phi^+}\bra{\Phi^+}_{A_jB_j}\otimes\id_{\bar{A_j}\bar{B_j}}$ for $C_{\text{LHST}}^{(j)}(U,V_{\vec{k}}(\vec{\alpha}))$. Finally, using
\begin{align}
    &\frac{\partial F(\vec{\alpha})}{\partial\alpha_\ell}=\Tr\left[H\left(U\otimes\left(\frac{V_{\vec{k}}(\vec{\alpha})}{\partial\alpha_\ell}\right)^*\right)\rho(U^\dagger\otimes V_{\vec{k}}(\vec{\alpha})^T)\right]\nonumber\\
    &+\Tr\left[H(U\otimes V_{\vec{k}}(\vec{\alpha})^*)\rho\left(U^\dagger\otimes\left(\frac{\partial V_{\vec{k}}(\vec{\alpha})}{\partial\alpha_\ell}\right)^T\right)\right],
\end{align}
substituting \eqref{eq-HST_grad_pf1} into this expression, and using \eqref{eq-HST_grad_pf2} to simplify, we obtain \eqref{eq:HST_gradient} and \eqref{eq:local_HST_gradient}.

\begin{algorithm}
    \DontPrintSemicolon 
    \KwIn{
    Unitary $U$ to be compiled; a trainable unitary $V_{\vec k}(\vec \alpha)$ of a given structure,
    where $\vec \alpha$ is a continuous
    circuit parameter of dimension $L$; maximum number of iterations $N$; gradient tolerance $\varepsilon'\in (0,1)$; sample precision $\delta >0$; cost function $C\in\{C_{\text{HST}},C_{\text{LHST}}\}$.}
    \KwOut{Parameters $\vec{\alpha}_{\text{opt}}$ such that at best $\norm{\nabla_{\vec{\alpha}}C(U,V_{\vec{k}}(\vec{\alpha_{\text{opt}}}))}^2\leq\varepsilon'$.}
    \kwInit{$\vec{\alpha}_{\text{opt}} \gets 0$; $\texttt{cost}\gets 0$; $\texttt{grad} \gets \infty$; $\tau~\gets~0$; $\texttt{gradCount}\gets 0$; $\eta\gets 1$}
    choose initial parameters $\vec {\alpha}^{(0)}$ at random\;
    $\texttt{cost}\gets C(U,V_{\vec{k}}(\vec{\alpha}^{(0)}))$\;
    
    \While{$\normalfont{\texttt{count}}<N$ and $\normalfont{\texttt{gradCount}}<4$}{
        $\tau \gets \tau+1$\;
        \For{$i=1,2,\dotsc,L$}{
            Calculate $\frac{\partial C}{\partial\alpha_i}$ using either \eqref{eq:HST_gradient} or \eqref{eq:local_HST_gradient}, taking approximately $\frac{1}{\delta^2}$ samples for each circuit.}
        
        $\texttt{grad}\gets\norm{\nabla_{\vec{\alpha}}C(U,V_{\vec{k}}(\vec{\alpha}^{(\tau-1)}))}^2$\;
        
        \If{$\normalfont{\texttt{grad}}\leq\varepsilon'$}{$\texttt{gradCount}\gets\texttt{gradCount}+1$}
        
        $\vec{\alpha}_1^{(\tau-1)}\gets\vec{\alpha}^{(\tau-1)}-\eta\nabla_{\vec{\alpha}}C(U,V_{\vec{k}}(\vec{\alpha}^{(\tau-1)}))$\;
        
        $\vec{\alpha}_2^{(\tau-1)}\gets\vec{\alpha}_1^{(\tau-1)}-\eta\nabla_{\vec{\alpha}}C(U,V_{\vec{k}}(\vec{\alpha}^{(\tau-1)}))$\;
        
        \If{$\normalfont{\texttt{cost}}-C(U,V_{\vec{k}}(\vec{\alpha}_2^{(\tau-1)}))\geq \eta\cdot\normalfont{\texttt{grad}}$}{$\eta\gets 2\eta$\; $\alpha^{(\tau)}\gets\alpha_2^{(\tau-1)}$}
        
        \ElseIf{$\normalfont{\texttt{cost}}-C(U,V_{\vec{k}}(\vec{\alpha}_1^{(\tau-1)}))<\frac{\eta}{2}\cdot\normalfont{\texttt{grad}}$}{$\eta\gets\frac{\eta}{2}$\;$\alpha^{(\tau)}\gets\alpha_1^{(\tau-1)}$}
        \Else{$\alpha^{(\tau)}\gets\alpha_1^{(\tau-1)}$}
        
        $\texttt{cost}\gets C(U,V_{\vec{k}}(\vec{\alpha}^{(\tau)}))$\;
        
        $\vec{\alpha}_{\text{opt}}\gets \vec{\alpha}^{(\tau)}$
    }
    
    \Return{$\vec\alpha_{\textnormal{opt}},\normalfont{\texttt{cost}}$}\;
    \caption{\sc Gradient-based Continuous\newline Optimization for QAQC via the HST and LHST}
    \label{algo:GbQC_2}
\end{algorithm}

%\section{Numerical study with a fixed input state}

%\subsection{Ansatz-based approach}

\end{document}